\documentclass[aps,twocolumn,nofootinbib]{revtex4}
\usepackage{graphicx} 
\usepackage{caption,subcaption}
\usepackage{amsmath}
\usepackage{csquotes}
\usepackage{comment}
\usepackage{bbm}
\usepackage[version=4]{mhchem}
\usepackage{tikz-cd} 
\usepackage{amssymb}

\usepackage{physics}
\usepackage{xcolor}
\usepackage{bm}
\usepackage{mathtools}
\usepackage[hidelinks]{hyperref}
\usepackage{algorithm}
\usepackage{algpseudocode}
\usepackage{xcolor}
\usepackage{xfrac}
\usepackage{bm}
\usepackage[T1]{fontenc}
\usepackage{upquote}

\usepackage[utf8]{inputenc}
\usepackage[english]{babel}

\begin{document}

\title{\bf Action functional gradient descent algorithm for estimating escape paths in stochastic chemical reaction networks}

\author{Praful Gagrani$^{1,2}$, Eric Smith$^{1,3,4,5,6}$}


\affiliation{$^1$Department of Physics, University of Wisconsin–Madison, Madison, WI 53706, USA\\
$^2$Wisconsin Institute for Discovery, University of Wisconsin-Madison, Madison, WI 53715, USA\\
$^3$ Department of Biology, Georgia Institute of Technology, Atlanta, GA 30332, USA\\
$^4$ Earth-Life Science Institute, Tokyo Institute of Technology, Tokyo 152-8550, Japan\\
$^5$ Santa Fe Institute, 1399 Hyde Park Road, Santa Fe, NM 87501, USA\\
$^6$ Ronin Institute, 127 Haddon Place, Montclair, NJ 07043, USA}
\date{\today}
\begin{abstract}
We first derive the Hamilton-Jacobi theory underlying continuous-time Markov processes, and then use the construction to develop a variational algorithm for estimating escape (least improbable or first passage) paths for a generic stochastic chemical reaction network that exhibits multiple fixed points. The design of our algorithm is such that it is independent of the underlying dimensionality of the system, the discretization control parameters are updated towards the continuum limit, and there is an easy-to-calculate measure for the correctness of its solution. We consider several applications of the algorithm and verify them against computationally expensive means such as the shooting method and stochastic simulation. While we employ theoretical techniques from mathematical physics, numerical optimization and chemical reaction network theory, we hope that our work finds practical applications with an inter-disciplinary audience including chemists, biologists, optimal control theorists and game theorists.  
\end{abstract}


\maketitle
\begingroup
  \hypersetup{hidelinks}
  \tableofcontents
\endgroup

\section{Introduction}
Stochastic modelling has played an increasingly central role in science since the advent of high speed computing. There are, however, many categories of events that are vital to the organization of long-term dynamics for which the rate diminishes exponentially with system size (`rare' events) and computer simulations become unaffordable. A typical approach in these scenarios is to employ importance sampling approaches in the stochastic simulation to efficiently simulate the rare event of interest (see \cite{hartmann2012efficient,cao2013adaptively,biondini2015introduction}). In this work, for the particular case of stochastic chemical reaction networks, we provide a deterministic alternative for estimating the likelihood of rare events.    

Stochastic nonlinear dynamical processes exhibiting multistability, or multiple coexisting attractors, have found several scientific applications ranging from modelling climate change to population biology and the origin of life \cite{benzi1983theory,dyson1982model,smith2015symmetry,smith2016origin}, and are an active subject of interdisciplinary research. A question of critical importance for any practical application of such a system is, how often does it transition out of a stable attractor and what are the least-improbable paths by which such a transition occurs? In literature, the dynamics that arise due to the system's transitioning from one stable attractor to another and the optimal path of transition is referred to as `switching dynamics' and `escape path', respectively.

Many phenomena in biology, chemistry, physics or engineering can be modelled as a chemical reaction network (CRN). It is also well known that CRNs can exhibit a host of dynamics, including limit cycles and multistability \cite{yu2018mathematical}. Perhaps less well-known is the role of Hamilton-Jacobi ray theory underlying stochastic CRN, and the interpretation of escape paths as particular characteristic curves that arise as a solution to the associated Hamilton-Jacobi equation. In our work, we review the necessary formalism and employ techniques from mathematical physics and numerical optimization to estimate the escape paths for a multistable CRN. More precisely, we first recognize that the escape paths are variational (locally least-action) solutions to the action functional, and subsequently use the functional to perform a gradient descent that converges on the desired path. 

It must be noted that neither the recognition of the escape path as a variational solution nor using the action functional to perform gradient descent is novel to us (for instance, see \cite{berne1998classical}), however, we differ from the earlier work in a few ways. In the past, the use of Hamilton-Jacobi theory to find the escape path for CRN has only been made using the `shooting-method', which relies on integrating Hamilton's equations of motion, rather than a functional gradient descent approach that we present here (see \cite{dykman1994large}). On the other hand, while the action functional has been used for a gradient descent, to our best knowledge, the form of the Hamiltonian assumed has always been separable in momentum and position (see \cite{weinan2004minimum,bouchet2019rare}). This form of the Hamiltonian, with other constraints on the functional form, makes it amenable to finding an analytic form for the associated Lagrangian which drastically simplifies the gradient descent. While this assumption is typical for a Hamiltonian describing mechanical energy in physics, it is far too restrictive for a generic Hamiltonian in chemistry, as can be seen below: 
\begin{align}
    H_\text{ME}(p,q) 
    &= 
    K(p) + U(q), \nonumber\\
        H_\text{CRN}(p,q)
    &= 
    \sum_{y_{\alpha},y_{\beta}}\big( e^{(y_{\beta}-y_{\alpha})\cdot p} - 1\big) k_{y_{\alpha}\to y_{\beta}}q^{y_{\alpha}}, \label{eq:H_ME}
\end{align}
where $p$ and $q$ are the momentum and position coordinates, $K(p)$ and $U(q)$ are the kinetic and potential energy, and $H_\text{ME}$ and $H_\text{CRN}$ (Eq.\ \ref{eq:Ham_CRN_rxn}) are the Hamiltonians for mechanical energy and chemical reaction networks respectively. In our algorithm, however, we start from the Hamiltonian, numerically solve for the Lagrangian at each iteration and use it to calculate the descent direction. While our algorithm is designed for stochastic CRNs, it is amenable to generalization for other classes of stochastic Hamiltonian dynamical systems.

The layout of the paper is as follows. In Section \ref{sec:theory}, we derive the Hamilton-Jacobi theory for stochastic processes starting from a master equation and apply it to stochastic CRN. We show that the Hamilton-Jacobi theory of the non-equilibrium potential (NEP) arises naturally as a result of a variational principle applied to the \textit{action functional}. In Section \ref{sec:Algortithm}, we propose an \textit{action functional gradient descent} (AFGD) algorithm that finds the variational solution in the space of paths constrained at end points for a given Hamiltonian value constraint. We also explain how the algorithm can be used to find least-improbable escape paths out of a stable attractor and assign a value to the NEP along them. The details of the implementation are provided in Appendix \ref{app:Details_AFGD}, and a MATLAB implementation is made available at \cite{Gagrani_AFGD-for-CRN-escapes_2022}. In Section \ref{sec:Results}, we demonstrate the applications of the AFGD algorithm on several CRNs. We first consider the Selkov model, and compare the result of the algorithm against the escape trajectory found by the `shooting-method' (Figure \ref{fig:Selkov_NEP}). We then define a class of high dimensional birth-death models, namely `N-Schl{\"{o}}gl model', and compare the results of the algorithm against a stochastic simulation for the 2-Schl{\"{o}}gl model (Figure \ref{fig:2-Schlogl_Gillespie}). We also use the algorithm on the six dimensional 6-Schl{\"{o}}gl model, and compare the result against the integration of Hamilton's equations of motion (Figure \ref{fig:6-Schlogl_HamEoM}). Finally, in Section \ref{sec:FutureResearch}, we conclude with a discussion of our contribution and potential avenues of future research.

\section{From master equation to Hamilton-Jacobi equation for CRN}

\label{sec:theory}
The application of Hamilton-Jacobi theory to chemical reaction networks (CRNs) has a long history \cite{gang1987stationary,dykman1994large}. In this section, we review the necessary formulation needed to understand the switching dynamics of stochastic chemical reaction networks (Section \ref{sec:switching}) and to devise a variational algorithm for predicting the transitions (Section \ref{sec:Algortithm}). In Section \ref{sec:HJ_theory} we derive the non-equilibrium potential for continuous time Markov population processes and explain its role in estimating the probability of stochastic events. In Section \ref{sec:Ham_CRN}, we derive the Hamiltonian for a CRN and prove that the NEP is a Lyapunov function along the deterministic trajectories under mass-action kinetics. In Appendix \ref{app:App_nontech}, we investigate the relevance our work might have to a stochastic modelling practitioner by posing a general practical problem, giving an overall picture of the solution and explaining where our algorithm fits in. In Appendix \ref{sec:rederiv_HJ}, we employ the Hamilton-Jacobi formalism to recover the ACK theorem and NEP for complex-balanced systems. In Appendix \ref{app:App_A}, we define a `non-equilibrium action', of which both the master equation and Schr\"{o}dinger equation can be seen as a variational solution, derive the path integral formula and action functional for stochastic population processes, and calculate the first and second variational derivatives of the action functional. There are many resources that provide an alternative treatment of the same subject, for instance see \cite{snarski2021hamilton,smith2020intrinsic} and references therein.

\subsection{Hamilton-Jacobi equation and Non-Equilibrium Potential (NEP) for stochastic dynamics}
\label{sec:HJ_theory}
A continuous time Markovian population process is specified by its \textit{master equation}, 

\begin{align}
    \frac{\partial}{\partial t} \rho(n,t) 
    &=
    \sum_{n'} \mathbb{T}_{nn'}\rho(n',t) \nonumber\\
    \sum_n \rho(n,t) 
    &= 
    1 \hspace{1em} \text{for all }  \hspace{0.25em} t \nonumber\\
    \sum_n \mathbb{T}_{nn'} 
    &= 
    0 \hspace{1em} \text{for all } \hspace{0.25em} n' 
    \label{eq:MasterEquation}
\end{align}

where $t$ is the time variable, $n$ is a $D-$dimensional discrete vector in the positive integer lattice $\mathbb{Z}_{\geq 0}^D$ denoting a position in the state space (to simplify notation we drop the arrow in $\vec{n}$) and $\rho(n,t)$ is a time-evolving probability distribution function (PDF) over the state space. $\mathbb{T}$ is referred to as a transition operator, and in this paper we only consider time independent transition operators. 

For what follows, it is useful to recast Eq.\ \ref{eq:MasterEquation} into the following form and define a Hamiltonian operator $\hat{\underbar{H}}$ that acts on the PDF $\rho$,
\begin{align}
    \frac{\partial}{\partial t} \rho(n,t) 
    &=
    \hat{\underbar{H}}\left(-\pdv{}{n},n\right)\rho(n,t) \label{eq:TimeEvoln}
\end{align}
where $\partial/\partial n$ is the infinitesimal-shift operator\footnote{
A function in $x$ can be shifted by $y$ by the application of a \textit{shift operator} given by $e^{y\pdv{}{x}}$, as can be seen by
\begin{align*}
    f(x+y) 
    &= \sum_{n=0}^\infty \frac{y^n}{n!}\frac{\partial^n f(x)}{\partial x^n}
    = e^{y\pdv{}{x}}f(x).
\end{align*}
\label{footnote:shift}}. 

The resemblance of Eq.\ \ref{eq:TimeEvoln} to the Schr\"{o}dinger equation is not a mere coincidence, and we show in Section \ref{app:NEA} how they can both be derived from a common variational problem of extremizing, what is termed as, the `non-equilibrium action' (NEA) in \cite{eyink1996action}. It should come as no surprise then that we can employ the same machinery developed in mathematical physics for quantum mechanics to derive results about stochastic dynamics. This line of reasoning has a long history and we refer the readers to \cite{doi1976second,baez2012quantum,smith2015symmetry,smith2019information} for a `second-quantization' treatment via abstract linear algebra.

We can integrate Eq.\ \ref{eq:TimeEvoln} starting from a PDF $\rho(n_0,0)$ at time $0$ to get a distribution $\rho(n_T,T)$ at time $T$, indexed by $n_0$ and $n_T$ respectively. 
\begin{align}
    \rho(n_T,T) 
    &= 
    e^{\int_0^T \hat{\underbar{H}} \,dt} \rho(n_0,0) \nonumber \\
    &= \int \left[\,d n \right] \int \left[\,d p \right] e^{-\mathcal{A}[n(t),p(t)]}\rho(n_0,0)
    \label{eq:PathIntegral}
\end{align}
where $\mathcal{A}$ is the \textit{action functional}
\begin{align}
    \mathcal{A}\left[n(t),p(t)\right] 
    &=
    \int_0^T \left[ p\cdot \dv{n}{t} - \underbar{H}(p,n)\right]\,dt. \label{eq:Ham_np}
\end{align}
and $p$ is a momentum variable canonically conjugate to the position variable $n$. The second line in Eq.\ \ref{eq:PathIntegral} is the path-integral formula, a proof of which can be found in Section \ref{app:PathIntegral}, and  $\left[\,d n \right] \left[\,d p \right]$ is the path-integral measure. Note that in Eq.\ \ref{eq:Ham_np}, the Hamiltonian function $\underbar{H}(p,n)$ has the same functional form as the Hamiltonian operator $\hat{\underbar{H}}$ except the operator $-\partial/\partial n$ is replaced by the variable $p$.

In the first line of Eq.\ \ref{eq:PathIntegral}, the exponentiation of the Hamiltonian operator is to be understood as a time-ordered matrix product over small intervals $\,dt$. Its role is to time-evolve the initial distribution $\rho(0)$ and accumulate probability through all possible chains of states until time $T$ so as to obtain a new distribution $\rho(T)$. In the second line, upon taking appropriate limits signifying a continuous time parameter, we obtain another description of the same process where we sum over all the paths starting at $n_0$ at time $0$ and ending at $n_T$ at time weighted with an appropriate measure. The negative log of the measure of a particular path is given by the action functional (in other words, $\mathcal{A}$ takes as input a path and returns its log-improbability), and integrating under the path-integral measure amounts to summing over all paths. For a didactic introduction to the topic and its application to stochastic dynamics, we refer the reader to \cite{baez2012quantum}.       

It is useful to pass from a discrete to a continuous state space by descaling $n$ with a scale factor $V$ and considering the large $V$ limit. In order to align our presentation here with the mathematical physics literature, we denote the continuous variable by $q=n/V$ and refer to the continuous state space as configuration space. There are several interpretations the variables can take; the one of interest to us is where $n$ is a population vector (thus $n\geq 0$), $q$ is the concentration vector and $V$ is a scale of the the total population. Following Eq.\ \ref{eq:PathIntegral}, we can write the time evolution of a distribution in the transformed coordinate $q$ as   
\begin{align}
    &\rho(q,T) 
    = \int \left[\,d q \right] \int \left[\,d p \right] e^{-V\mathcal{A}\left[q(t),p(t)\right]}\rho(q(0),0) \nonumber\\
    &   \mathcal{A}\left[q(t),p(t)\right] 
    =
    \int_0^T \left[ p\cdot \dv{q}{t} - H(p,q)\right]\,dt, \label{eq:PathIntegral_q_coord}
\end{align}
where $\mathcal{A}$ is the action functional in the new coordinates and has a different Hamiltonian $H(p,q)$ which relates to $\underbar{H}(p,n)$ by \footnote{We say $f(V)\asymp g(V)$ (read as $f(V)$ is asymptotic to $g(V)$) if 
\begin{align*}
    \lim_{V\to\infty}\frac{f(V)}{g(V)}&=1.
\end{align*}}
\begin{align}
       V H(p,q) &\asymp \underbar{H}(p,qV)= \underbar{H}(p,n). \label{eq:Extensivity_H}
\end{align}
It must be noted that Eq.\ \ref{eq:Extensivity_H} is an additional constraint on the form of the Hamiltonian, but is one that is satisfied by the CRN Hamiltonian which we are interested in for the scope of this paper. Moreover, if the asymptotic equality was an equality then the condition is often referred to by saying `the Hamiltonian is extensive in the position coordinate', or `the Hamiltonian is a homogenous function of degree one in the position coordinate'. We will return to this point in the next subsection, once we write the explicit Hamiltonian for a CRN. 

Together the position and momentum coordinates $(q,p)$ constitute a point in the \textit{phase space} \cite{arnol2013mathematical}. In order to evaluate the distribution at time $T$ using Eq.\ \ref{eq:PathIntegral_q_coord}, in principle we need to sum over all the phase space paths that end at $q$ at time $T$. Evaluating the full sum, however, is not necessary to obtain the leading large-$V$ asymptotics. Through Laplace's or saddle-point method, for large $V$ the path integral is dominated by the saddle point of the functional and the distribution at time $T$ can be approximated by
\begin{align}
        \rho(q,T) & 
     \asymp   N(V)e^{-V\mathcal{A}\left[q^*(t),p^*(t)\right]} \rho(q^*(0),0) \nonumber\\
    \text{where }&(q^*(t),p^*(t)) \text{ is such that } \nonumber \\
    &\delta\mathcal{A}\left[q^*(t),p^*(t)\right] 
    = 
    0 \text{ and } q^*(T)= q, \label{eq:StationaryAction}
\end{align}
  and $N(V)$ is a normalization factor expounded upon in the subsequent paragraph. The constraint in the last line specifies the stationarity or optimality condition on the saddle point path $(q^*(t),p^*(t))$, also referred to as the path of stationary action, and we henceforth refer to it simply as the \textit{optimal path} (for a MinMax formulation of the saddle-point see Section \ref{sec:MinMax}). We require here that the Hamiltonian function is convex in the momentum variable $p$, which we show is the case for CRN Hamiltonians with mass-action kinetics in Section \ref{sec:Ham_CRN}.

There are two caveats to Eq.\ \ref{eq:StationaryAction} that we now point out. First, the normalization factor $N(V)$ in the first line depends not only on the scale factor $V$ but also on the optimal path. There is an in-principle method, although costly in practice, to obtain $N(V)$ from the action functional itself, outlined in \cite{kirsten2003functional}, Ch. 7 of \cite{coleman1988aspects} and Ch. 4 of \cite{smith2015symmetry} (for an application, see \cite{manikandan2017asymptotics}). Second, since the path integral is a sum over all paths, there are several stationary paths starting from different $q^*(0)$ that reach $q$ at $T$, and one must perform a sum over all of them. This becomes increasingly relevant in the limit $T\to \infty$, where the system can bounce back and forth multiple times between a $q^*(0)$ and $q$ (see \cite{smith2015symmetry,coleman1988aspects}). There is, however, a two-fold resolution to this problem. First, the normalization factor $N(V)$ is sub-exponential in $V$, and can be well-approximated by unity large for $V$ (also see the introduction to Section \ref{sec:Algortithm}). Second, the initial distribution $\rho(q,0)$ is typically taken to be peaked at a given value, thus yielding negligible contribution from every $q^*(0)$ that is not near the peak of the initial condition. Thus we only pick the least-improbable direct paths that start from near the peak of the initial distribution and reach $q$ in Eq.\ \ref{eq:StationaryAction}.

In the first line of Eq.\ \ref{eq:StationaryAction}, the only free variables are configuration $q$ and final time $T$, which together characterize a time evolving distribution. The optimality condition picks a path $(q^*(t),p^*(t))$, and it is only the probability mass at $q^*(0)$ in the initial distribution at time $0$ i.e. $\rho(q^*(0),0)$, weighted appropriately by the action functional, that contributes to the final distribution at configuration $q$ and time $T$. The weight factor $e^{-V\mathcal{A}[q^*,p^*]}$ accounts for the fact that not all the probability mass from $q(0)$ goes to $q$ in time $T$, and in this sense takes the interpretation of a conditional probability.    

Since the conditional probability term is only in the exponential, it is useful to write the probability distribution itself as an exponential function $\rho(q,t)=e^{-V S(q,t)}$. Here $S(q,t)$ is the \textit{action function} (not be confused with $\mathcal{A}$ which is the action functional) and its time evolution is given as follows 
\begin{align}
    & S(q,T) 
    = 
    \mathcal{A}\left[q^*(t),p^*(t)\right] + S(q^*(0),0) \nonumber\\
    &= 
    \int_0^T \left[ p^*\cdot \dv{q^*}{t} - H(p^*,q^*)\right]\,dt + S(q^*(0),0)\label{eq:Action_func_time_evol}
\end{align}
where in the second line, we expand the action functional using Eq.\ \ref{eq:PathIntegral_q_coord}. Thus the action function evolves by the optimal value of the action functional, and since it is a function we can also consider its total differential
\begin{align}
    & S(q,T) 
    = \int_0^T \,dS + S(q^*(0),0) \nonumber \\
    &= 
    \int_0^T \left[ \pdv{S(q^*,t)}{q^*}\cdot \dv{q^*}{t} + \pdv{S(q^*,t)}{t}\right]\,dt + S(q^*(0),0). \label{eq:diff_Action}
\end{align}

Comparing equations \ref{eq:Action_func_time_evol} and \ref{eq:diff_Action}, we get  
\begin{align}
    \pdv{S(q^*,t)}{q^*} 
    &=
    p^* \nonumber\\
    \pdv{S(q^*,t)}{t}
    &=
    -H(p^*,q^*)  \nonumber \\
    \pdv{S(q^*,t)}{t}
    &=
    -H\left( \pdv{S}{q^*},q^* \right)
    \label{eq:HamiltonJacobiEqns}
\end{align}
where the first two lines are obtained by comparison and the last line is obtained by substituting the first line in the second line. The non-linear PDE in the last equation is called the \textbf{Hamilton-Jacobi equation}. For an introduction to the subject and its history, we refer the readers to \cite{goldstein2002classical,arnol2013mathematical} (what we call as the action function is also called Hamilton's principal function). We briefly remark that the definite integral of the action functional along the optimal path $\int_0^T \,d S$ also acts as a divergence function in information geometry (see \cite{leok2017connecting,smith2019information}) and is denoted as $S(q^*(T)||q^*(0))$ yielding the relation $S(q^*_T) = S(q^*_T||q^*_0) + S(q^*_0)$, as mentioned in Eq.\ \ref{eq:S_div}. 

The role of \textit{momentum} for stochastic systems can be best understood by the Hamilton Jacobi equation. The first line in Eq.\ \ref{eq:HamiltonJacobiEqns} shows that momentum is the gradient of the action function, which in turn is descaled negative log probability of the time evolving distribution given some initial conditions. Using this, we see that $p^*(0)$ must be the gradient of the action function $S$ at $q^*(0)$, and that an optimal path in phase space corresponds to a contour in the probability distribution arising by scaling and exponentiating the action.  

The third line in Eq.\ \ref{eq:MasterEquation} ensures that the transition operator always has a cokernel, namely the row vector consisting of all ones $[1,\ldots,1]$. The Perron-Frobenius theorem for stochastic matrices proves that the operator also has a unique right kernel, referred to as the stationary distribution which we denote here by $\pi$. $\pi$ thus satisfies $\sum_n \mathbb{T}_{n'n}\pi(n,t)=0$ or equivalently
\begin{align*}
    \hat{\underbar{H}}\pi(n,t) &= 0. 
\end{align*}
From Eqs.\ \ref{eq:MasterEquation}, \ref{eq:TimeEvoln} one can see that the stationary distribution is indeed time independent as $\partial \pi(n,t)/\partial t=0$, thus justifying its name. 

The descaled log-improbability or action function of the stationary distribution is commonly referred to as the \textbf{non-equilibrium potential (NEP)} and denoted by $\mathcal{V}$ (see \cite{anderson2015lyapunov}),
\begin{align}
    \pi(n) \asymp & \phantom{.} e^{-V \mathcal{V}(q)},\label{eq:NEP_1}
\end{align}
where the normalization constant has been set to unity following the discussion below Eq.\ \ref{eq:StationaryAction}. As the action of a stationary distribution, it must satisfy
\begin{align}
    \pdv{\mathcal{V}}{t} 
    &=
    0,  \nonumber \\
    H\left( \pdv{\mathcal{V}}{q},q\right) &= 0 \label{eq:HJ_NEP}
\end{align}
where the second line is the Hamilton-Jacobi equation from Eq.\ \ref{eq:HamiltonJacobiEqns} that must be satisfied by the NEP.

In the remainder of this section, we will develop and demonstrate the applications of Hamilton-Jacobi theory for CRNs. Finding the NEP $\mathcal{V}$ for a general CRN is of importance for getting any numerical estimates on its behavior, and it is the task that we will concern ourselves with in this paper starting Section \ref{sec:Algortithm}. As explained above, to calculate the difference of the action function or NEP between two points we first need to find the optimal path connecting them, which is precisely what the Action Functional Gradient Descent (AFGD) algorithm is designed to do.

\subsection{Hamilton-Jacobi theory for stochastic Chemical Reaction Networks (CRN)}
\label{sec:Ham_CRN}

In this subsection, we start by introducing the mathematical formulation of CRNs and derive its Hamiltonian function in the concentration coordinate. Next, we show that the Hamiltonian function is convex in the momentum coordinate, a necessary condition for the existence of optimal paths, and write the equations that the optimal paths must satisfy. Then we classify the optimal paths constituting the stationary distribution into two categories, namely relaxation paths and escape paths, and show that the former yield the deterministic mass-action kinetics (MAK) and the latter yields the non-equilibrium potential (NEP). Finally we show that, for a general CRN with multiple fixed points, the NEP is always a Lyapunov function with respect to MAK. 

A CRN is defined by the triple $\{\mathcal{S},\mathcal{C},\mathcal{R}\}$, where $\mathcal{S},\mathcal{C},\mathcal{R}$ are the set of species, complexes and reactions respectively.

\begin{align*}
    \mathcal{S} &= \{S_1, ... , S_i,..., S_{|\mathcal{S}|}\}\\
    \mathcal{C} & = \{y_1,...,y_\alpha,...,y_{|\mathcal{C}|} :y_\alpha \in \mathbb{N}^{|\mathcal{S}|}\}\\
    \mathcal{R} &= \{ \ce{ y_{\alpha} ->[k_{y_{\alpha} \to y_{\beta}}] y_{\beta} } : k_{y_{\alpha} \to y_{\beta}} \geq 0\},   
\end{align*}
where Roman letters ($i,j$) and Greek letters ($\alpha,\beta$) are used to denote species and complex indices, respectively. A complex is a multi-set of species, and is denoted by the column vector $y_\alpha$ representing the stoichiometry of the multi-set. The state of a CRN is characterized by a population vector $n \in \mathbb{Z}^{|\mathcal{S}|}_{\geq 0}$, and the time evolution of a probability distribution over the state space is given by (for a rigorous microphysical derivation, see \cite{gillespie1992rigorous} )
\begin{widetext}
\begin{align}
    	\frac{\,d\rho(n,t)}{\,dt} 
    	&=
    	\sum_{y_{\alpha},y_{\beta}}\frac{k_{y_{\alpha}\to y_{\beta}}}{V^{y_{\alpha}-1}}\cdot 	\bigg(\prod_{i=1}^{|\mathcal{S}|}\frac{(n_{i}-y_{\beta,i}+y_{\alpha,i}))!}{(n_{i}-y_{\beta,i})!}\rho(n-y_{\beta}+y_{\alpha},t)  - \prod_{i=1}^{|\mathcal{S}|}\frac{n_{i}!}{(n_{i}-y_{\alpha,i})!}\rho(n,t)\bigg) \nonumber\\
    &=
    \sum_{y_{\alpha},y_{\beta}}\frac{k_{y_{\alpha}\to y_{\beta}}}{V^{y_{\alpha}-1}}\bigg(\frac{(n-y_{\beta}+y_{\alpha}))!}{(n-y_{\beta})!}\rho(n-y_{\beta}+y_{\alpha},t) 
    -\frac{n!}{(n-y_{\alpha})!}\rho(n,t)\bigg) \label{Master_eqn_2}
\end{align}
\end{widetext}
where we get the last equation by making use of the multi-index notation.

 \begin{figure*}[!t]
\centering
    \begin{subfigure}{0.4\textwidth}
  \includegraphics[width=\linewidth]{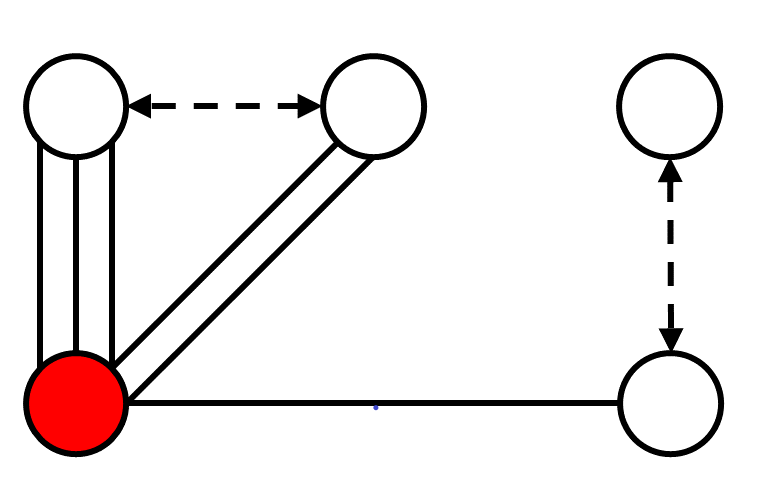}
    \end{subfigure}\hfil 
    \begin{subfigure}{0.5\textwidth}
  \includegraphics[width=\linewidth]{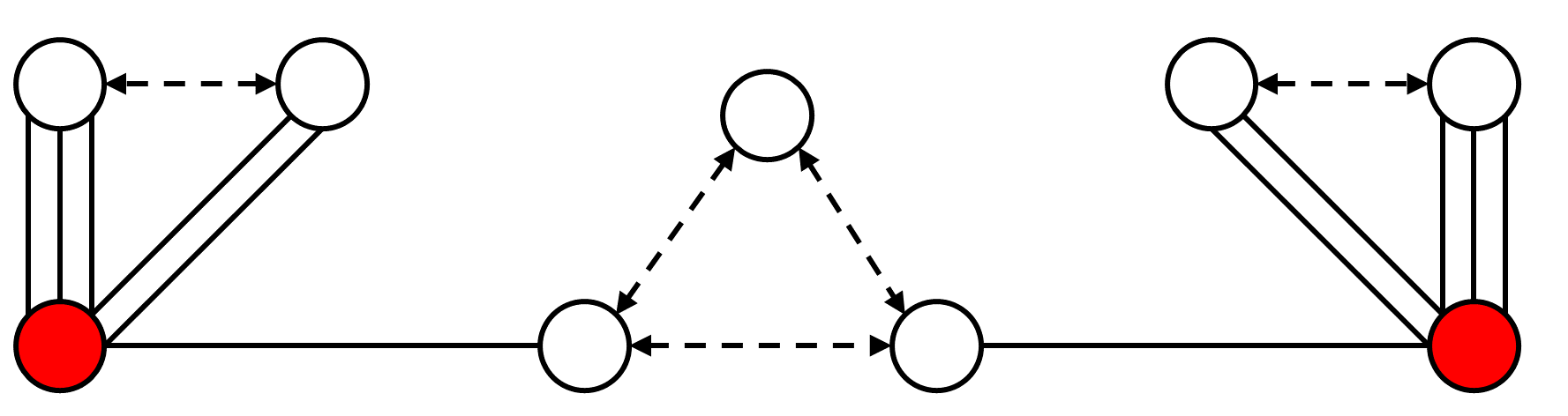}
    \end{subfigure}\hfil 
    \medskip
    \begin{subfigure}{0.5\textwidth}
  \includegraphics[width=\linewidth]{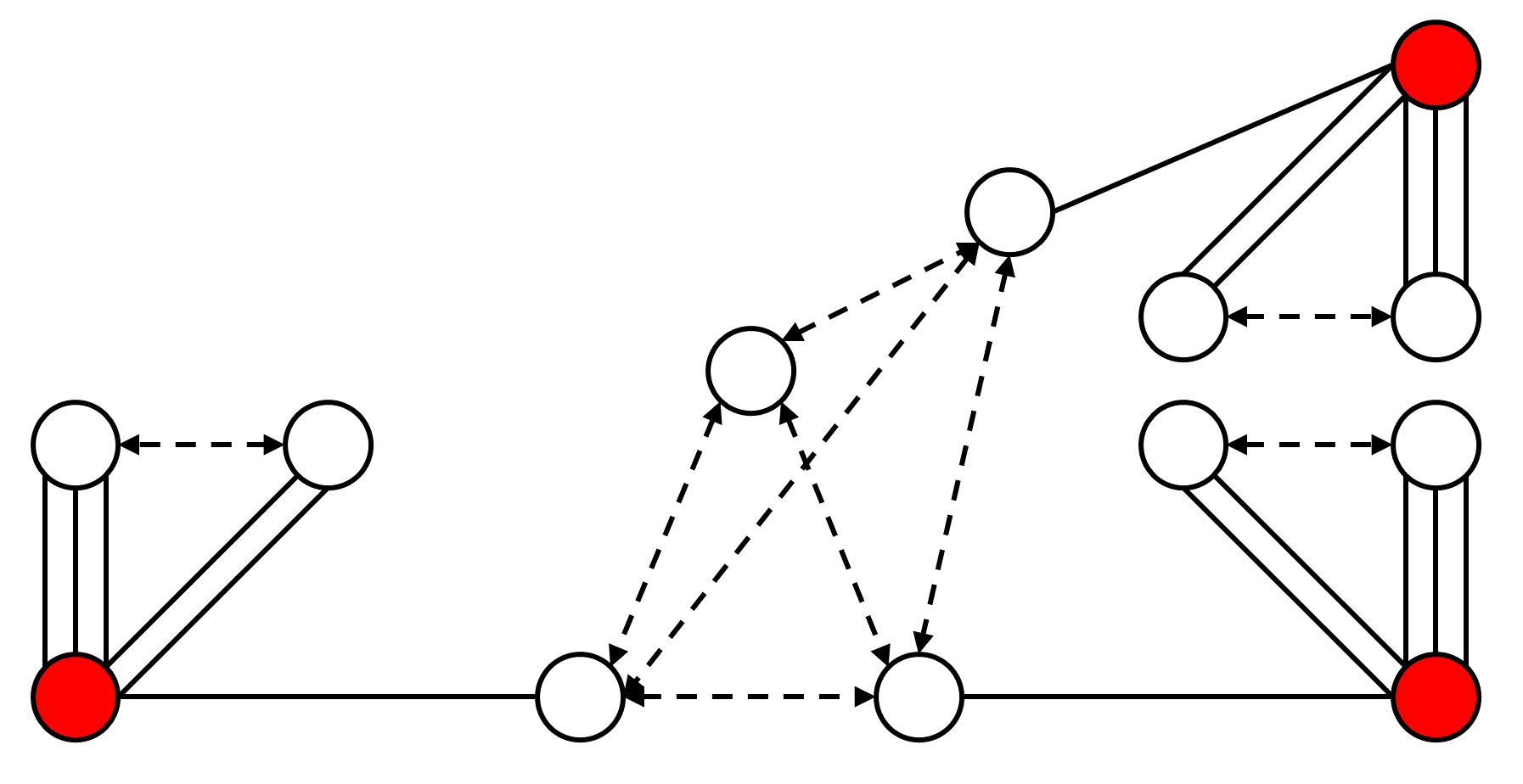}
    \end{subfigure}\hfil 
    \begin{subfigure}{0.5\textwidth}
  \includegraphics[width=\linewidth]{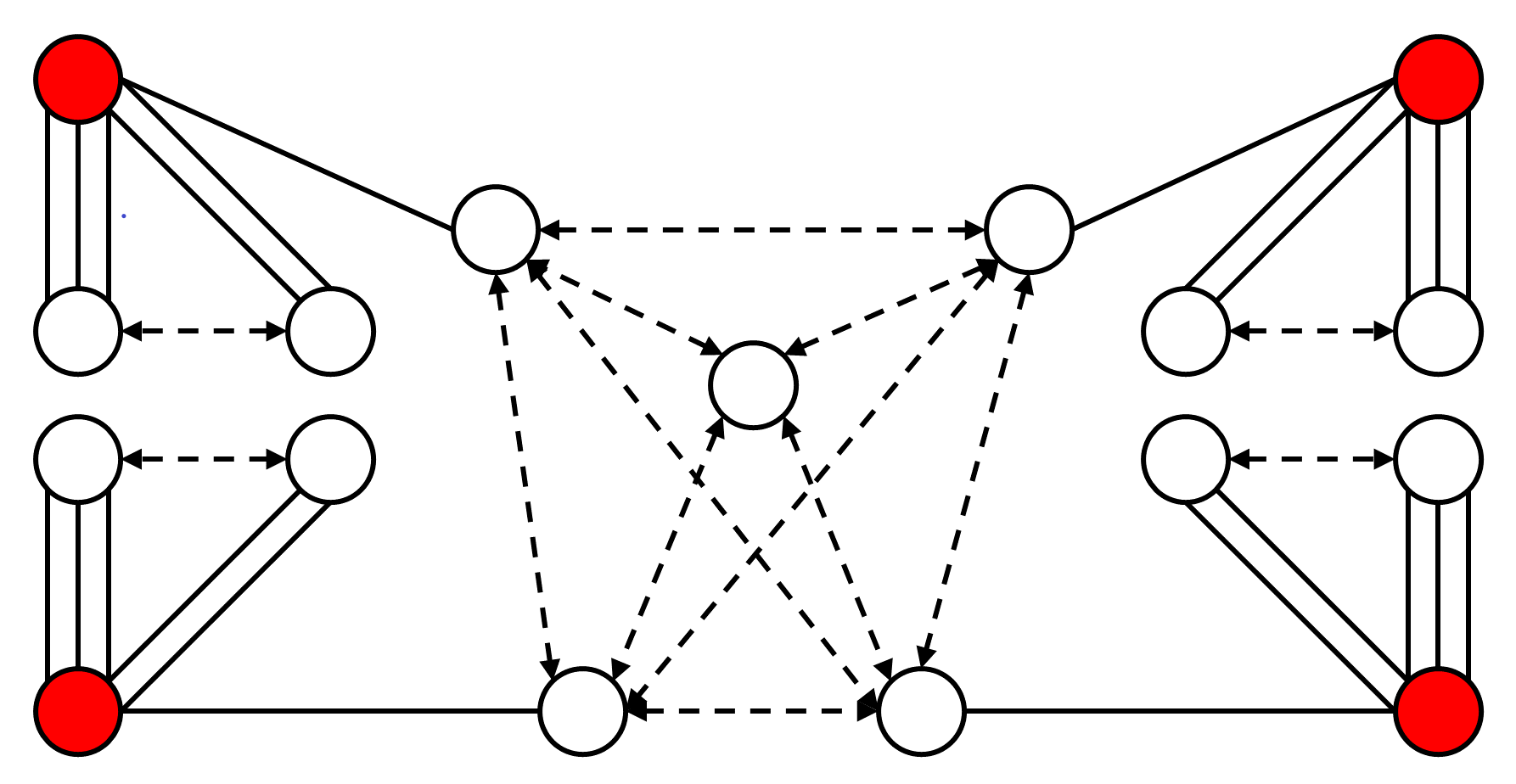}
    \end{subfigure}\hfil 
\caption{Diagrammatic representation of 1-Schlogl, 2-Schlogl, 3-Schl{\"{o}}gl and 4-Schl{\"{o}}gl model CRNs from top left to bottom right. Species, complexes, stoichiometry and transitions are represented with solid circles, empty circles, solid lines and dashed arrows, respectively. }
\label{fig:N-D_Schlogl_CRN}
\end{figure*}

Writing Eq.\ \ref{Master_eqn_2} in the form of equation Eq.\ \ref{eq:TimeEvoln} by making use of the shift operator (see footnote \ref{footnote:shift}), we get
\begin{align*}
       	&\frac{\,d\rho(n,t)}{\,dt} \\
       	&=\sum_{y_{\alpha},y_{\beta}}\big( e^{-(y_{\beta}-y_{\alpha})\cdot\frac{\partial}{\partial n}} - 1\big)
       \frac{k_{y_{\alpha}\to y_{\beta}}}{V^{y_{\alpha}-1}}\frac{n!}{(n-y_{\alpha})!}\rho(n,t) \\
       &\equiv \hat{\underbar{H}}_{\text{CRN}}\left(-\pdv{}{n},n\right) \rho(n,t)
\end{align*}
and obtain the following Hamiltonian operator for CRN 
\begin{align}
    \hat{\underbar{H}}_{\text{CRN}} 
    &=
    \sum_{y_{\alpha},y_{\beta}}\left( e^{-(y_{\beta}-y_{\alpha})\cdot\frac{\partial}{\partial n}} - 1\right) 
    \frac{k_{y_{\alpha}\to y_{\beta}}}{V^{y_{\alpha}-1}}\frac{n!}{(n-y_{\alpha})!}. \label{eq:Ham_CRN_op}
\end{align}
Following the derivation in Section \ref{app:PathIntegral}, we can replace the differential operator $-\partial/\partial n$ with momentum variable $p$ to obtain the Hamiltonian function 
\begin{align}
    \underbar{H}_\text{CRN}(p,n) 
    &= 
    V \sum_{y_{\alpha},y_{\beta}}\big( e^{(y_{\beta}-y_{\alpha})\cdot p} - 1\big) \frac{k_{y_{\alpha}\to y_{\beta}}}{V^{y_{\alpha}-1}}\frac{n!}{(n-y_{\alpha})!}. \nonumber
\end{align}
Finally, using the observation $\frac{(qV)!}{(qV-y)!}\asymp (qV)^y$, we can pass to the Hamiltonian function in the concentration variable $q=n/V$ using Eq.\ \ref{eq:Extensivity_H} to get
\begin{align}
    H_\text{CRN}(p,q)
    &= 
    \sum_{y_{\alpha},y_{\beta}}\big( e^{(y_{\beta}-y_{\alpha})\cdot p} - 1\big) k_{y_{\alpha}\to y_{\beta}}q^{y_{\alpha}}. \label{eq:Ham_CRN_rxn}
\end{align}

For the Hessian of the Hamiltonian function in the momentum variable $p$, we get
\begin{align}
    \pdv{H_{\text{CRN}}}{p^2} 
    &= \sum_{y_{\alpha},y_{\beta}} (y_{\beta}-y_{\alpha}) (y_{\beta}-y_{\alpha})^T e^{(y_{\beta}-y_{\alpha}) \cdot p}k_{y_{\alpha}\to y_{\beta}}q^{y_{\alpha}}.
\label{eq:Convexity_of_H}
\end{align}
It can be seen that Hessian of the Hamiltonian in $p$ is positive definite as it is a sum of symmetric dyadics, each of which is positive definite. Thus, as commented in the discussion below Eq.\ \ref{eq:StationaryAction}, the Hamiltonian function for a CRN is indeed convex in the momentum variable.

As noted in the previous subsection, the convexity of the Hamiltonian in the momentum coordinates allows us to find stationary or optimal paths where the variation of the action functional vanishes. The optimal phase space paths $(q(t),p(t))$ satisfy the Hamilton's equations of motion (EoM)
\begin{align}
    \dv{q}{t} &= \phantom{-}\pdv{H}{p}, \nonumber\\
    \dv{p}{t} &= - \pdv{H}{q} \label{eq:Ham_EoM}
\end{align}
where we have dropped the $^*$ in denoting the optimal path for simplifying notation. For a derivation, see Section \ref{sec:Equiv_HJ_HamEoM}, where we also show that the Hamiltonian is constant along the optimal path and the equivalence of Hamilton's EoM with the Hamilton-Jacobi equations. Intuitively, the ODEs of Hamilton's EoM are the characteristic curves of the Hamilton-Jacobi PDE, and thus the union of all solutions to Hamilton's EoM form the complete solution to the Hamilton-Jacobi PDE. 

Substituting the CRN Hamiltonian from Eq.\ \ref{eq:Ham_CRN_rxn} into Eq.\ \ref{eq:Ham_EoM} and omitting the species index for notational clarity, we get
\begin{align}
\dv{q}{t} 
    &= \phantom{-}
    \sum_{y_{\alpha},y_{\beta}}(y_{\beta}-y_{\alpha})\big( e^{(y_{\beta}-y_{\alpha})\cdot p} \big) k_{y_{\alpha}\to y_{\beta}}q^{y_{\alpha}} \nonumber\\
\dv{p}{t} 
    &= 
    -\sum_{y_{\alpha},y_{\beta}} \big( e^{(y_{\beta}-y_{\alpha})\cdot p} - 1\big) k_{y_{\alpha}\to y_{\beta}}\pdv{q^{y_{\alpha}}}{q}, \label{eq:Ham_EoM_CRN}
\end{align}
which are the set of ODEs satisfied by the optimal paths for a CRN.

Let us begin investigating the optimal paths specified by the above equations. For reasons mentioned in Eq.\ \ref{eq:HJ_NEP}, we only consider the subclass of optimal paths in the $H_\text{CRN}=0$ submanifold. For the remainder of this section, we drop the subscript CRN and refer to the CRN Hamiltonian simply by $H$.

Notice that for $p=0$, the Hamiltonian is identically zero i.e. $H(p=0,q) = 0 \hspace{1em}\text{for all } q$. Substituting this assignment in Eq.\ \ref{eq:Ham_EoM_CRN}, we get
\begin{align}
\dv{p}{t}\bigg|_{p=0} 
    &= 
    0, \nonumber\\
    \dv{q}{t}\bigg|_{p=0} 
    &=
    \sum_{y_{\alpha},y_{\beta}}(y_{\beta}-y_{\alpha})k_{y_{\alpha}\to y_{\beta}}q^{y_{\alpha}}.     \label{eq:MassActionKinetics}
\end{align}
The first line ensures that, since the time derivative of $p$ vanishes, $p=0$ is a consistent assignment everywhere along the optimal path. The second line is nothing but the set of ODEs corresponding to the law of mass-action kinetics for a CRN. Thus the optimal paths with $p=0$ correspond to the deterministic trajectories of a CRN, which we term \textbf{relaxation trajectories} and denote as $q_\text{rel}(t)$.

To understand why the $p=0$ solution must correspond to the deterministic behaviour of CRN in the Hamilton-Jacobi formalism, let us calculate the change in action along the relaxation trajectory. From Eq.\ \ref{eq:Action_func_time_evol}, 
\begin{align}
    \int_0^T \,d S[q_\text{rel}] 
    &= 
    \int_0^T \left[ 0\cdot \dv{q_\text{rel}}{t} - H(0,q_\text{rel})\right]\,dt \nonumber\\
    &= 0.
\end{align}
Recall from Eq.\ \ref{eq:StationaryAction} that the exponential of the scaled change in action acted as a conditional probability up to a normalization factor. Since the change in action is identically zero, almost all of the probability measure without any suppression evolves along the relaxation trajectory yielding 
\begin{align*}
    \rho(q_\text{rel}(T),T) &= N \rho(q_\text{rel}(0),0).
\end{align*}

It is quite common for the deterministic dynamics of a CRN, Eq.\ \ref{eq:MassActionKinetics}, to exhibit multiple fixed points. Let us denote a fixed point by $\underline{q}$. Then each fixed point must satisfy
\begin{align}
    \dv{\underline{q}}{t} = 0 &= \sum_{y_{\alpha},y_{\beta}}(y_{\beta}-y_{\alpha})k_{y_{\alpha}\to y_{\beta}}\underline{q}^{y_{\alpha}} \nonumber\\
    &= \sum_{y_{\alpha}}y_{\alpha}\sum_{y_{\beta}}(k_{y_{\beta}\to y_{\alpha}}\underline{q}^{y_{\beta}}-k_{y_{\alpha}\to y_{\beta}}\underline{q}^{y_{\alpha}}).
\label{eq:fixed_points}
\end{align}
The existence of multiple steady states can be ruled out in many cases by certain topological considerations and is studied in deficiency theory. For an introduction and detailed derivation of some results, we refer the reader to \cite{feinberg1974dynamics,feinberg2019foundations,smith2017flows,smith2021eikonal}.


\begin{figure}[!t]
    \includegraphics[scale=0.5]{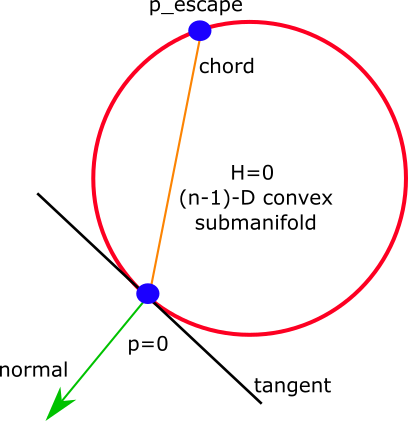}
    \caption{Diagrammatic proof of the Lyapunov property of the non-equilibrium potential with respect to mass action kinetics.}
    \label{fig:NEP_Lyapunov}
\end{figure}

For our purposes in this paper, we restrict our attention to generic CRNs with multiple fixed points. These points can be indexed by the number of repelling directions, which can easily be determined using the eigenvalues of the Jacobian of the flow field in Eq.\ \ref{eq:MassActionKinetics}. A fixed point with no repelling or all attracting directions is referred to as a \textit{stable attractor}. In a purely deterministic setting, one would expect any evolving probability distribution to end up concentrated at the multiple stable attractors with ratios determined by the initial conditions. In a stochastic setting however, there can be large fluctuations that take the system out from the vicinity of a stable attractor to a generic point $q \neq \underline{q}$ in its basin of escape. The path along which the system arrives at $q$ with a leading exponential probability will be the optimal path that starts from $\underline{q}$ and ends at $q$, and the probability of this event is precisely given by the action function along that path using \ref{eq:StationaryAction}. We call the optimal paths out of a fixed point in the $H(p,q)=0$ submanifold as \textbf{escape trajectories}, denoted by $q_\text{esc}$, and from eqs \ref{eq:NEP_1}, \ref{eq:HJ_NEP} it is precisely the action function along paths that determine the value of the NEP,
\begin{align}
    \mathcal{V}(q)-\mathcal{V}(\underline{q}) 
    &= 
    \int_{q_\text{esc}} p_\text{esc}(q')\dv{q'}{t} \,dt \nonumber\\
    &= \int_{\underline{q}}^q p_\text{esc}(q')\,d q' \label{eq:NEP_2}
\end{align}
where $(q_\text{esc},p_\text{esc})$ is the optimal path that connects $\underline{q}$ and $q$. Note that the momentum assignment along the escape path must necessarily be non-zero, except at the fixed points, and the Hamiltonian function must evaluate to zero everywhere along the optimal path. 
\begin{align}
    p_\text{esc}(q) \equiv p(q_\text{esc}) &\neq 0 \nonumber\\
    H(p_\text{esc},q_\text{esc}) 
    &= 
    0 \label{eq:p_neq_0} 
\end{align}

To understand why the momentum assignment along escapes must be non-zero everywhere except the fixed points and to prove that the NEP is a Lyapunov function with respect to the relaxation trajectories (mass-action kinetics), let us investigate the geometry of the $H(p,q)=0$ submanifold in the $2|\mathcal{S}|$ dimensional phase space. Since we are considering a CRN with fixed points, there must exist shift vectors $y_\beta-y_\alpha$ such that $H\to \infty$ when $|p|\to \infty$ in all directions. Also, recall from Eq.\ \ref{eq:Convexity_of_H} that for a fixed concentration $q$, the Hamiltonian is a convex function in $p$. This means the $H(p,q)=0$ submanifold at a given $q$ must be a closed surface and bounded in the interior with $H<0$. Next, let us consider the outward normal to the surface given by $\partial H / \partial p$. Recall from Hamilton's equations from Eq.\ \ref{eq:Ham_EoM_CRN} that the outward normal is equal to the rate of change of concentration. Since by the fixed point condition 
\begin{align}
    \pdv{H(0,\underline{q})}{p} &= 0, \label{eq:pinch_H_0}
\end{align}
this means that the outward normal to the surface vanishes and thus the $H(p,q)=0$ submanifold must pinch off at the fixed points yielding $p=0$ as the only solution in the submanifold.

\begin{figure*}[t]
\hspace*{-2cm}  
    \centering 
\begin{subfigure}{\textwidth}
  \includegraphics[width=1.05\linewidth]{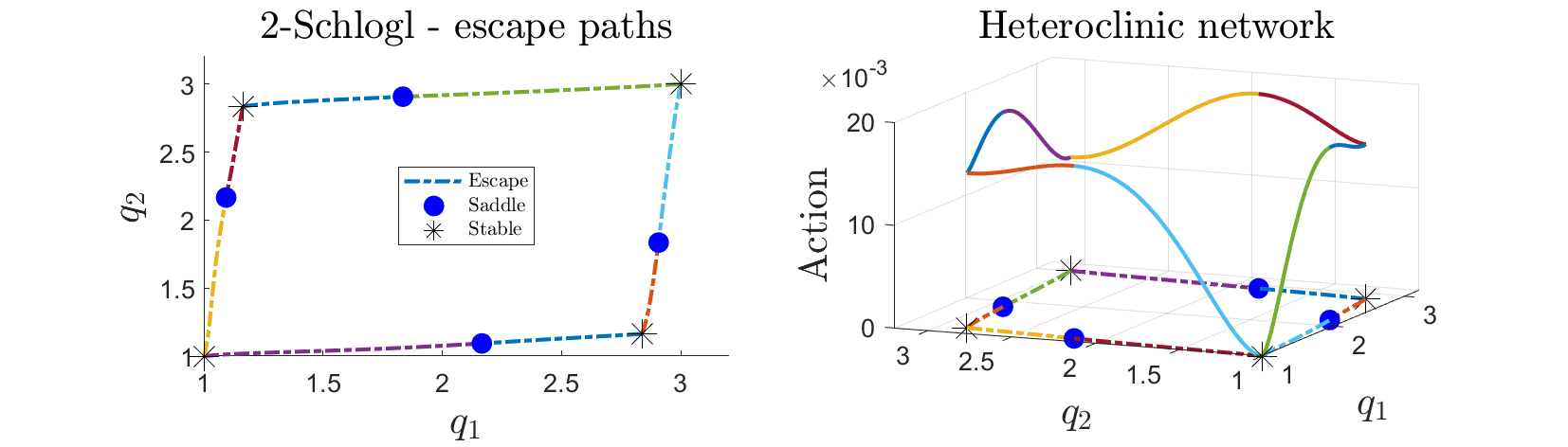}
\end{subfigure}\hfil 
\caption{Escape paths and heteroclinic network for the 2-Schl{\"{o}}gl model.}
\label{fig:2-Schlogl_hc_net}
\end{figure*}

\begin{figure*}[t]
    \centering 
\begin{subfigure}{\textwidth}
  \includegraphics[width=\linewidth]{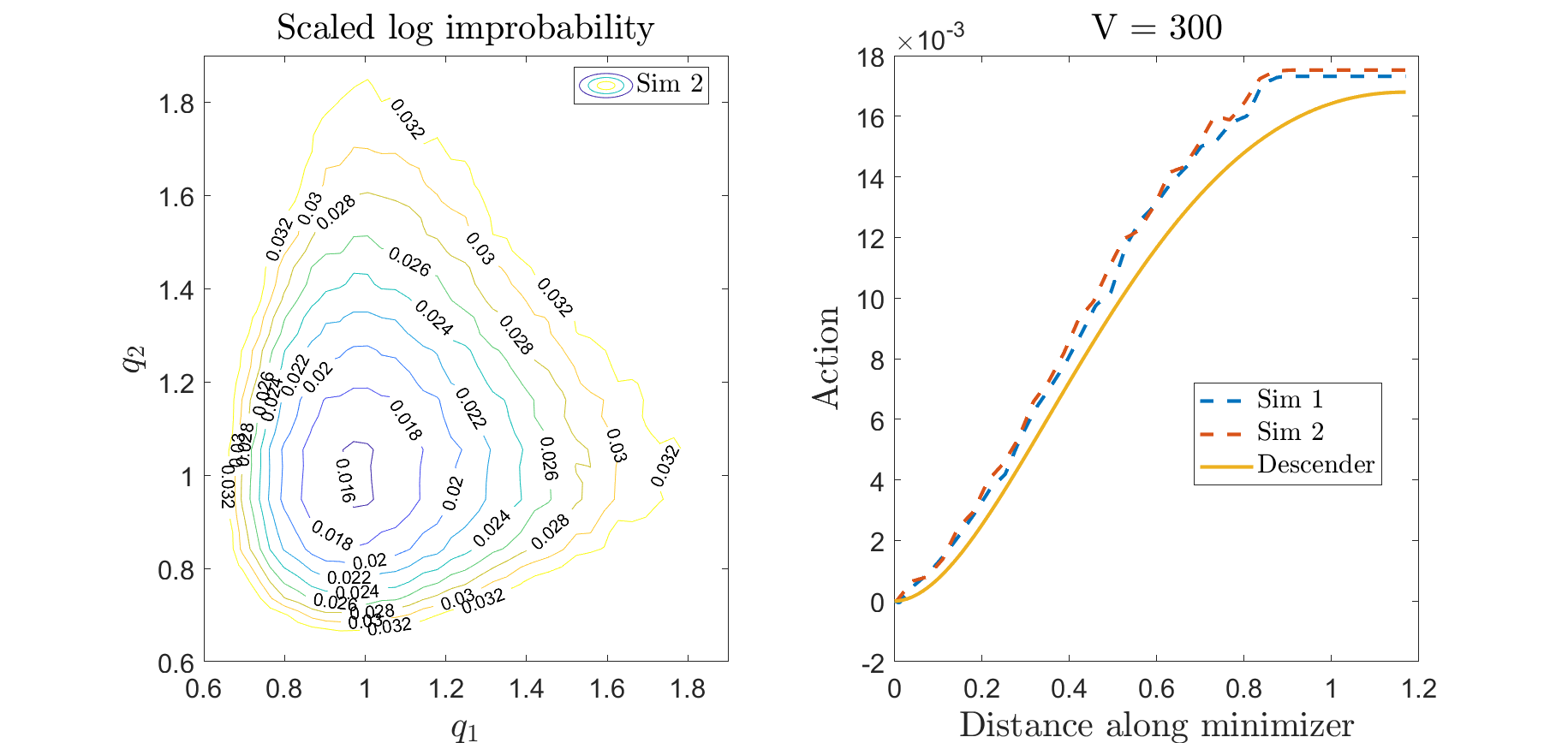}
\end{subfigure}\hfil 
\caption{Comparing the algorithm against Gillespie stochastic simulation for the 2-Schl{\"{o}}gl model (Section \ref{sec:2Schlogl_model}). Scaled log-improbability of the stationary distribution $\pi$ is defined to be $(-1/V)\log(\pi)$.}
\label{fig:2-Schlogl_Gillespie}
\end{figure*}

Finally, to show that the NEP is a Lyapunov function along the deterministic MAK, let us consider its time derivative along a relaxation trajectory.
\begin{align}
    \dv{\mathcal{V}(q_{\text{rel}}(t))}{t}
    &= 
        \pdv{\mathcal{V}(q)}{q}\cdot\dv{q}{t}\bigg|_{\text{rel}} \nonumber\\
    &=
    p_\text{esc}(q) \cdot \dv{q}{t}\bigg|_{p=0} \nonumber\\
    &= (p_\text{esc}-0)\cdot\pdv{H}{p}\bigg|_{p=0} \nonumber \\
    &\leq 0 \label{eq:NEP_Lyapunov}
\end{align}
where the first line follows from chain rule, the second line uses the Hamilton-Jacobi relation of momentum and NEP from Eq.\ \ref{eq:NEP_2}, and the third line follows from Eq.\ \ref{eq:MassActionKinetics}. The last line follows from the fact that the gradient of a convex function along a chord in the convex subset (secant) is non-positive, or equivalently the outward normal always makes more than $90^\circ$ with a secant, shown diagramatically in Figure  \ref{fig:NEP_Lyapunov}.

\subsection{Switching dynamics for a multistable stochastic CRN}
\label{sec:switching}

Let us refer to the set of attractors (fixed points) of a CRN, when taken under MAK, by $\underline{q}\equiv \{ \underline{q}_1, \underline{q}_2,\ldots,\underline{q}_N\}$. Now, label the attractors by the number of positive eigenvalues of the Jacobian of the MAK rates or the mixed-Hessian of the Hamiltonian $\partial^2 H(p,\underline{q})/ \partial p \partial q$ evaluated at $p=0$. Recall that the positive eigenvalues give the number of repelling directions around the attractor. Let us refer to the fixed points with zero and one repelling directions as \textit{stable} and \textit{saddle} attractors, respectively.

For large scale factor $V$, the switching dynamics of the stochastic CRN will be generally governed by the stable and saddle attractors. Recall from Section \ref{sec:HJ_theory} that the most likely or optimal paths for a stochastic trajectory lie in the $H=0$ submanifold. Moreover, in Section \ref{sec:Ham_CRN}, we showed that there are both trajectories leading into a stable attractor (termed relaxation trajectories) and coming out of a stable attractor (escape paths). In particular then, corresponding to relaxation trajectories that emanate nearby a saddle attractor into a stable attractor, there will also be escape paths emanating from the stable attractor and terminating at the saddle attractor. The data obtained by collecting all the stable and saddle attractors and the relaxation and escape paths joining them is called as the \textit{heteroclinic network} for the CRN. For example, the heteroclinic network obtained for the 2-Schl{\"{o}}gl model from Figure \ref{fig:N-D_Schlogl_CRN} is shown in Figure \ref{fig:2-Schlogl_hc_net} (also see Section \ref{sec:2Schlogl_model}). 

Generically, one expects a stochastic CRN to spend a long time near a stable attractor before transitioning to a subsequent one by passing a saddle attractor in between. Recall that the NEP or action along an escape trajectory (see Section \ref{sec:HJ_theory}) quantifies the scaled log-improbability of the escape event. Moreover, as we have shown in Section \ref{sec:Ham_CRN}, the NEP is a Lyapunov function along MAK. This means that the NEP must monotically decrease along the relaxation trajectory, or correspondingly monotonically increase along the escape trajectory. For example, the NEP for the 2-Schl{\"{o}}gl model along all the escape heteroclinic orbits is shown in the right panel of Figure \ref{fig:2-Schlogl_hc_net}. Moreover, in Figure \ref{fig:2-Schlogl_Gillespie}, the NEP is shown for a particular escape and compared against the occupation log-improbability probability obtained by a Gillespie simulation. 

We briefly want to remark that since the NEP increases along an escape trajectory, the quantity must not be seen as an \textit{entropy} for a stochastic system. This viewpoint will naturally lead to paradoxical violations of the second law of thermodynamics. However, in \cite{smith2020intrinsic}, it is shown that for an appropriate definition of entropy for stochastic CRN, its rate of change for any system can be asymptotically estimated by the rate of change of the NEP along a relaxation trajectory, thus yielding a rigorous \textit{second law} for CRN.


\section{Action Functional Gradient Descent (AFGD) algorithm  }
\label{sec:Algortithm}

\begin{table*}[]
    \centering
    \begin{tabular}{|c|c|}
    \hline
     Object & Notation  \\
     \hline
     Configuration space & $Q\subset \mathbb{R}^{D}$ \\ 
     Space of curves & $\mathcal{P}_Q = \left\{ q: [0,1] \to Q, q(0)=q_I, q(1) = q_F \right\}$ \\
     Curve & $q \in \mathcal{P}_{Q}$  \\ 
     Phase space & $T^*Q \subset \mathbb{R}^{2D}$\\
     Space of trajectories & $\mathcal{P}_{T^*Q} = \left\{ (q,p): [0,T] \to T^*Q, q^*(0)=q_I, q^*(T) = q_F \right\}$\\
     Trajectory & $(q,p) \equiv \gamma \in \mathcal{P}_{T^*Q}$\\
     \hline
    \end{tabular}
    \caption{Notation used for denoting paths in configuration and phase space.}
    \label{tab:notation}
\end{table*}

A central property of chemical reaction networks (CRNs) that makes them particularly useful for modeling complex systems is their ability to exhibit multiple attractors when taken under deterministic mass-action kinetics. Moreover, in a stochastic CRN, due to fluctuations in population sizes, it is possible for the system to transition out of a stable attractor into another with some probability. It is of practical value to quantify the transition events, termed as \textit{escapes}, and can help in experiment design as well as network inference \cite{langary2019inference}. However, the escapes are rare, scaling exponentially in $-V$, compared to fluctuations in population that are polynomial in $1/V$, making stochastic simulation an inefficient approach to estimate the leading consequences of escapes, which dominate basin-switching.

In this section, we provide a deterministic alternative to estimating the escape paths between appropriate start and end points in chemical concentration space using a variational method. Specifically, we present a functional gradient descent algorithm that finds the optimal trajectory, that minimizes the action functional while satisfying the boundary conditions. A mathematical formulation of the optimization problem is posed in Section \ref{sec:MinMax} and the algorithm is explained in Section \ref{subsec:algorithm}, while relegating the technical details to Appendix \ref{app:Details_AFGD}. A discussion of the algorithm's performance costs and its relation to other methods is presented in Section \ref{subsec:addn_remarks}.

Finally, we will briefly remark on obtaining the probability of system transition from the escape path. First, notice that if the boundary conditions are such that the trajectory emanates from a stable fixed point and terminates in a saddle fixed point, then the resulting escape trajectory does not lie in the solution set of the deterministic mass-action kinetics, and can only be interpreted in the stochastic framework (see Section \ref{sec:Ham_CRN}). Let the phase space trajectory of such an escape be denoted by $\gamma^*$. For reasons explained below Eq.\ \ref{eq:StationaryAction}, the likelihood of a system of scale factor $V$ starting from the stable fixed point and escaping to the saddle fixed point can be well-approximated simply by $e^{-V \mathcal{A}[\gamma^*]}$ (see Eq.\ \ref{eq:NEP_1}). For example, we choose $V=300$ in Figure \ref{fig:2-Schlogl_Gillespie}, the authors consider values of $V$ as low as $10$ in Figure 4 of \cite{smith2017flows}, and in both cases the occupation probability from the stochastic simulation data is well in agreement with the exponential of the scaled action along the escape trajectory.

\subsection{Formulation as a MinMax problem}
\label{sec:MinMax}

We refer to the space of states in which we can find our chemical system as \textit{configuration space}, denoted by $Q$. Suppose we have $|\mathcal{S}|=D$ species in our network, then the state of a system at a given time is a vector of concentration of each species $\in \mathbb{R}_{\geq 0}^D$. The space of paths that a system can take in configuration space is denoted by $\mathcal{P}_Q$, and we denote a path by $q(s)$. Notice that although the space of parametrized differentiable paths lies in the tangent bundle of $Q$ (see \cite{milnor1963morse}), the parametrization is arbitrary and we do not yet have a notion of physical time and velocity. To introduce a notion of physical time, we have to introduce a Hamiltonian function (see Eqs.\ \ref{eq:PathIntegral_q_coord}, \ref{eq:Ham_CRN_rxn}) defined on the cotangent bundle of $Q$, commonly referred to as the \textit{phase space}. The coordinates canonically conjugate to configuration coordinates $q$ in the cotangent space are also called as momentum coordinates and denoted by $p$. Moreover, since the Hamiltonian is convex in momentum, every path in the configuration space has a unique lift in phase space parametrized by time variable $t$, which we denote by $\gamma \equiv (q(t),p(t))$. To avoid confusion, henceforth we refer to a path in configuration space in $\mathcal{P}_Q$ as a \textit{curve}, and phase space path in $\mathcal{P}_{T^*Q}$ as a \textit{trajectory}.

In the notation described in Table \ref{tab:notation}, we can succinctly formulate the problem that we wish to devise an algorithm to solve. We wish to find the optimal phase space trajectory, i.e.\ a trajectory that minimizes the action functional, constrained such that its projection to the configuration space curve begins at configuration $q_I$ and ends at $q_F$, and the phase space trajectory is along the $H(p,q)=0$ submanifold (reason for the $H(p,q)=0$ constraint can be found in discussion around Eq.\ \ref{eq:HJ_NEP} in Section \ref{sec:HJ_theory}). 
\begin{align}
    \text{Find } & 
    \gamma^* \equiv(q^*,p^*) \in \mathcal{P}_{T^*Q} \nonumber\\
    \text{such that } &
    \mathcal{A}\left[\gamma^*\right] 
    = 
    \min_{q(t)}\max_{p(t)} \int_0^T \left(p\cdot\dv{q}{t}-H(p,q)\right)\,dt, \nonumber\\
    & 
    H(p^*(t),q^*(t))=    0, \nonumber\\
    & q(0) = q_I, q(T) = q_F. \label{eq:Problem}
\end{align}

Once the escape path from $q_I$ to $q_F$ is identified, the difference in the non-equilibrium potential between the two configurations is given by value of the action along the path, i.e.
\begin{align}
    \mathcal{V}(q_F) - \mathcal{V}(q_I) &= \mathcal{A}\left[\gamma^*\right].
\end{align}

For reasons discussed in Section \ref{sec:switching}, we are only interested in finding escape paths from a stable to a nearby saddle fixed point, thus we will choose the end points to be in the set of fixed points of the mass-action kinetics $\underline{q}$, as defined in Eq.\ \ref{eq:fixed_points}. Moreover, from Eq.\ \ref{eq:pinch_H_0}, we know that the $H(p,q)=0$ phase space submanifold pinches off and the only solution to $H(p,\underline{q})=0$ is $p(\underline{q}) = 0$. Thus, for our purposes here, the momentum end point constraints are $p(q_I)=0=p(q_F)$.

\subsection{Algorithm and pseudocode}
\label{subsec:algorithm}
\begin{figure}[!t]
    \includegraphics[scale=0.5]{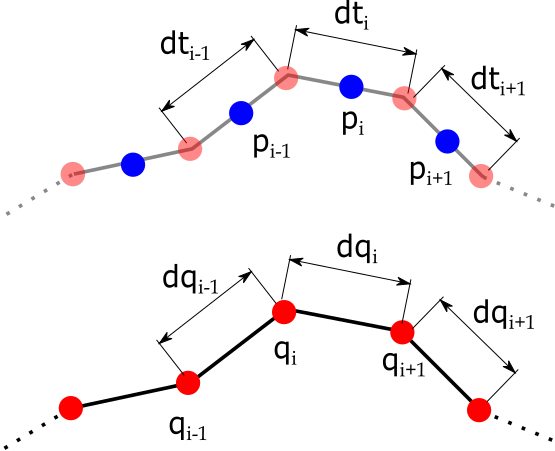}
    \caption{Lifting a curve in configuration space to a trajectory in phase space by assigning a momentum and time difference to each segment.} 
     \label{fig:Leg_trans}
\end{figure}

\begin{figure*}[!t]
    \includegraphics[scale=0.5]{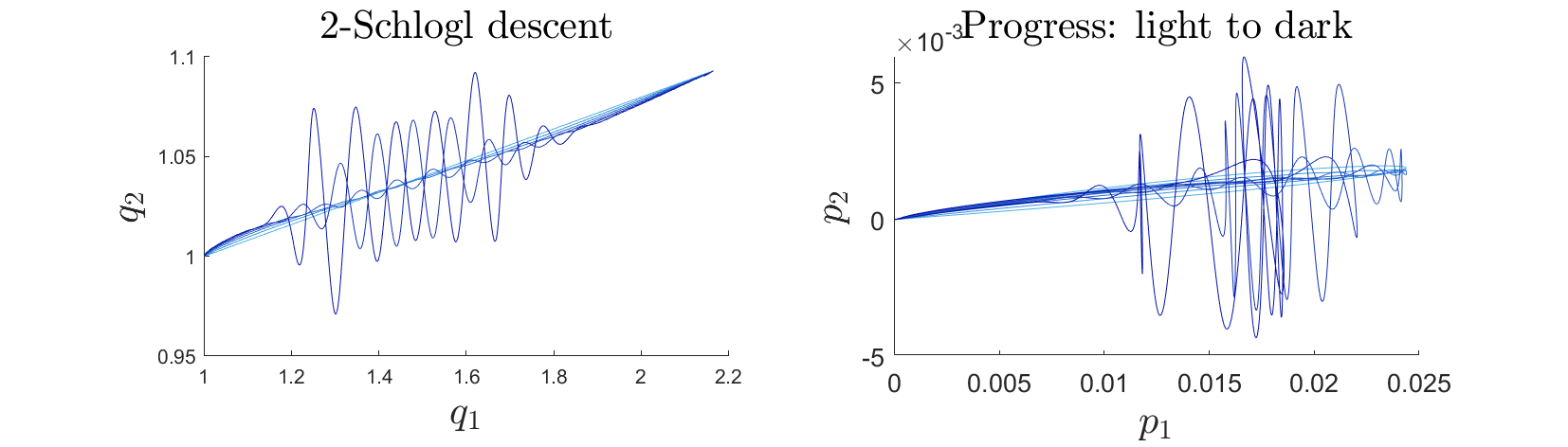}
    \caption{Functional gradient descent for 2-Schl{\"{o}}gl model without filtering. Note that the low-frequency waves first amplify and then concentrate at higher frequencies near a central, badly-performing place on the trajectory. } 
     \label{fig:unfiltered_2Schlogl}
\end{figure*}

\subsubsection{Main idea}
\label{subsubsec:idea}
For a CRN exhibiting multiple fixed points, we start by picking a stable and a nearby saddle fixed point. The fixed points are determined by finding roots of the polynomial rate equations of mass-action kinetics, and can be found for high dimensional systems using a numerical algebraic geometry software such as Bertini \cite{BHSW06}. Once the end points have been determined, we obtain a configuration space \textit{curve} by joining the end points with a straight line, which serves as an \textit{initial condition} for the algorithm (see App.\ \ref{app:initial_cond}). While in the actual algorithm the initial curve will be uniformly discretized at finitely many points, in this subsection, for the sake of explaining the main idea, consider that the curve is continuous.

Let the continuous configuration space curve at iteration $I$ be denoted by $q^I$. Then, as explained in App.\ \ref{sec:lift_curve_to_traj} and shown in Figure \ref{fig:Leg_trans}, we \textit{lift} the curve to a phase space \textit{trajectory} by assigning a scalar $\,dt$ and vector $p^I$ at each point, such that the variation of the action functional in momentum is zero while the phase space trajectory is constrained to the $H(p,q)=0$ submanifold,
\begin{align*}
    \dv{q^I}{t} &= \pdv{H}{p}\bigg|_{(q^I,p^I)}\\
      0 &= H(p^I,q^I).
\end{align*}
Assigning momentum and time coordinates at each point along the discrete curve (see Figure \ref{fig:Leg_trans}) can be implemented using standard constraint optimization solvers (see \ref{sec:lift_curve_to_traj}) and is the \textit{costliest step} in the algorithm. Notice that since the Hamiltonian is convex in momentum, the solution exists. However, recall from Eq.\ \ref{eq:H_ME} that the CRN Hamiltonian, as opposed to the generic Hamiltonian for mechanical energy, is not quadratic in momentum. Thus, the momentum coordinate generally cannot be solved for analytically and the difficulty of assigning it will depend on both the dimensionality of the system as well as the number of reactions in the network.

From Eq.\ \ref{eq:1st_variation_Action}, the variation of the action functional along a phase space trajectory $(q,p)$ is
\begin{align}
    \delta \mathcal{A}[q,p]  &= \int \,dt \left(\frac{\delta \mathcal{A}}{\delta q}\cdot\delta q + \frac{\delta \mathcal{A}}{\delta p}\cdot\delta p \right) \text{ where} \nonumber\\
    \frac{\delta \mathcal{A}}{\delta q} &= -\left( \dv{p}{t} + \pdv{H}{q}\right),\nonumber\\
        \frac{\delta \mathcal{A}}{\delta p} &= \phantom{-} \left( \dv{q}{t} - \pdv{H}{p}\right). \label{eq:variation_action_iteration}
\end{align}
Thus, as explained in App.\ \ref{sec:func_gradient}, we update the new configuration curve in the next iteration $q^{I+1}=q^I + \delta q^I$, where a variation $\delta q^I$ is chosen to be in the \textbf{steepest descent direction} $g^I$ (to first order), where 
\begin{align*}
    g^I :=-\frac{\delta \mathcal{A}}{\delta q}  &= \left( \dv{p}{t} + \pdv{H}{q}\right)\bigg|_{(q^I,p^I)},\\
    \text{and }\delta q^I &= \epsilon \phantom{.} g^I
\end{align*}
with a step-size $\epsilon>0$ picked by employing a backtracking line search (see \ref{sec:step_size}, \cite{wright1999numerical}). This assignment of $\delta q^I$ ensures that, using Eq.\ \ref{eq:1st_variation_Action}, the variation of the action is negative semidefinite
\begin{align}
    \delta \mathcal{A} &= -\epsilon \frac{\delta \mathcal{A}}{\delta q} \cdot \frac{\delta \mathcal{A}}{\delta q} \nonumber\\
    &\leq \phantom{-} 0, \label{eq:grad_desc}
\end{align}
with equality only at the optimal curve.

The algorithm for finding the solution to Eq.\ \ref{eq:Problem} can thus be seen as performing a $\max$ by assigning optimal momentum values, followed by moving towards the $\min$ by taking a gradient in the descent direction in each iteration. The convergence of the algorithm relies on the convexity of the action functional around the optimal solution, on which we comment in App.\ \ref{app:convexity}.

\subsubsection{Noise and discretization}

\begin{algorithm*}[!t]
\caption{Pseudocode for Action Functional Gradient Descent (AFGD) algorithm (implementation \cite{Gagrani_AFGD-for-CRN-escapes_2022})}\label{alg:AFGD}
\begin{algorithmic}
\State $I\gets 1;\epsilon \gets \epsilon_\text{IC} ;f_c \gets f_\text{IC} ; \Delta^1 \gets \infty$
\State $q_{\text{IC}} \gets \text{Initial\_condition} $ 
\State $q^1 \gets \text{Space\_uniform\_sampling}[q_\text{IC}]$
\State $(q,p,t)^I \gets \text{Lift\_curve}[q^I]$  \Comment{Lift curve to phase space trajectory}
\State $S^I \gets \text{Action}[(q,p)^I]$ \Comment{Calculate action}
\State $g^I = \text{Functional\_gradient}[(q,p,t)^I]$ \Comment{Calculate functional gradient}
\State $g^I_s = \text{Filter\_gradient}[g^1,f_c]$ \Comment{Low-pass filter gradient}
\State $\Delta S \gets 0$
\While{$I \leq I_\text{Max} \textbf{ or } f_c \leq f_\text{Max} \textbf{ or } \Delta^I > \Delta_\text{thresh}$}
\State $I\gets I+1$
\State $\epsilon^I \gets \text{Step\_Size}(q^{I-1},g_s^{I-1},\epsilon^{I-1})$ \Comment{Pick step size}
\State $q^{I} \gets q^{I-1} + \epsilon^I \times g^{I-1}_s$
\State $(q,p,t)^{I} \gets \text{Lift\_curve}[q^{I}]$
\State $S^{I} \gets \text{Action}[(q,p)^{I}]$
\State $\Delta S \gets S^{I}-S^{I-1}$
\If{$|\Delta S| < \Delta S_\text{thresh} \textbf{ or } \epsilon < \epsilon_\text{thresh}$}
\State $f_c \gets f_c + \Delta f; \epsilon^I \gets \epsilon_\text{IC}$       \Comment{Loosen pass-band; refresh step-size}
\State $q_s^{I} \gets \text{Filter\_in\_time\_uniform}[q^{I}]$ \Comment{Optional: increase sampling/anneal}
\State $(q,p,t)^{I} \gets \text{Lift\_curve}[q_s^{I}]$
\EndIf
\State $\Delta^I \gets \text{Distance\_Hamilton's\_equations}[(q,p)^{I}]$ \Comment{Verification: Least distance from integrated EoM}
\State $g^{I} \gets \text{Functional\_gradient}[(q,p,t)^{I}]$ \Comment{Calculate functional gradient}
\State $g^{I}_s \gets \text{Filter\_in\_space\_uniform}[g^{I},f_c]$ \Comment{Low-pass filter gradient}
\EndWhile
\end{algorithmic} 
\end{algorithm*}

\begin{figure*}[htb]
    \centering 
\begin{subfigure}{\textwidth}
  \includegraphics[width=\linewidth]{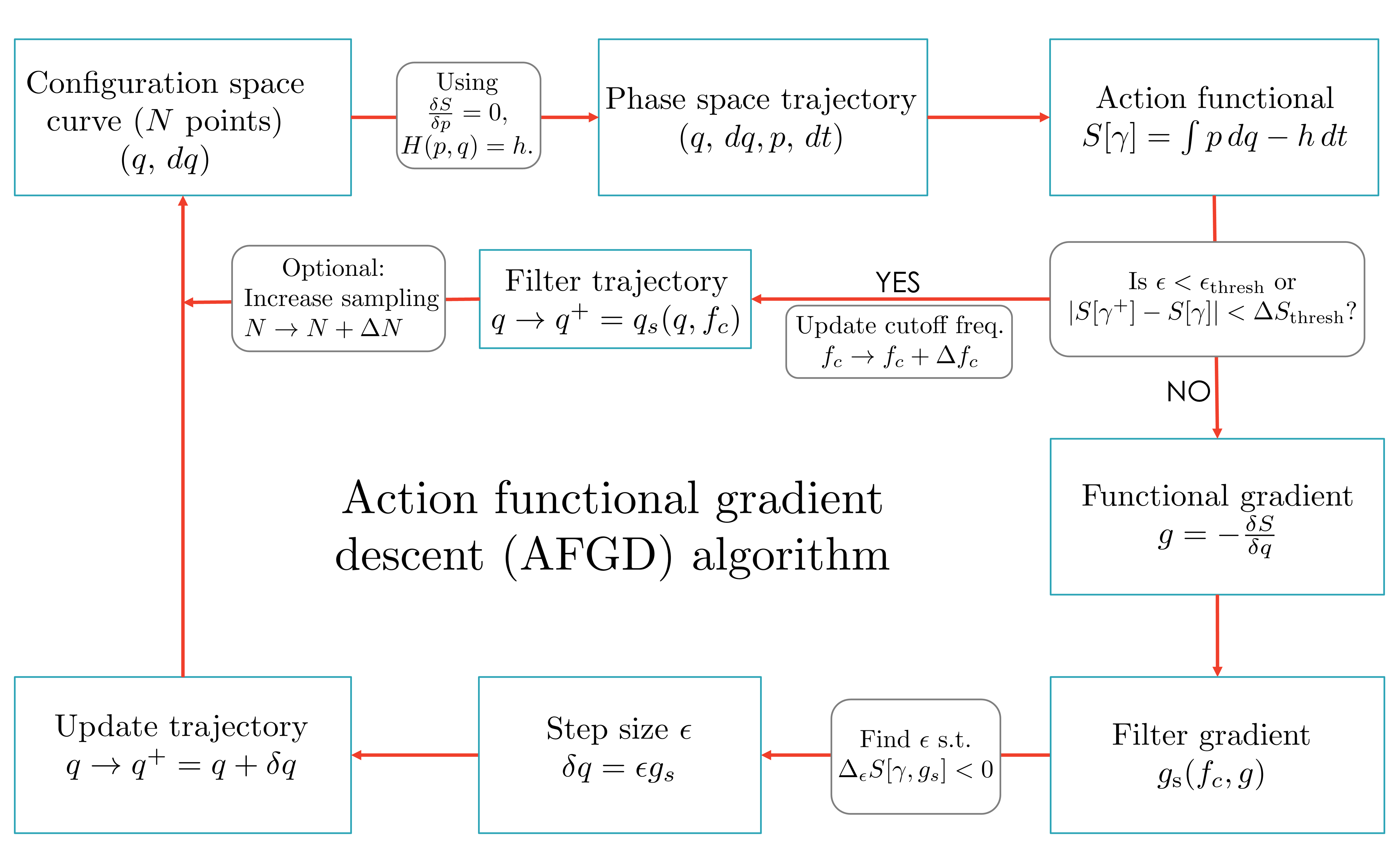}
\end{subfigure}
\caption{The AFGD algorithm in a nutshell. For the CRN Hamiltonian, when the end points are fixed points of the system then $h=0$ must be chosen. For a detailed description, see Section \ref{subsec:algorithm}.}
 \label{fig:AFGD_pseudocode}
\end{figure*}

Since the optimal trajectory is a solution to a variational problem, it must be smooth. A naive implementation along the above account on a finitely sampled or discretized curve will fail on two ends. Firstly, there are two major sources of noise that self-amplify along the iterations, namely due to numerical solving of assigning momentum values (lifting the curve) and discretization. Secondly, any information in the optimal curve below the length scale of the discretization length will not be able to be captured. We elaborate on this in App.\ \ref{sec:Filter}, and for an illustration of the results with such a naive implementation, see Figure \ref{fig:unfiltered_2Schlogl}. \textbf{Our contribution} in this work then is to provide a principled algorithm for controlling these sources of noise and providing a recipe for how, in principle, one could approach the continuum limit in a controlled fashion where the algorithm is guaranteed to converge to the unique solution provided the action functional is convex.

\paragraph{Stuck in a local minima:} The objective of the algorithm is to converge upon a phase space trajectory where the value of the action functional along it is minimized. If the difference in the action calculated in two subsequent iterations is below a threshold, or the step size is below a certain threshold, the descent is considered stuck. 

\paragraph{Controlling noise:} 

Firstly, in order to control the noise in the functional gradient $g$, we employ a low-pass filter and obtain a smoothed gradient $g_s$, as explained in App.\ \ref{sec:Filter}. To begin the descent, we choose a small low-pass frequency and allow the algorithm to proceed until it is stuck in a local minimum. The descent gets stuck away from the global minimum whenever the remaining frequencies to be relaxed in the trajectory fall above the pass-band shoulder. Thus we slightly increase the cut-off frequency of the low-pass filter, allowing more meaningful frequencies to go through and at the same time keeping noise from creeping in, and let the algorithm converge to a new curve (explained in App.\ \ref{sec:cutoff_update}). Moreover, it must be noted that the algorithm uses a single step size for the complete curve, i.e. $q^{I+1}=q^I + \epsilon g_s$, where $\epsilon$ is a scalar. Thus we find it best to filter the gradient in a space-uniform parametrization of the curve (see Eq.\ \ref{eq:space_uni}), which ensures that the magnitude of the gradient affects all the discrete segments in a uniform fashion.

Secondly, we also low-pass filter the configuration curve $q^I$ in a time-uniform sampling (see Eq.\ \ref{eq:time_uni}) each time we update the cutoff frequency. The reason for this is, in practice, although we smooth the gradient, some noise might still accumulate in the descended curve. This step ensures that the descended curve is smooth, and in practice, can often take the curve closer to the optimal curve even in the absence of a descent step (for instance, see Figure \ref{fig:2-Schlogl_descent_progress} and the discussion in \ref{sec:int_Ham_EoM}). 

\paragraph{Annealing/Reducing discretization interval:}
If the algorithm is stuck at a curve where no further descent is possible by increasing the low-pass frequency or filtering the trajectory, we increase the sampling of the curve by linearly interpolating on a smaller discretization interval. We also refer to this process as \textit{annealing}, the details of which can be found in App.\ \ref{subsec:anneal}. One might consider adding $\geq 1000$ points in the trajectory, and restarting the algorithm from a new value of the low-pass cut-off frequency at this finely sampled trajectory. A typical reason why the algorithm does not descend further, despite not finding the optimal trajectory, is that the gradient might have meaningful information at length scale smaller than the discretization interval. Increasing the number of sample points then ensures that the meaningful information is not cancelled by the low-pass filter. Through examples considered in the next section, we show that this step can take the curve arbitrarily close to the optimal curve, as it theoretically should since we are approaching the continuum limit of the trajectory.

\subsubsection{Verification protocol}

At the optimal trajectory, we know that the functional gradient of the action in both position and momentum must be zero, in other words, the phase space trajectory must satisfy Hamilton's equations of motion. This gives a canonical verification method for the algorithm, namely numerical integration of Hamilton's equations forwards and backwards from each point in the phase space trajectory and looking for how closely they approach the boundary points. The algorithm halts when we have an exact trajectory that starts and ends at the fixed points, and the deviation of the numerical integration from the end points gives us a way to quantify convergence. We explain the process by which one can obtain such a quantity in App.\ \ref{sec:int_Ham_EoM}. However, it must be noted that symplectic integration in high dimensions is often numerically unstable and we refer readers to \cite{hairer2006geometric, mclachlan2009linearization}, for their careful and efficient implementation.

\subsubsection{Algorithm in a nutshell}
Starting with a uniformly discretized straight line between the end points, the Action Functional Gradient Descent (AFGD) algorithm consists of three nested subroutines:
\begin{enumerate}
    \item Lift the curve to a trajectory and calculate the functional gradient. Filter the gradient in space-uniform parametrization and calculate a step size to descend. Descend to obtain a new curve.
    \item If the descent is stuck, filter the curve in time-uniform parametrization and advance the space-uniform filter parameters. Go to routine $1$.
    \item If the descent is stuck, anneal or add more sample points and refresh the filter parameters. Go to routine $1$.
\end{enumerate}
Optionally, to quantify a distance from the global minimum, one can also repeatedly employ the verification protocol.

A pseudocode is presented in Algorithm \ref{alg:AFGD} and the algorithm is diagrammed in Figure \ref{fig:AFGD_pseudocode}. The mathematical and implementation details of the subroutines can be found in App.\ \ref{app:Details_AFGD} and the code can be found in \cite{Gagrani_AFGD-for-CRN-escapes_2022}.

\subsection{Performance costs and relation to other methods}
\label{subsec:addn_remarks}             

From the last subsection, recall that the costliest step of our algorithm is the constrained optimization function needed to lift the curve. Let us refer to the cost of solving the optimization problem as $C_\text{c.opt.}$. Also, recall that the algorithm consists of three nested subroutines, namely obtaining the gradient and descending, advancing the filter-parameter when stuck, and advancing the number of sample points when stuck in the second routine. Let us refer to the number of points in the curve to be $N_P$, and the number of times the second and third routines are called as $N_2$ and $N_3$, respectively. Moreover, suppose that each time the third routine is performed, $\Delta N_P$ points are added to the curve. The cost of running the algorithm then can simply be estimated as
\begin{align*}
    \text{Cost of algorithm} &\approx  \sum^{N_3}_{i=0} \left( N_P+ i \Delta N_P \right) C_{\text{c.opt.}} \times N_2 .
\end{align*}
As commented upon in Section \ref{sec:lift_curve_to_traj}, a parallel implementation of calculating the lift of a curve can reduce the cost of computation by a factor proportional to the number of cores.

The proposed AFGD algorithm is a novel functional gradient descent method for solving two-point boundary value problems for a Hamiltonian system. Since it is known that the solutions satisfy the Hamilton equations of motion, another popular method for solving the same type of problem is the \textit{shooting method} (\cite{press2007numerical}). In the shooting method, after selecting the two boundary points, one searches for the trajectory connecting them by integrating a brush of trajectories forwards and backwards from the starting and end points, respectively, in order to reach one or more fitting points in between. The technical difficulties associated with the shooting method method are two-fold. First, the search cost of the method increases exponentially with system dimension, as the (cross-sectional) brush of trajectories that must be sampled and refined has codimension 1 in the configuration space (starting position) and also codimension 1 in the tangent space (direction). Second, the method as a single algorithm can also fail for CRNs because the Hamiltonian diverges exponentially in the conjugate momentum variables (see Eq.\ \ref{eq:Ham_CRN_rxn}).  Stabilizing numerical integrators with diverging exponentials is challenging \cite{hairer2006geometric,duruisseaux2021adaptive}, and if floating-point precision is magnified so that no trajectory can converge to a fixed point, the shooting method must be performed recursively along a path, introducing a further step of error bounding and path updating (for example, see \cite{smith2021eikonal}).

\begin{figure*}[t]
\centering
    \begin{subfigure}{0.4\textwidth}
  \includegraphics[width=0.9\linewidth]{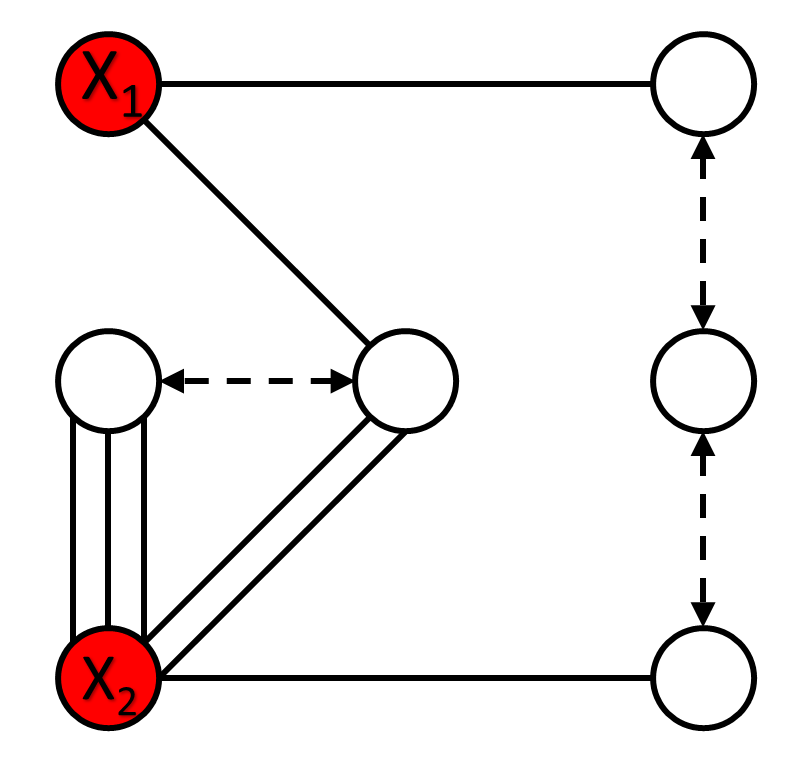}
  \label{fig:4}
\end{subfigure}\hfil 
\begin{subfigure}{0.55\textwidth}
  \includegraphics[width=0.9\linewidth]{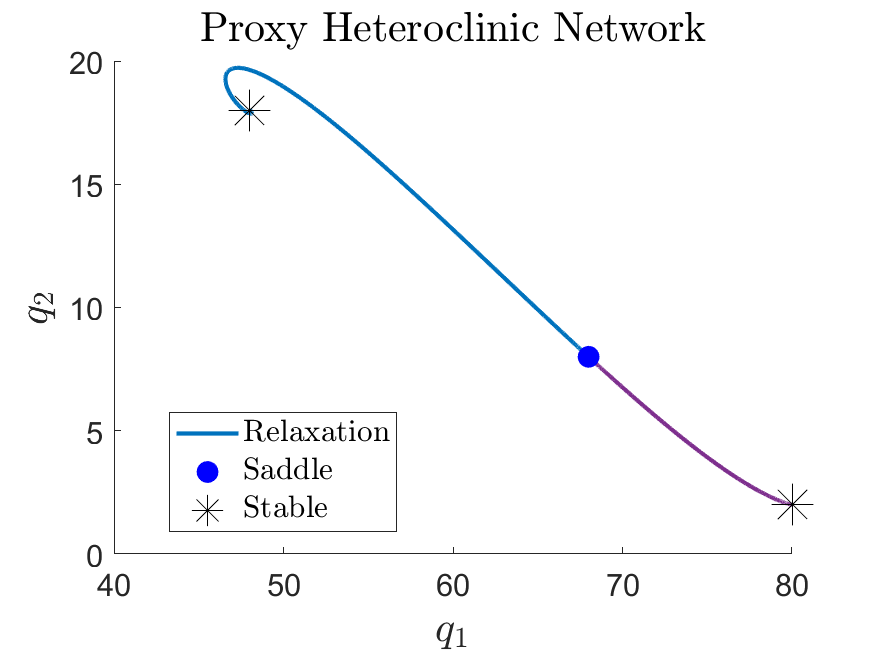}
\end{subfigure}
\caption{Diagrammatic representation of the reaction network (left) and a proxy-heteroclinic network using relaxation trajectories for the Selkov model.}
\label{fig:Selkov_details}
\end{figure*}

\begin{figure*}[htb]
    \centering 
\begin{subfigure}{\textwidth}
  \includegraphics[width=\linewidth]{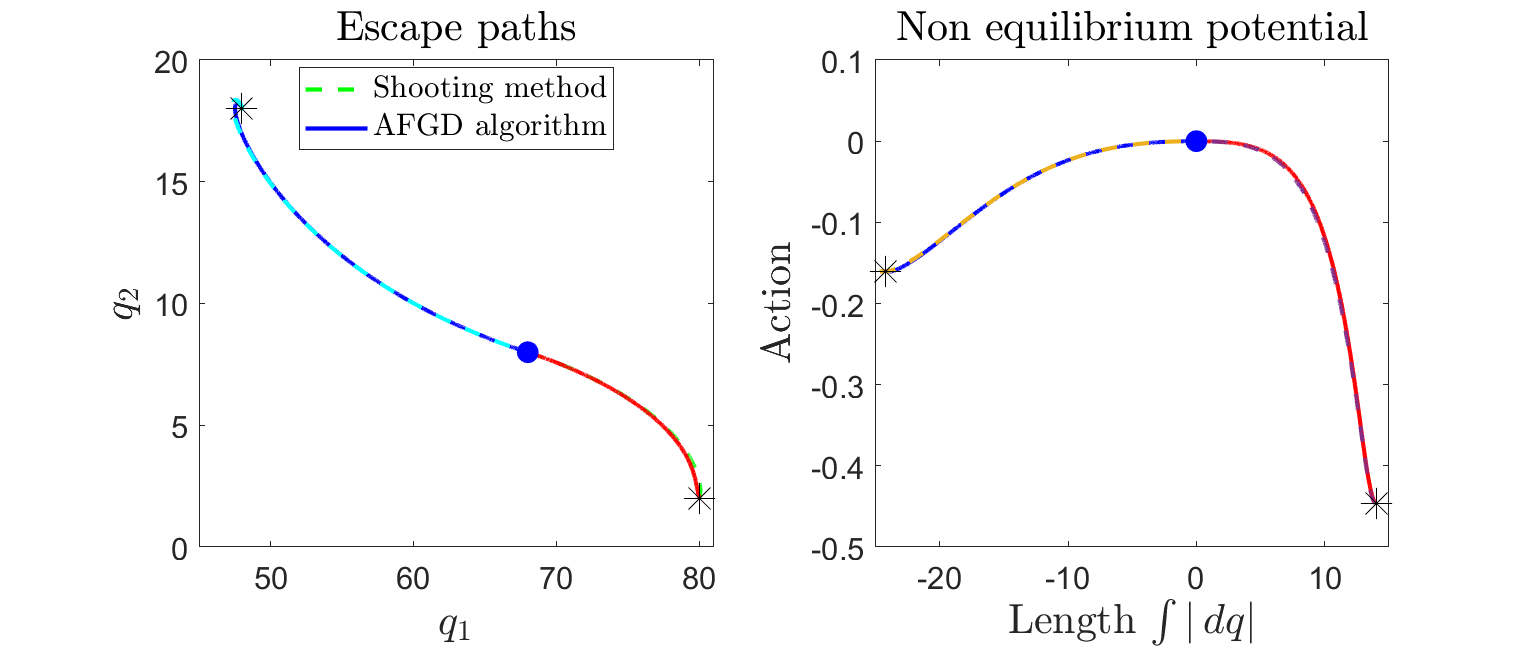}
\end{subfigure}
\caption{Comparing escape paths obtained from AFGD algorithm against the shooting method (left panel) and plotting the non equilibrium potential along the escape (right panel). The data is taken from \cite{smith2021eikonal}, and can be compared against Figures 9 and 15 of the same reference. (For details on the differences between the two methods, see the last paragraph in Section \ref{subsec:addn_remarks}.)}
 \label{fig:Selkov_NEP}
\end{figure*}

The AFGD algorithm, on the other hand, does not suffer from the same issues. Firstly, it relies on computing derivatives, minimizing objective functions, and filtering, all of which are significantly cheaper and more robust than numerical integration. Secondly, due to the above mentioned reasons, it is also feasible in large dimensions. However, it must be noted that the computation time scales with proportionally to the number of points in the trajectory, and in high dimensions a rather fine sampling of the curve might be needed to get a meaningful escape. The demerit of the AFGD algorithm is that it does not find the exact escape path, as the action can reach its minimum value within tolerance without having completely descended upon the solution to the equations of motion. An example of this can be seen towards the ends of the trajectory near the stable fixed points in Figure \ref{fig:Selkov_NEP}, where the trajectory converged upon by the AFGD does not agree completely with that from the shooting method. This suggests that, even in high dimensions, some hybrid of the shooting method and AFGD can be used if the exact escape trajectory is desired. However, if simply a reliable estimate of the action along the escape trajectory is required (to estimate the escape probability, for example), then the AFGD algorithm should be preferred.

\section{Application and results}
\label{sec:Results}

In this section, we will demonstrate the applicability of the AFGD algorithm using three models with varying features. The first application we consider will be to determine the escape paths and NEP of the Selkov model. The Selkov model, first introduced in 1968 by E. Sel'kov  \cite{sel1968self} to model self-oscillations in glycolisis, exhibit relaxation curves that spiral into the stable attractors. Since a spiralling curve, in principle, needs an exponential number of sample points near the fixed point, it poses a difficult challenge for the AFGD algorithm. The other two applications that we consider are higher dimensional analogs of the Schl{\"{o}}gl model. The Schl{\"{o}}gl model was introduced by F. Schl$\ddot{\text{o}}$gl in 1972 to understand non-equilibrium phase transitions in chemical reaction systems \cite{schlogl1972chemical}. The Schl{\"{o}}gl model is an example of a 1-dimensional birth-death process, for which the NEP can be analytically found using Hamilton-Jacobi theory, as we show in Section \ref{sec:1D_bd}. We then define N dimensional analogs of the Schl{\"{o}}gl model, that we term as N-Schl{\"{o}}gl models, for which no analytic results are yet known. We then make use of the AFGD algorithm for the 2-Schl{\"{o}}gl and 6-Schl{\"{o}}gl model, and verify it against Gillespie simulation and Hamilton's equations of motion, respectively. It must be noted that throughout the examples considered in this section, we tune only the algorithm parameters while keeping the underlying algorithm identical to the pseudocode \ref{alg:AFGD}, thus demonstrating that the algorithm is agnostic to the dimensionality of the model. For a MATLAB implementation of the AFGD algorithm, using which we obtain the figures displayed in this paper, see \cite{Gagrani_AFGD-for-CRN-escapes_2022}. 

\begin{figure*}[htb]
    \centering 
\begin{subfigure}{\textwidth}
  \includegraphics[width=0.9\linewidth]{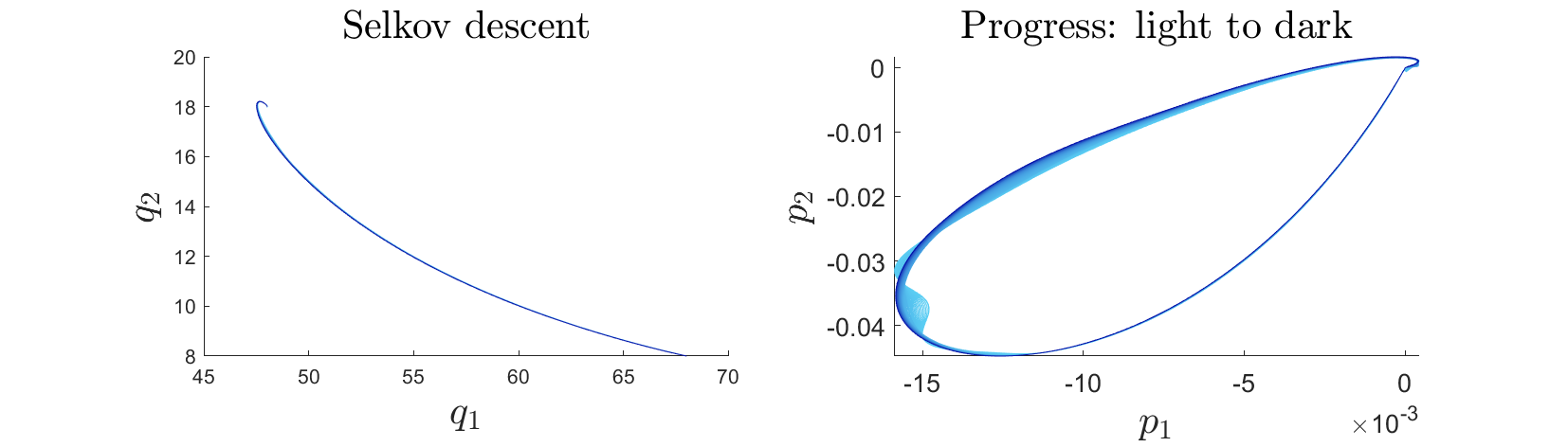}
\end{subfigure}
\medskip
\vspace{-1.35em}
\begin{subfigure}{\textwidth}
  \includegraphics[width=0.9\linewidth]{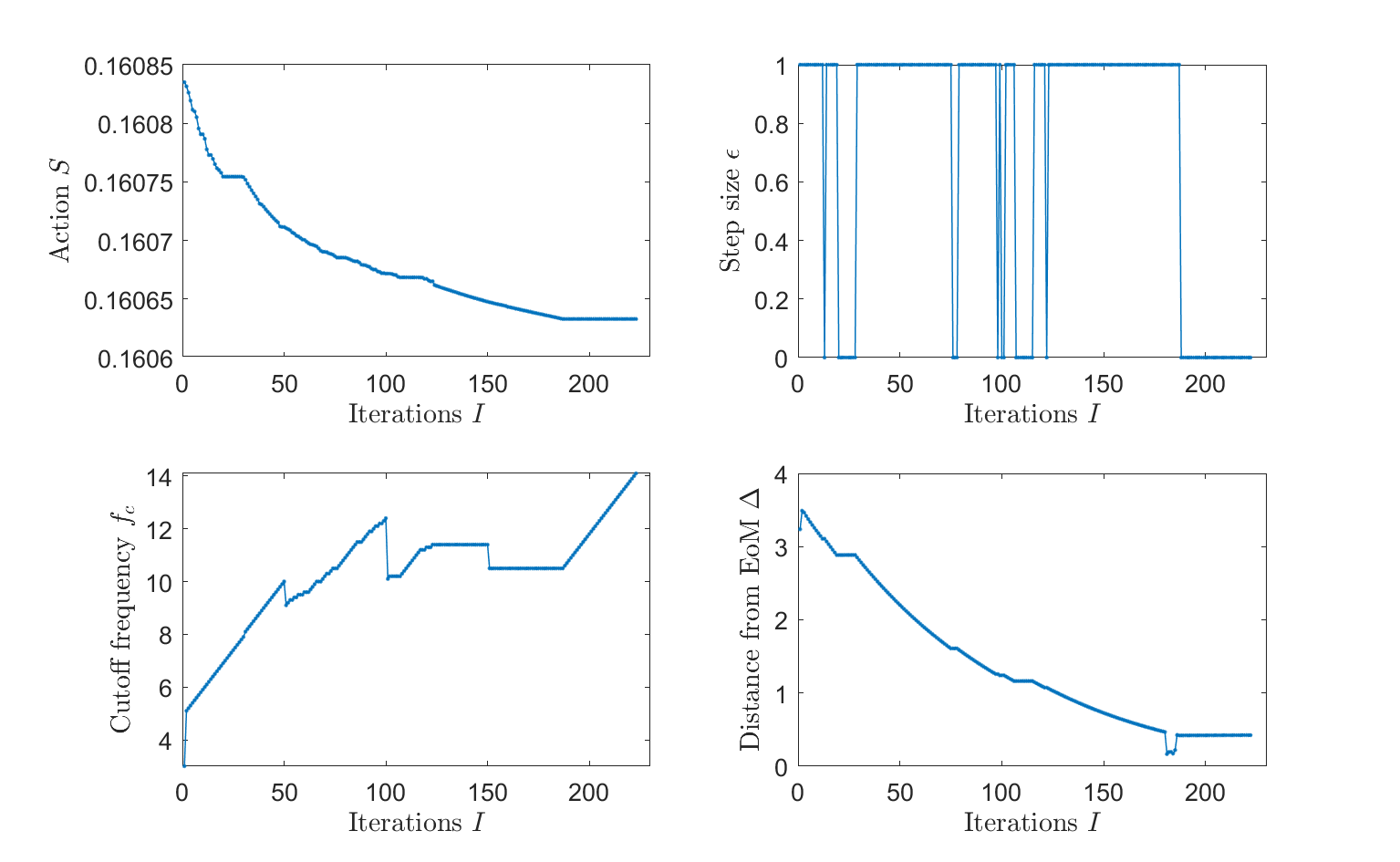}
\end{subfigure}\hfil 
\medskip
\begin{subfigure}{\textwidth}
  \includegraphics[width=0.9\linewidth]{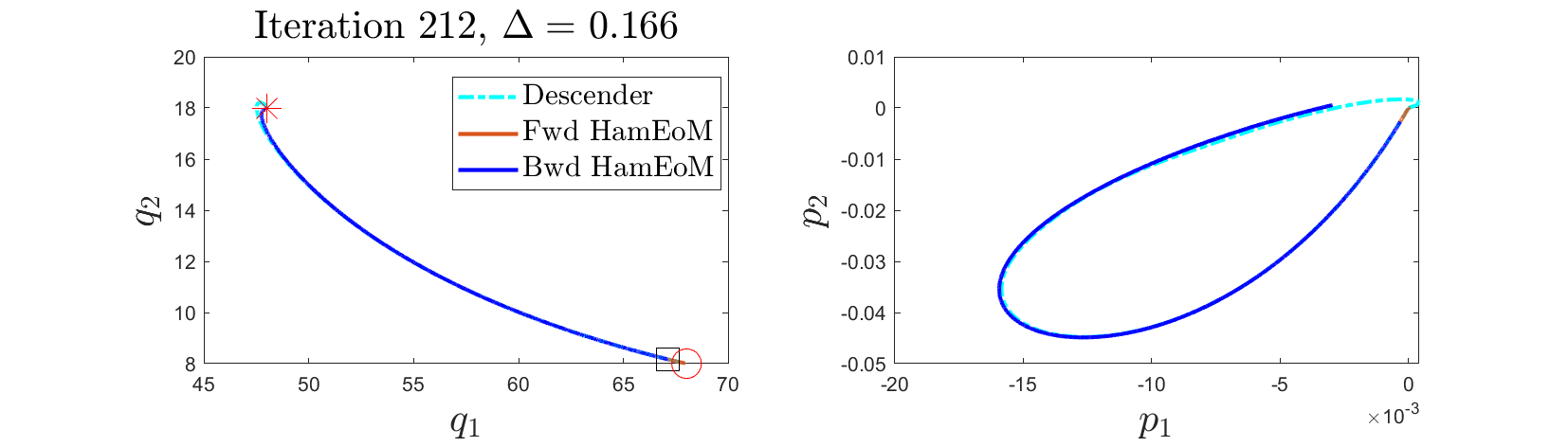}
\end{subfigure}\hfil 
\caption{Descent progress, summary and proof of convergence for the Selkov model after annealing to a trajectory with 4000 points. For initial descent with 2500 points, see Figure \ref{fig:Selkov_run_2500}.}
 \label{fig:Selkov_run_4000}
\end{figure*}

\subsection{Selkov model}

In this subsection we will consider the well-known Selkov model, which has been analyzed by eikonal methods in \cite{dykman1994large,smith2021eikonal}. A peculiarity of the model, that makes it both interesting and challenging for analysis, is that both relaxation and escape trajectories exhibit vorticity around the fixed points. In both \cite{dykman1994large,smith2021eikonal}, the authors identify the escape trajectories by using the \textit{shooting method} (see \cite{press2007numerical}), which amounts to integrating a brush of optimal trajectories in phase space emanating from a point, selecting the one that approaches the desired end point most closely, and repeating. The AFGD algorithm, however, identifies the optimal trajectory using a functional gradient descent, and we summarize the main results of the implementation as well as compare it to the optimal trajectory obtained via the shooting method in \cite{smith2021eikonal} here.

To explicitly define the model, our choice of rate constants is identical to \cite{smith2021eikonal}. The concrete model we consider is, 
\begin{align}
    \ce{$X_1$ <=>[2/3][2648/49] $\phi$ <=>[96/49][4/3] $X_2$ } \nonumber\\
    \ce{$X_1+2 X_2$ <=>[1/441][1/441] $3 X_2$}. \label{CRN:Selkov_model}
\end{align}
A diagrammatic representation of the reaction network can be seen in the left panel of Figure  \ref{fig:Selkov_details}. The resulting fixed points of the model are at $\underline{q}\in \{(80,2) , (68,8) , (48,18)\}$, which we first characterize using the number of repelling directions and then find the \textit{relaxation} trajectories that emanate close to a saddle and reach the stable point in the right panel of Figure  \ref{fig:Selkov_details}, which we refer to as as a `proxy-heteroclinic network'. Our goal is to find the \textit{escape} trajectories that emanate from each of the two stable points and reach the saddle point, and thus obtain the `true heteroclinic network'. For the sake of brevity we will only demonstrate the workings of the algorithm on the top left escape path starting from $(48,18)$ and ending at $(68,8)$, however the same procedure can be used to obtain the other escape as well.

The first feature that one might notice in the top left relaxation path is that it spirals inwards towards the stable fixed point. This is not uncommon for a CRN and we discuss how to obtain initial conditions for such trajectories in App.\ \ref{app:initial_cond}. We then sample the initial configuration space curve uniformly with $2500$ points and run the algorithm, the results of which are summarized in Figure \ref{fig:Selkov_run_2500}. The top row of the figure displays how the descent progresses across the iterations, and the lower two rows summarize how the action, step size, cutoff frequency and minimum distance from the integrated EoM change along the descent.

Although the action seems to stabilize and the step size is zero roughly beyond iteration $150$, the momentum assignment at the converged trajectory (top right panel) clearly looks unreliable. As a rule of thumb, we expect the optimal trajectory to be smooth and containing only low frequency modes. Thus, one can see that despite the AFGD algorithm's effort to smooth the cusp in the momentum initial conditions, it has not yet converged even close to the true solution. To remedy this, we sample the converged curve in $4000$ points, a process we call \textit{annealing} (explained in Section \ref{sec:Algortithm} and App.\ \ref{subsec:anneal}), run the algorithm again starting with this initial condition and plot the summary in Figure  \ref{fig:Selkov_run_4000}. This time we indeed find that the converged momentum assignment is without any cusp (right panel of first row), the action has converged to a yet lower value (second row) and the least distance from the integrated equations of motion $\Delta$ drops below $0.5$ (third and fourth row), guaranteeing convergence near the true solution.

\begin{figure*}[!t]
\centering
  \includegraphics[width=\linewidth]{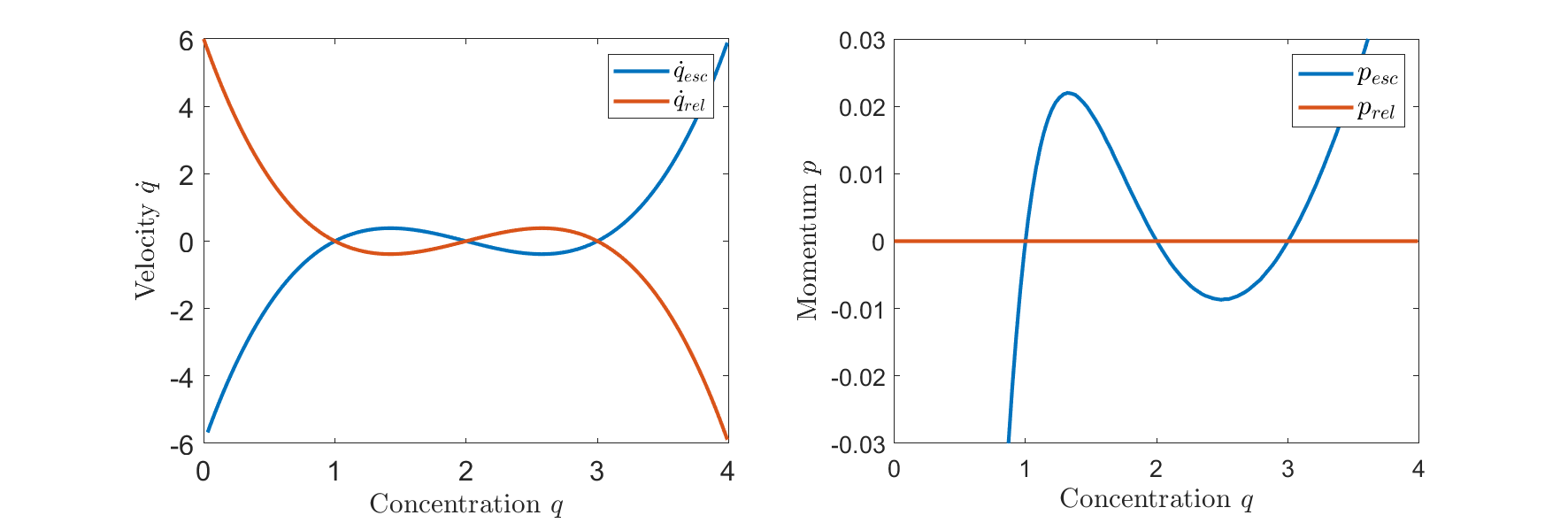}
\caption{$H(p,q)=0$ submanifold in the tangent and cotangent space for the 1-D Schl{\"{o}}gl model.}
\label{fig:1D-Schlogl}
\end{figure*}

We now proceed by the same method to find the other escape trajectory and plot them against the optimal trajectory found by the shooting method in \cite{smith2021eikonal} in the left panel of Figure  \ref{fig:Selkov_NEP}. Using Eq.\ \ref{eq:NEP_2}, we find the NEP along the escape trajectories and thus obtain the true heteroclinic network for the Selkov model, as displayed in the right panel of Figure \ref{fig:Selkov_NEP}.

\subsection{N-D Schl{\"{o}}gl model}
 
\label{subsec:N_Schlogl}

\begin{figure*}[!t]
\centering
\begin{subfigure}{0.5\textwidth}
  \includegraphics[width=\linewidth]{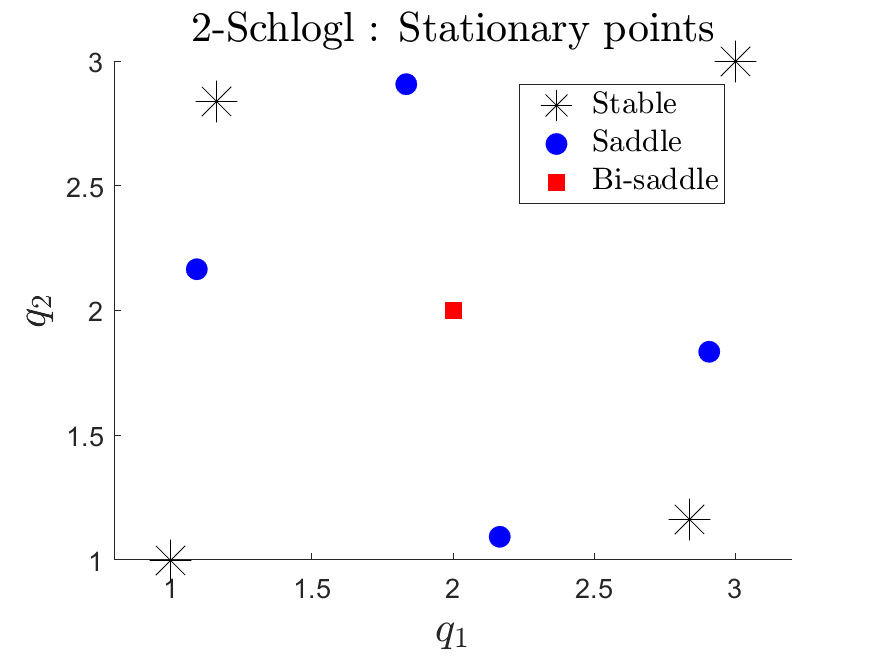}
\end{subfigure}\hfil 
\begin{subfigure}{0.5\textwidth}
\includegraphics[width=\linewidth]{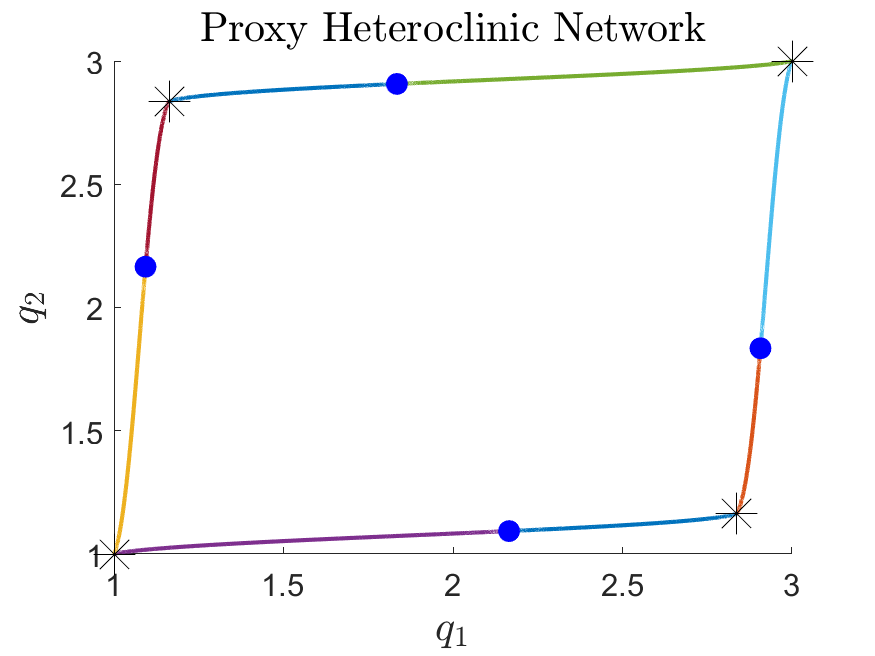}
\end{subfigure}\hfil 
\caption{Fixed points labelled by repelling directions (left) and proxy-heteroclinic network using relaxation trajectories (right) for the 2-Schl{\"{o}}gl model.}
\label{fig:2D-Schlogl-stability}
\end{figure*}

\begin{figure*}[!t]
\hspace*{-1cm}
    \centering 
\begin{subfigure}{\textwidth}
  \includegraphics[width=\linewidth]{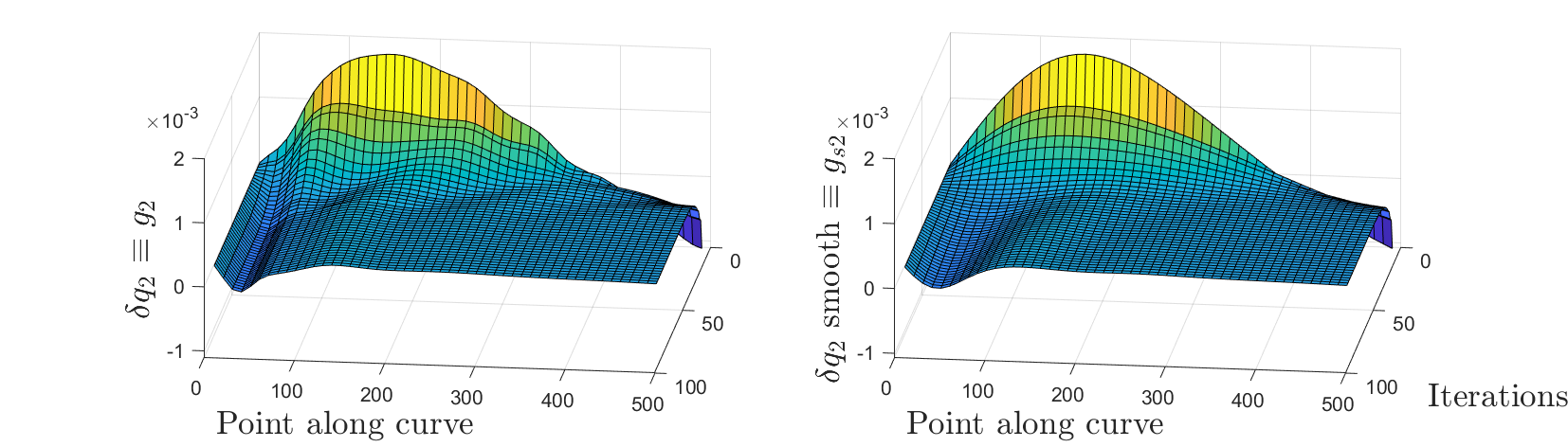}
\end{subfigure}\hfil 
\medskip
\begin{subfigure}{\textwidth}
  \includegraphics[width=\linewidth]{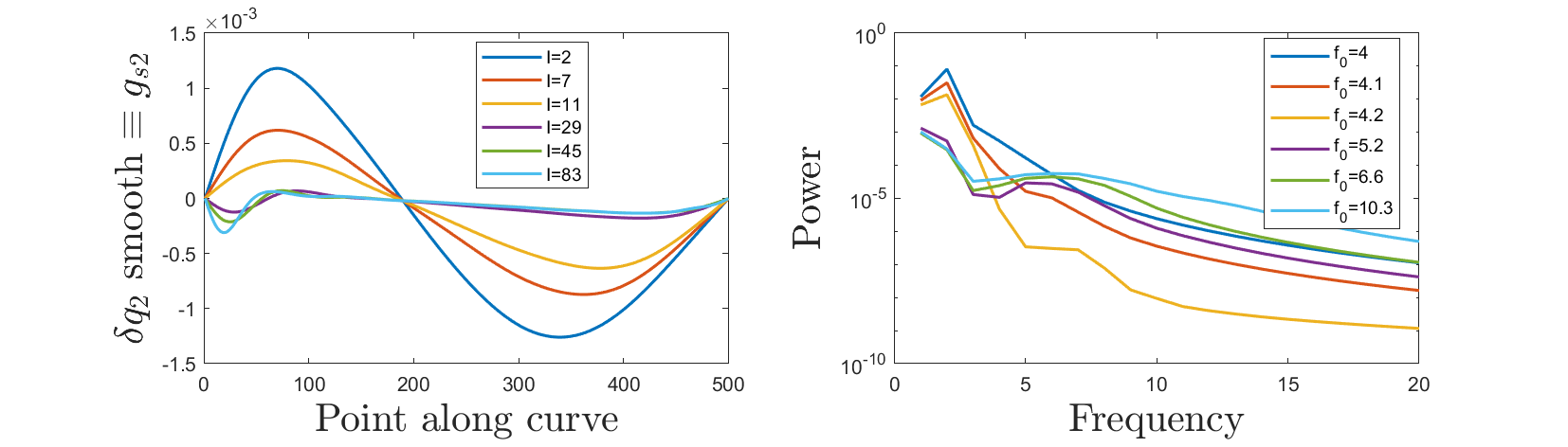}
\end{subfigure}
\caption{Extracting functional gradient from phase space trajectory and smoothing it (top panel). Smoothed gradient and power spectrum at selected iterations (bottom panel) for the 2-Schl{\"{o}}gl model.}
\label{fig:2-Schlogl_smooth_power}
\end{figure*}

\subparagraph{1-D Schl{\"{o}}gl model}
\label{sec:1D_bd}

We begin our discussion by first considering 1-D birth-death processes, of which the 1-D Schl{\"{o}}gl model is an example. A 1-species reaction network is called as a birth-death process if the difference of the vectors denoting the target and source complex for each reaction is either positive or negative one, that is
\begin{align*}
    |y_\beta - y_\alpha| &= 1 \text{ for all reactions in }    \mathcal{R}, \\
    \text{where } \mathcal{R} &= \{ \ce{ y_{\alpha} ->[k_{y_{\alpha} \to y_{\beta}}] y_{\beta} } : k_{y_{\alpha} \to y_{\beta}} \geq 0\}.
\end{align*}
The reactions where the stoichiometry of the target complex is one more or less than the source correspond to `birth' or `death' reactions, respectively \cite{anderson2015lyapunov}. 

Using the form of the CRN Hamiltonian in Eq.\ \ref{eq:Ham_CRN_rxn}, we can write the Hamiltonian function of a 1-D birth death process as
\begin{align}
        H_\text{1-b.d.}(p,q)&= \big( e^{ p} - 1\big)r_{+1}(q) + \big( e^{- p} - 1\big)r_{-1}(q), \label{H_1bd}
\end{align}
where $r_{+/- 1}$ are polynomials in $q$ with coefficients as the corresponding rate constants appearing in the birth/death reactions. From the Hamiltonian, one can read that the deterministic rate of growth in the concentration of the species is $\dot{q}=r_{+1}(q)-r_{-1}(q)$ and the roots of $\dot{q}$ correspond to fixed points of the system.

We can now proceed to find the NEP for such processes by solving for an escape momentum assignment in the $H_\text{1-b.d.}=0$ submanifold. Since we have only one species and a two-dimensional phase space manifold, for a birth-death process this constraint uniquely picks a $p_\text{esc}\neq 0$,   
\begin{align}
     H_\text{1-b.d.}(p_\text{esc},q)&=0,\nonumber\\
     p_\text{esc}(q) &= \ln\left(\frac{r_{-1}(q)}{r_{+1}(q)} \right). \label{eq:p_esc_1bd}
\end{align}
Following the discussion on Hamilton-Jacobi theory in Section \ref{sec:Ham_CRN}, integrating the escape momentum in Eq.\ \ref{eq:p_esc_1bd} yields the NEP $\mathcal{V}$. Notice that at the fixed points, $p_\text{esc}=0$, and correspondingly the NEP is at a local extrema. To ensure that the NEP is always greater than zero, we find a possibly non-unique fixed point $\underline{q}$, such that 
\begin{align*}
    \underline{q} &= \arg \min_q \int_0^q \ln\left(\frac{r_{-1}(q)}{r_{+1}(q)} \right) \,dq,
\end{align*}
using which the NEP is defined as  
\begin{align}
     \mathcal{V}_\text{1-b.d.}(q) &= \int_{\underline{q}}^q \ln\left(\frac{r_{-1}(q)}{r_{+1}(q)} \right) \,dq. \label{eq:NEP_1bd}
\end{align}
The ratio of the value of the stationary distribution at points $n_2$ and $n_1$ in a stochastic simulation of scaling volume $V$ is then given by 
\begin{align}
 \frac{\pi(n_2)}{\pi(n_1)} &= \exp\left(-V \int_{q_1}^{q_2} \ln\left(\frac{r_{-1}(q)}{r_{+1}(q)} \right) \,dq\right).
\end{align}

\begin{figure*}[htb]
    \centering 
\begin{subfigure}{\textwidth}
  \includegraphics[width=\linewidth]{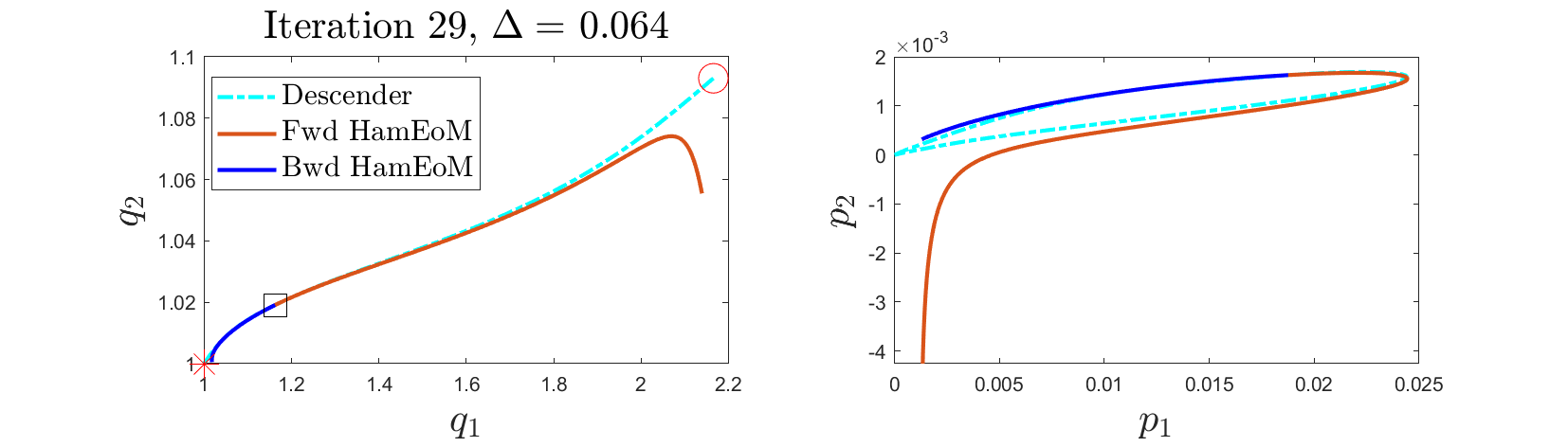}
\end{subfigure}\hfil 
\medskip
\begin{subfigure}{\textwidth}
  \includegraphics[width=\linewidth]{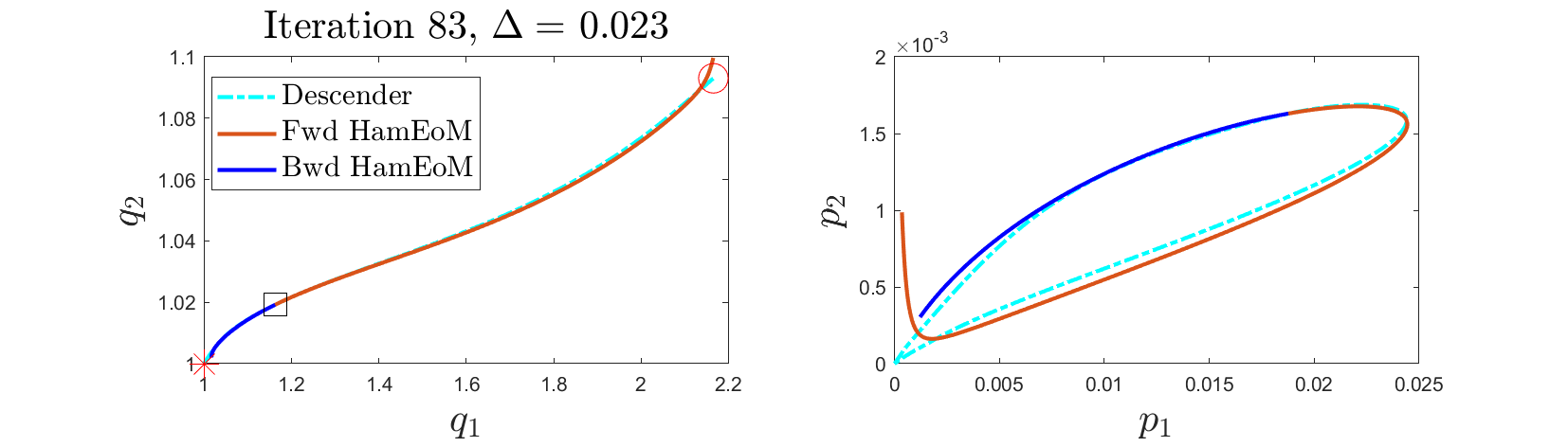}
\end{subfigure}
\caption{Progress of integrated Hamilton's equations of motion against the AFGD algorithm at selected iterations for the 2-Schl{\"{o}}gl model.}
\label{fig:2-Schlogl_HamEoM}
\end{figure*}

The results concerning the NEP of 1-D birth death systems are well known, and an alternative derivation can be found in \cite{anderson2015lyapunov}. A particular example of a birth-death process relevant to our purposes here is the \textbf{Schl{\"{o}}gl model} \cite{schlogl1972chemical}. For a pedagogical exposition of an application of Hamilton-Jacobi theory to Schl{\"{o}}gl model and its generalizations, see \cite{lazarescu2019large,smith2020intrinsic}. For this section, we use the concrete 1 species reaction network 
\begin{align}
    \ce{$\phi$ &<=>[6][11] $X$ } \nonumber\\
    \ce{$2 X$ &<=>[6][1] $3 X$}, \label{CRN:Schlogl_1d}
\end{align}
where the rate constants are identical to example 10 of \cite{anderson2015lyapunov}. For a diagrammatic representation of the reaction network, see top left panel in Figure  \ref{fig:N-D_Schlogl_CRN}. For the choice of reaction network in Eq.\ \ref{CRN:Schlogl_1d}, we have 
\begin{align}
    r_{-1}(q) &= 11 q + q^3 \nonumber \\
    r_{+1}(q) &= 6 + 6 q^2 \nonumber\\
    \dv{q_\text{rel}}{t} = \pdv{H}{p}\bigg|_{p=0} &= \phantom{-} r_{+1}(q) - r_{-1}(q) \nonumber\\
    &= - (q-1)(q-2)(q-3). \label{eq:1Sch_rate}
\end{align}

Notice that the relaxation flow field has three roots, which means that the model exhibits three fixed points for the rate constants chosen $\underline{q}\in\{(1),(2),(3)\}$. We display the velocity and momentum assignment of the variational solution along the $H(p,q)=0$ submanifold for this model in Figure \ref{fig:1D-Schlogl}.

\subparagraph{2-D Schl{\"{o}}gl model}
\label{sec:2Schlogl_model}

We define the 2-D Schl{\"{o}}gl model to be a two species reaction network with 1-Schl{\"{o}}gl reactions in each species and diffusion between the two species. For the results in this section and App.\ \ref{app:Details_AFGD}, we use the concrete network 
\begin{align}
    \ce{$\phi$ &<=>[6][11] $X_1$ } & \ce{$\phi$ &<=>[6][11] $X_2$ } & \ce{$X_1$ &<=>[0.15][0.15] $X_2$ } \nonumber\\
    \ce{$2 X_1$ &<=>[6][1] $3 X_1$} & \ce{$2 X_2$ &<=>[6][1] $3 X_2$}, \label{CRN:Schlogl_2d}
\end{align}
which we represent diagrammatically in the top right panel in Figure  \ref{fig:N-D_Schlogl_CRN}. For the choice of rate constants, the relaxation flow field has nine fixed points, out of which four are stable (with no repelling direction), four are saddle (with one repelling direction) and one is unstable (with all repelling directions). 

As explained in Section \ref{sec:HJ_theory}, in order to understand the switching dynamics from one stable fixed point to another, we need to only consider the least-improbable path of escape. We know that the least-improbable path between two nearby fixed points is through the saddle point between them, and thus we look for the equation of motion that emanates from a stable attractor and ends at any of the closest saddle points using the AFGD algorithm. Before looking for escape paths however, we find it useful to find the relaxation trajectories that emanate close to a saddle point and reach a nearby stable point. This information yields a `proxy-heteroclinic network' and tells us whether or not the system exhibits any vorticity in the areas of interest. For the position and classification of the fixed points and a `proxy-heteroclinic network' for the 2-Schl{\"{o}}gl model, see Figure  \ref{fig:2D-Schlogl-stability}.


We now proceed to identifying all the relevant escape paths to create a heteroclinic network for the 2-Schl{\"{o}}gl model. As an illustrative example of an application of the AFGD algorithm, we first focus our attention to the bottom left escape path from the stable root $\underline{q_I}=(1,1)$ to the saddle root $\underline{q_F}=(2.16,1.09)$. From the criteria specified in App.\ \ref{app:initial_cond}, we begin the descent from a straight line initial condition as shown in Figure  \ref{fig:initial_conditions}. We run our algorithm starting from a curve consisting of $500$ points, and we summarize the progress of the descender in Figure  \ref{fig:2-Schlogl_descent_progress}. The topmost row depicts how the the phase space trajectory changes in the configuration space (left) and momentum space (right) as the algorithm progresses. The last two rows summarize the progress of the action along the escape, step size, cutoff frequency and least distance from the end points of the integrated EoM during descent.

One can observe that the descent in the action has a step-like behavior which is mirrored in the step size as well as cutoff frequency. What is happening is that the algorithm descends at a given setting of a cutoff frequency as much as it can, by reducing the step size as long as it is above a certain threshold and the value of the action function is strictly decreasing. Once no further step can be taken because the step size has decreased below the minimum threshold of $\epsilon_\text{thresh}$, the cutoff frequency is slightly raised to allow the pass-band to inject more meaningful signal into the gradient. It can be seen in Figure \ref{fig:2-Schlogl_smooth_power} that the overall magnitude of the gradient decreases and its power spectrum becomes flatter as the descent proceeds through the iterations of the algorithm. For more details, see App.\ \ref{sec:cutoff_update}.

The bottom right panel in Figure \ref{fig:2-Schlogl_descent_progress} shows how the minimum distance $\Delta$ of the integrated Hamilton's EoM, as explained in App.\ \ref{sec:int_Ham_EoM}, changes along iterations of the algorithm. Curiously, the distance continues to decrease even when the step size is zero for $I>55$. The reason for this is that our algorithm also smooths the configuration space curve in a time-uniform sampling, which for the 2-Schl{\"{o}}gl demonstrably takes the trajectory towards the optimal solution. The distance, however, ceases to decrease after a point, and it is a signal that the number of points in the trajectory needs to be increased if further progress is to be made. We show the closest integrated trajectory overlaid on the descender trajectory for a few iterations in Figure  \ref{fig:2-Schlogl_HamEoM}. Since, in the previous example, we have already discussed the process of descending further towards the true solution by `annealing' (increasing the number of sample points in the trajectory), we will not pursue it here. 

\subparagraph{Validation against Gillespie algorithm}
Now that we have verified the correctness of the descended trajectory against Hamilton's EoM, we can ask how well does it perform against a stochastic modelling method such as the Gillespie algorithm \cite{gillespie2007stochastic}? To compare the two, we run two simulations with scaling volume $V=300$ and find the scaled-log improbability of their stationary distribution $\pi$, i.e. we calculate $(-1/V)\log(\pi)$. Since the two simulations give similar results, we only display the contour plot resulting from the second simulation for clarity in the left panel of Figure  \ref{fig:2-Schlogl_Gillespie}. We then proceed to find the scaled log-improbability along only the escape trajectory output by the AFGD algorithm and compare it against the action or NEP found using Eq.\ \ref{eq:NEP_2}. The result is displayed in the right panel of Figure \ref{fig:2-Schlogl_Gillespie}. We can see that the scaled NEP obtained from the simulation is higher than the log-improbability obtained from the algorithm, as it theoretically should because the variational solution provides a lower bound, becoming exact in the $V\to \infty$ limit. For a similar plot for the 1-Schl{\"{o}}gl model, see Figure 1 in \cite{anderson2015lyapunov}.

Having convinced ourselves of the correctness of the algorithm by two different means, i.e.\ via integrating Hamilton's EoM and verifying against stochastic simulation, we proceed to find the true heteroclinic network for the 2-Schl{\"{o}}gl model. The heteroclinic network consists of all the escape trajectories from every stable attractor, as well as the log improbability or NEP along each escape. We obtain the escape paths by running the algorithm for every pair of nearby stable and saddle points, and display the resulting heteroclinic network in Figure \ref{fig:2-Schlogl_hc_net}.

\subparagraph{N-D Schl{\"{o}}gl model}

The AFGD algorithm is defined independently of the dimensions of the system, and thus works equally well for dimensions higher than 2. To demonstrate this point, we consider an N-dimensional generalization of the Schl{\"{o}}gl model. Analogous to the 2-Schl{\"{o}}gl model, we define a system with $N$ species and include the Schl{\"{o}}gl reactions for each species, as well as diffusion amongst the species.

For $N>2$ we have to make a choice regarding the underlying diffusion network, and to simplify considerations we choose a fully connected diffusion network, i.e. each species is diffusing with all species. Other choices could also be made, such as the species could form a 1-D or a 2-D lattice with diffusion only between the nearest neighbors, but this choice is beside the point of our purposes here. A diagrammatic representation of the 3-D and 4-D Schl{\"{o}}gl model can be seen in the lower line of Figure \ref{fig:N-D_Schlogl_CRN}. 

For an application of the AFGD algorithm to higher dimensional systems, we consider the 6-Schl{\"{o}}gl model. To demonstrate the efficacy of the algorithm, the descended phase space trajectory is plotted against Hamilton's EoM in Figure \ref{fig:6-Schlogl_HamEoM}. The output trajectory is obtained using $2000$ sample points and running for under $300$ iterations at a relatively low cutoff frequency. Since other details are not particularly illuminating, we omit the summary plots for brevity, but make it available in \cite{Gagrani_AFGD-for-CRN-escapes_2022}.

\section{Discussion}

\label{sec:FutureResearch}

Chemical reaction networks (CRNs) are essential for modeling a wide range of natural phenomena such as star formation, the origin of life, spatial or ecological patterns in living organisms, and climate (\cite{smolin1996galactic,smith2016origin,turing1990chemical,benzi1983theory}). The widespread utility of CRNs stems from their ability to exhibit dynamic equilibria which, unlike the equilibrium at the top or bottom of a potential well where the velocities of objects are zero, are states of a system where the composition remains unchanging although the constituent species are being dynamically exchanged. The state of a star, organism, ecosystem or climate, when modeled as a stochastic CRN, can undergo transitions from one dynamic equilibrium to another, and it is the probability of such a transition occurring that we give an algorithm to numerically estimate in this work. It must be noted that since the probability of transitions is exponentially suppressed in the number of simulated species, finding cheaper ways of estimating them is of practical importance. 

Our main contribution has been to employ the Hamilton-Jacobi formalism to rigorously formulate the problem of finding transitions between steady states (or fixed points) of a CRN as a MinMax problem, and to construct a principled algorithm to solve it. The functional whose value along its optimal points are the desired transitions is called the \textit{action} functional, due to which we name our algorithm as the \textit{Action Functional Gradient Descent} (AFGD) algorithm. Our algorithm only requires computation of derivatives, solving function optimization problems, and basic tools from signal processing. Moreover, while the algorithm itself does not rely on numerical integration, its validity can be readily verified by integrating the equations of motion starting anywhere along the converged output (for more detailed, see Section \ref{sec:Algortithm}). Finally, in Section \ref{sec:Results}, we explore applications of the algorithm on several high dimensional problems and validate them against other methods of obtaining transition paths and probabilities. 

While in this work, we only use the algorithm to calculate transitions between two fixed points, in principle it can be used to find a transition between any two points which are guaranteed to have a direct optimal phase space trajectory connecting them. In particular, rather than escaping from a stable fixed point to a nearby saddle point, one might be interested in finding the probability of an escape to any point within the basin of escape of the stable point in the long time limit. As explained in Section \ref{sec:theory}, the optimal trajectories that connect the stable fixed point to the desired point will also be in the $H(p,q)=0$ submanifold, and can readily be found (by leaving the momentum at the end point unspecified) by the AFGD algorithm. In this way, by connecting arbitrary points to their nearest stable fixed points, in principle, one can assign a transition probability to each point on the state space and recover the occupation probability distribution that one would otherwise obtain by running a stochastic simulation of a CRN for a very long time. We will then leave is as future work to take as input the time-series data or occupation probability distribution obtained from a stochastic simulation and to learn the CRN from which it was generated. Due to the widespread utility of CRN in modeling real-world phenomena, a machine learning algorithm to infer the CRN from its simulated data would have several significant scientific applications.        

\section*{Acknowledgements}
This work was partially funded by the National Science Foundation, Division of Environmental Biology under the grant titled `Collaborative Research: From ecological dynamics to adaptive evolution during the origin of life' (Grant No: 2218817). PG wants to thank Ahmet Alacaoglu and Stephen Wright for discussion and references on optimization, David Anderson, Gheorghe Craciun and Tung Nguyen for explanations and references on mathematical theory of CRN, Melvin Leok for references on symplectic integrators, Alex Levchenko for clarifications and references regarding spectrum of the action functional, Dave Auckly for clarifying the relationship between Hamilton's EoM and Hamilton-Jacobi PDE, friends Abhimanyu Dubey, Asvin G., Merritt Losert, Rahul Parhi, John Podczerwinski, Tymofii Sokolskyi and Vladimir Sotirov for useful technical discussions, and advisors David Baum and Sridhara Dasu for financial and professional support. 

\appendix

\section{Experimental applications of AFGD}
\label{app:App_nontech}

Consider a dynamical experiment where the individual populations of a list of different types or `species' of objects is tracked through time. Supposing all objects are discrete and objects of the same species are indistinguishable, we can label the different species as $\{S_1,S_2,\ldots\}\equiv \mathcal{S}$ and obtain a population vector $(n_1,n_2,\ldots)\equiv \vec{n}$ which represents the state of the experiment. Furthermore, we will denote the entry in the time series data at time $t$ as $(\vec{n}_t,t)$ or simply $\vec{n}_t$. 

In a stochastic experiment, starting from the same initial condition does not guarantee the same observation after some time $t$ has elapsed. In this case, if we are to completely understand the dynamics of the system under investigation, we have to consider a collection or `ensemble' of experiments starting from some initial condition $(\vec{n}_0,0)$ and see how the relative frequencies of the different observations in the experiment ensemble changes in time. More precisely, from the data of the experiment ensemble, we get the probability that the experiment is found in $\vec{n}_t$ at time $t$ given that it was in $\vec{n}_0$ at time $0$, denoted as $\mathbb{P}[(\vec{n}_t,t) \cap (\vec{n}_0,0)]$. 

Recall from the definition of conditional probability that
\begin{align}
    \mathbb{P}[\vec{n}_t \cap \vec{n}_0] &= \mathbb{P}[\vec{n}_t | \vec{n}_0] \mathbb{P}[\vec{n}_0], \label{eq:cond_prob}
\end{align}
where $\mathbb{P}[\vec{n}_t | \vec{n}_0]$ denotes the probability of observing $\vec{n}_t$ conditioned on having observed $\vec{n}_0$. This formula can be useful to us in two ways. First, if the conditional probability $\mathbb{P}[\vec{n}_t | \vec{n}_0]$ is known, given an observation at the time of initialization (detection), we can make a probabilistic prediction (retrodiction) of what will be (was) the state of the system at a future (past) time $t$. Second, if we have access to both $\mathbb{P}[\vec{n}_t \cap \vec{n}_0]$ and $\mathbb{P}[\vec{n}_0]$, then we can improve our model of the system. In the remainder of the paper we focus on the first of these and comment on the second use in Section \ref{sec:FutureResearch}.

In Section \ref{sec:HJ_theory} (Eq.\ \ref{eq:StationaryAction}), we give a detailed exposition of how the Hamilton-Jacobi formalism yields asymptotic estimates to the conditional probability term $\mathbb{P}[\vec{n}_t | \vec{n}_0]$. The basic idea can be summarized as follows. Whenever the dynamics of the ensemble of experiments can be modelled as a Hamiltonian dynamical system satisfying certain conditions, in the limit of a large number of objects in each experiment (making sample fluctuations negligible) the conditional probability $\mathbb{P}[\vec{n}_t | \vec{n}_0]$ can be estimated by finding the optimal path that takes the system from $\vec{n}_0$ to $\vec{n}_t$. The optimality condition can take various interpretations such as most probable, least improbable, least costly, etc.\ \cite{snarski2021hamilton}, but in all these interpretations there is a variational principle at play which derives from the Hamiltonian structure of the dynamics. Instead of yielding the conditional probability, Hamilton-Jacobi theory gives a descaled conditional log improbability that is referred to as the action function and denoted by
\begin{align}
 S(\vec{q}_t||\vec{q}_0)\equiv -\frac{1}{V}\log\left(\mathbb{P}[V\vec{q}_t | V\vec{q}_0]\right), \label{eq:S_div}   
\end{align}
where $\vec{q} = \vec{n}/V$. In technical terms, the action function is a solution to the Hamilton-Jacobi PDE, and the optimal paths given by ODEs of Hamilton's equations of motion are its characteristic curves. While historically this formalism has been used to describe motion of celestial or terrestrial objects and is the workhorse of classical mechanics, it can indeed also be used to quantify population space dynamics such as in chemical reactions, population genetics or economics \cite{baez2012quantum,smith2016origin,smith2021beyond,shubik2016guidance}.

A particular type of measurement procedure that is of practical relevance is letting the stochastic experiment evolve under its own rules for an extremely long time i.e. several orders of magnitude more time than any characteristic time scale in the system. Due to a theorem by Perron and Frobenius \cite{pillai2005perron}, the probability distribution of the ensemble of experiments will always approach a stationary distribution independent of the initial conditions, which we denote by $\pi(\vec{n})$. By definition then, $\mathbb{P}[(\vec{n}^\prime,\infty) \cap (\vec{n},0)] = \pi(\vec{n}^\prime)$, and the ratio of the stationary distribution at two population states $\vec{n}$ and $\vec{n}^\prime$, will be given by 
\begin{align}
    \frac{\pi(\vec{n}^\prime)}{\pi(\vec{n})} &= \mathbb{P}[(\vec{n}^\prime,\infty)|(\vec{n},0)]\equiv \pi(\vec{n}^\prime|\vec{n}). \label{eq:stat_dist_ratio}
\end{align}
Thus, for the stationary distribution, the conditional probability between two events $\pi(\vec{n}^\prime|\vec{n})$ is simply equal to the ratio of their stationary probabilities $\pi(\vec{n}^\prime)/\pi(\vec{n})$. Using Hamilton-Jacobi theory, we will show in Section \ref{sec:HJ_theory} that $\pi(\vec{n}^\prime|\vec{n})$ is asymptotically estimated by finding the optimal path connecting $\vec{n}$ and $\vec{n}^\prime$ in the $H(p,q)=0$ submanifold of the phase space. Thus, the complete solution to the Hamilton-Jacobi PDE in the $H(p,q)=0$ manifold, also referred to in literature as Non-Equilibrium Potential (NEP), plays a central role for a certain set of questions. We will return to this point and explain its relevance to our algorithm after we introduce the basic idea behind stochastic chemical reaction network theory. 

To explain the type of systems that can be modelled using stochastic chemical reaction networks, we first need an understanding of a `reaction'. In a system described by the population of its individual species, a reaction is a rule for a transition that replaces a collection or multiset of objects with a different one. For example, suppose in a system consisting of only two species, a reaction 
\begin{align*}
    S_1 \to S_2
\end{align*}
will transition the population vector $(n_1,n_2)$ to $(n_1-1,n_2+2)$. A `reaction network' is a set of such reactions, each of which are assigned a rate constant determining a proportionality factor in the propensity or how often these reactions are to occur at random. In the particular case of a stochastic `chemical' reaction network (CRN), the reaction rates are also proportional to the concentrations of the species and model proportional sampling without replacement. Although the dynamics of chemical reaction networks can be seen rigorously as arising from an underlying physical process \cite{gillespie1992rigorous}, we will concern ourselves only with the modelling aspects of a stochastic CRN.

Chemical reaction network theory has a long history, and for an excellent review we point the readers to \cite{yu2018mathematical}. The most notable feature of chemical reaction systems is that they can exhibit a wide array of dynamics, with single or multiple attractors, limit cycles, etc. Although there are a host of applications that CRN have, to continue along our practical problem let us pick a particular one. Supposing that our system under investigation can be modelled as a multi-attractor CRN, we want to design an experiment to physically observe the different attractors. This amounts to finding how many experiments our ensemble should consist of so as to detect a significant amount in each attractor, which, as we will explain, is a type of problem that our algorithm can help get numerical estimates to.

Let us refer to the set of stable attractors in the population space by $\underline{\vec{n}}\equiv \{ \underline{\vec{n}}_1, \underline{\vec{n}}_2,\ldots \}$. To estimate the number of experiments our ensemble must consist of, it is sufficient to estimate the ratio of the stationary distribution $\pi(\underline{\vec{n}}_k|\underline{\vec{n}}_j)$ at any two stable attractors $\underline{\vec{n}}_j$ and $\underline{\vec{n}}_k$. From CRN theory and other considerations mentioned in the main text, any optimal path in the $H(p,q)=0$ submanifold that emanates from the stable attractor $\underline{\vec{n}}_j$ must go through an adjoining saddle point, before going on to the next stable attractor, and so on in the process of reaching some $\underline{\vec{n}}_k$. The optimal paths that take the system out of a stable attractor to an adjoining saddle attractor are called `escape paths', and their collection consists of what we call a `heteroclinic network'. The heteroclinic network then determines the asymptotic stationary distribution of the stochastic system at the stable attractors, and the escape paths that constitute the heteroclinic network is precisely what our algorithm is designed to find (for e.g.\ see Figure  \ref{fig:2-Schlogl_hc_net}).  

We conclude this section by explaining how one can also use the algorithm to estimate the ratio $\pi(\vec{n}|\underline{\vec{n}}_1)$ for a general multi-attractor CRN at any point $\vec{n}$. First, consider the case where the system exhibits a unique stable attractor $\underline{\vec{n}}_1$. In that case, the ratio $\pi(\vec{n}|\underline{\vec{n}}_1)$ is found simply by determining the optimal path that connects the stable attractor $\underline{\vec{n}}_1$ to the point $\vec{n}$, which is how we recover the Horn-Jackson potential using Hamilton-Jacobi theory in Section \ref{sec:rederiv_HJ}. Next, in the case where the system has multiple attractors, we can use the algorithm to determine the stable attractor $\underline{\vec{n}}_k$ from which there is a direct escape path (that does not go through any other stable attractor) to $\vec{n}$. Once the optimal path joining $\underline{\vec{n}}_k$ and $\vec{n}$, or equivalently $\pi(\vec{n}|\underline{\vec{n}}_k)$ is found, we can use the heteroclinic network (as discussed in the preceding paragraph) and

\begin{align*}
    \pi(\vec{n}|\underline{\vec{n}}_1)
    &=
    \pi(\vec{n}|\underline{\vec{n}}_k)\pi(\underline{\vec{n}}_k|\underline{\vec{n}}_1)
\end{align*}
to determine the desired ratio. In technical terms, since the stationary distribution is asymptotically estimated by the solution of the Hamilton-Jacobi PDE in the $H(p,q)=0$ submanifold, the conditional probability between any two points $\vec{n}_1$ and $\vec{n}_2$ is obtained by finding the difference in the values of the solution while traversing only along any of the optimal paths or characteristic curves. The geometry of the optimal paths in the $H(p,q)=0$ submanifold is such that each stable attractor has a region within which each point is connected to the attractor by an escape path (escape basin \cite{smith2021eikonal}), and neighboring regions are connected by a saddle point which lies on the boundaries of two adjacent basins (for a detailed exposition, see \cite{smith2020intrinsic}). Thus one can can use our algorithm to find any number of desired characteristic curves of the Hamilton-Jacobi PDE in the $H(p,q)=0$ submanifold and construct the complete stationary distribution for a chemical reaction network with multiple attractors.

\section{Rederiving previous results in the Hamilton-Jacobi formalism}
\label{sec:rederiv_HJ}

In this subsection we illustrate the usage of the formalism developed up to this point by rederiving some well known results in CRN theory. We start by giving a diagrammatic representation of CRNs, through the help of which we define the complex-balanced condition. Then we find the stationary distribution for such CRN by finding zero-eigenvectors of the corresponding Hamiltonian operators, which recovers the Anderson-Craciun-Kurtz (ACK) theorem \cite{anderson2010product}. Next, we find the NEP for such a distribution using Hamilton-Jacobi theory which recovers the Horn-Jackson potential and finally show that the negative descaled logarithm of the stationary distribution is indeed the NEP by comparison. We return to the techniques in this section by defining and finding the  analytic NEP of another class of models, namely one dimensional (1-D) birth-death models in Section \ref{subsec:N_Schlogl}. There we also consider its generalization to N-dimensional (N-D) birth-death models, and use our algorithm to numerically estimate the corresponding NEPs.

Recall from Section \ref{sec:Ham_CRN} that a CRN is defined by the triple consisting the set of species $\mathcal{S}$, set of complexes $\mathcal{C}$ and set of reactions $\mathcal{R}$. For a diagrammatic representation, we represent each species with a solid circle and each complex with an empty circle. Since a complex is a multi-set of species, we connect each complex to its constituting species with solid lines, where the number of lines denote the stoichiometry of the complex (denoted by column vector $y$). Finally, a reaction is a directed edge with a pair of complexes as source ($y_\alpha$) and target ($y_\beta$), which we denote through a dashed line with an arrow indicating direction of reaction. For examples of such a representation, see Figures \ref{fig:N-D_Schlogl_CRN} and \ref{fig:Selkov_details} or diagrams in \cite{krishnamurthy2017solving,smith2021eikonal}.

In this diagrammatic representation, the deterministic flow of mass-action-kinetics can be visualized by assigning weights to the directed edges (dashed lines) equal to $k_{y_\alpha \to y_\beta}q^{y_\alpha}$. An equilibrium is said to be \textit{complex balanced} \cite{horn1972general} at $\underline{q}$ if the total flow directed out of each complex equals the total flow inwards or the net flow out of each complex is zero, i.e. 
\begin{align}
    \sum_{y_{\beta}}\bigg(k_{y_{\beta}\to y_{\alpha}}(\underline{q})^{y_{\beta}}-k_{y_{\alpha}\to y_{\beta}}(\underline{q}){}^{y_{\alpha}}\bigg)&=0 \text{ for all } \alpha. \label{eq:Comp_bal_cond}
\end{align}

Given a CRN that exhibits a complex-balanced steady state, we will now find the stationary distribution $\pi(n)$, such that $\hat{H}(-\partial/\partial n,n)\pi(n)=0$. To identify the stationary distribution, we first need to make the same change of coordinates as we did for deriving the path integral formula in Section \ref{app:PathIntegral} Eq.\ \ref{eq:Z_ham_CRN}, in which the Hamiltonian operator takes the form   

\begin{align}
    &\phantom{==} \hat{\tilde{\underbar{H}}}_{\text{CRN}}\left(z,\frac{\partial}{\partial z}\right) \nonumber\\ 
    &=
    \sum_{y_{\alpha},y_{\beta}}\left(z^{y_{\beta}}-z^{y_{\alpha}}\right) 
\frac{k_{y_{\alpha}\to y_{\beta}}}{V^{y_{\alpha}-1}}\left(\pdv{}{z}\right)^{y_{\alpha}}\nonumber\\
    &=
        \sum_{y_{\alpha}}z^{y_{\alpha}} \sum_{y_{\beta}}\left[\frac{k_{y_{\beta}\to y_{\alpha}}}{V^{y_{\beta}-1}}\left(\pdv{}{z}\right)^{y_{\beta}} -\frac{k_{y_{\alpha}\to y_{\beta}}}{V^{y_{\alpha}-1}}\left(\pdv{}{z}\right)^{y_{\alpha}}\right] 
\label{eq:Ham_z_coord}
\end{align}
where in the last line we have switched indices to rewrite the Hamiltonian in a form similar to Eq.\ \ref{eq:Comp_bal_cond}, also termed the \textit{complex representation} in \cite{smith2017flows}. As mentioned in Section \ref{app:PathIntegral}, the change of coordinates corresponds to evolving the $\mathcal{Z}$-transform or the moment-generating function (MGF) rather than the distribution itself. Observe that the coordinate change preserves the commutation relations before and after the transformation, \footnote{This transformation is analogous to the Dirac transformation for simple harmonic oscillator in quantum mechanics, where we change from position and momentum to raising and lowering operators. $z$ and $\partial / \partial z$ precisely play the role of the raising and lowering operators respectively, however, since stochastic dynamics preserves the $\ell_1$ norm, the normalization is different. For the relevance of these operators in statistics, see \cite{baez2012quantum}.}

\begin{align}
    \left[n,-\pdv{}{n} \right]  &= \mathbb{I} = \left[\pdv{}{z},z \right]. \label{eq:comm_relns}
\end{align}
We will refer to $z$ as the raising operator and $\partial / \partial z$ as the lowering operator. For a representation of the above in abstract linear algebra making use of Fock space operators $a,a^\dagger$ with $[a,a^\dagger]=1$, see Ch. 4 of \cite{smith2015symmetry}. 

Next, we find the eigenvectors of the lowering operator $\partial / \partial z$. Let us denote the eigenvector with eigenvalue $Vc$ by $C(z)$. Then we have,
\begin{align}
    \pdv{C(z)}{z} &= Vc \phantom{\cdot} C(z) \nonumber\\
    C(z) &= N e^{cz}, \nonumber
\end{align}
where $N$ is a normalization factor (determined in the next line). Recall that the $\mathcal{Z}$-transform or MGF of a probability distribution must be such that $C(z=1)=1$, which means that $N=e^{-c}$ yielding 
\begin{align}
    C(z) &= e^{-Vc+Vcz}= e^{Vc(z-1)}. \label{eq:coherent_state}
\end{align}

Let us denote the eigenvector of the lowering operator with eigenvalue $V \underline{q}$ by $\underline{Q}(z)$. Then using Eq.\ \ref{eq:Ham_z_coord}, 
\begin{align}
    &\phantom{=} \hat{\tilde{\underbar{H}}}_{\text{CRN}}\underline{Q}(z) \nonumber\\
    &=  \sum_{y_{\alpha}}z^{y_{\alpha}} \sum_{y_{\beta}}\left[\frac{k_{y_{\beta}\to y_{\alpha}}}{V^{y_{\beta}-1}}\left(\pdv{}{z}\right)^{y_{\beta}} -\frac{k_{y_{\alpha}\to y_{\beta}}}{V^{y_{\alpha}-1}}\left(\pdv{}{z}\right)^{y_{\alpha}}\right] \underline{Q}(z)\nonumber \\
    &= V\sum_{y_{\alpha}}z^{y_{\alpha}} \underline{Q}(z)     \sum_{y_{\beta}}\bigg(k_{y_{\beta}\to y_{\alpha}}(\underline{q})^{y_{\beta}}-k_{y_{\alpha}\to y_{\beta}}(\underline{q}){}^{y_{\alpha}}\bigg) \nonumber \\
    &= 0, \label{eq:ACK_thm}
\end{align}
where in second line we make use of the fact that $\underline{Q}$ is an eigenvector of $\partial / \partial z$ and in the last line we use the complex-balanced condition from Eq.\ \ref{eq:Comp_bal_cond}. 

Finally, to obtain the distribution $\pi(n)$ from $\underline{Q}(z)$, we can either take the inverse-$\mathcal{Z}$ transform or simply expand the exponential as a summation, 
\begin{align*}
    \underline{Q}(z) &=  \sum_{n=0}^\infty z^n e^{-V\underline{q}} \frac{(V\underline{q})^n}{n!}\\
        &= \sum_{n=0}^{\infty} z^n \pi(n).
\end{align*}
By reading the coefficients of the series, we obtain for the stationary distribution 
\begin{align}
    \pi(n) &= e^{-V\underline{q}} \frac{\left(V\underline{q}\right)^n}{n!} \label{eq:stat_dist_comp_bal}
\end{align}
which is the ACK or multi-Poisson distribution as derived in \cite{anderson2010product,smith2017flows}.

Having determined the stationary distribution using the Hamiltonian operator, we proceed to making use of Hamilton-Jacobi theory to find the NEP for complex-balanced system. Recall from \ref{sec:Ham_CRN} that to find the NEP, we need to find a momentum assignment $p_\text{esc}(q)\neq 0$ at every configuration $q$, such that $H(p_\text{esc}(q),q)=0$ and the Hamilton's equations are satisfied for any initial condition. For finding the momentum assignment, namely escape momentum, along the $H(p,q)=0$ submanifold it is easiest to recast the CRN Hamiltonian from Eq.\ \ref{eq:Ham_CRN_rxn} in the following form 
\begin{align*}
   & \phantom{-} H(p,q) \nonumber \\
   &= \sum_{y_{\alpha}}(e^{p})^{y_{\alpha}}\bigg[\sum_{y_{\beta}}\bigg(k_{y_{\beta}\to y_{\alpha}}(e^{-p}q){}^{y_{\beta}}-k_{y_{\alpha}\to y_{\beta}}(e^{-p}q){}^{y_{\alpha}}\bigg)\bigg].
\end{align*}

We will now show that $p_{\text{esc}}=\ln\left( q/\underline{q} \right)$ is such an assignment. Observe that,
\begin{align*}
    & \phantom{=} H(q,p_\text{esc}) \\
    &= \sum_{y_{\alpha}}\left(\frac{q}{\underline{q}}\right)^{y_{\alpha}} 
    \bigg[\sum_{y_{\beta}}\bigg(k_{y_{\beta}\to y_{\alpha}}(\underline{q})^{y_{\beta}}-k_{y_{\alpha}\to y_{\beta}}(\underline{q})^{y_{\alpha}}\bigg)\bigg]\\
    &= 0,
\end{align*}
where the last line follows from the complex-balanced condition. Next, we calculate the total time derivative of $p_\text{esc}$ and confirm the consistency of such a momentum assignment against the equations of motion.
\begin{align*}
    \dv{p_\text{esc}}{t} = \frac{1}{q}\dv{q}{t} &= \frac{1}{q} \pdv{H}{p} = - \pdv{H}{q},
\end{align*}
where the last equality can be verified by a simple calculation.

 Thus the NEP $\mathcal{V}$ for CRN that exhibit a complex-balanced steady state is given by
 \begin{align}
     \mathcal{V}(q) &= \int_{\underline{q}}^q p_\text{esc}\,dq\\
     &= q\ln \left(\frac{q}{\underline{q}}\right) - (q - \underline{q}), \label{eq:NEP_comp_bal}
 \end{align}
where an integration constant has been chosen such that $\mathcal{V}(\underline{q})=0$, and we recover the Horn-Jackson potential that appears in \cite{horn1972general}.

Finally, we show the equivalence of the stationary distribution in Eq.\ \ref{eq:stat_dist_comp_bal} and NEP in Eq.\ \ref{eq:NEP_comp_bal}. Using Eq.\ \ref{eq:stat_dist_comp_bal} and substituting $n=Vq$, we have
\begin{align*}
   -\frac{1}{V} \log( \pi(n)) &= -\frac{1}{V} \left(-V\underline{q}+ Vq\log{\left(V\underline{q}\right)}-\log{n!}\right) \\
   &\asymp  q\log\left(\frac{q}{\underline{q}}\right) - (q-\underline{q})
\end{align*}
where the last line is obtained by using Stirling's approximation. We recognize the last equation to be the same as the NEP in Eq.\ \ref{eq:NEP_comp_bal}, which completes our expository example of the equivalence of the Hamiltonian operator techniques and Hamilton-Jacobi theory for complex-balanced systems.

\begin{table*}[t]
    \centering
    \begin{tabular}{c|c|c}
         & Stochastic dynamics & Quantum dynamics \\
         \hline
        Space of states of system  &  Discrete  & Continuous   \\
        $q\in Q$  & $n\in \mathbb{Z}^D_{\geq 0}$ &$x\in \mathbb{R}^D$\\
        Space of states of ensemble  &  Probability distribution  & Wave-function  \\
        $\mathcal{F}\in L^p$ & $\rho^R(n)\in \ell^1$ & $\psi(x) \in L^2$ \\
        Dual state space &  Sampling protocol  & State of detector  \\
        $\mathcal{F}\in L^{(1-\frac{1}{p})^{-1}}$ & $\rho^L(n)\in \ell^\infty$ & $\psi(x) \in L^2$ \\
        Dual pairing or & $\langle \rho_1^L, \rho_2^R \rangle$ & $\langle \psi_1, \psi_2 \rangle$ \\
        Inner product & $\sum_n \rho_1^L(n) \rho_2^R(n)$  & $\int \psi_1^*(x) \psi_2(x) \,dx$\\
       Hamiltonian operator  &  $\ell^1$ preserving dynamics  & $L^2$ preserving dynamics   \\
                            & Infinitesimal stochastic  & Infinitesimal unitary \\
                            & $\sum_{n}\hat{H}_{nn'} = 0 \hspace{1em}\text{for all } n'$ & $\hat{H} = -\hat{H}^\dagger$\\
                            & & $\hat{H} = -\frac{i}{\hbar} \hat{M}, \hat{M}^\dagger = \hat{M}$\\
                            && ($\hat{M}$ is Hermitian)\\
        Forward equation & Master equation & Schr$\ddot{\text{o}}$dinger equation\\
            & $\pdv{\rho^R(n)}{t} = \hat{H}\rho^R(n)$ & $\pdv{\psi}{t} = -\frac{i}{\hbar} \hat{M}$ \\  
            & $\pdv{\rho^R(q)}{t} = V \hat{H}\rho^R(q)$\\
        Adjoint representation & $\mathcal{Z}$-transform  & Fourier transform \\
            $\langle p |\Psi\rangle = \tilde{\Psi}(p)$    & $\tilde{\rho}(z) = \sum_n z^n \rho(n)  $ & $\tilde{\psi}(p) = \int \,dx e^{-\frac{i}{\hbar}px} \psi(x) $\\
                & Laplace-transform & \\
                & $\tilde{\rho}_V(p) = \int_0^\infty e^{Vpq} \rho(q)\,dq  $ &\\
        Momentum operator &  $\hat{P} = -\pdv{}{n}$ & $\hat{P} = \frac{\hbar}{i} \pdv{}{x}$\\
        $\langle p| \hat{P}|\Psi\rangle = p \langle p|\Psi\rangle  $  &  $\hat{P} = -\frac{1}{V}\pdv{}{q}$ & \\
        Scaling limit & $V\to \infty$ & $\hbar\to 0$\\
        Large-deviation theory & $\rho^R(q,t) \asymp e^{-V S(q,t)}$ & $\psi(x,t) \asymp e^{\frac{i}{\hbar} S(x,t)}$
        \end{tabular} 
    \caption{Correspondence between stochastic and quantum dynamics}
    \label{tab:NEA_setup}
\end{table*}

\section{Non-equilibrium action, action functional and its first and second variational derivatives}
\label{app:App_A}

\subsection{Deriving stochastic and quantum dynamics from the non-equilibrium action (NEA)}
\label{app:NEA}
Hamiltonian dynamical systems is a mathematical framework for formulating and analyzing dynamics of physical systems defined through a variational principle. The space of states (state-space) of a physical system is the space of all possible values one can observe upon measurements at the finest resolution. In physics, the measurable can be position, spin, energy of a particle or a group of particles, while in chemistry or biology, the measurable can be the count, concentration, etc. of a given species of molecule, organelle or organism. An experiment consists of an ensemble of systems, each obeying the same set of rules. These systems can either evolve simultaneously (like particles in a fluid, molecules in a solution, organisms in an ecosystem) or can be spatio-temporally separated (like a series of independent quantum or biological experiments) or both. In all cases, one can assign a distribution over the state-space and refer to it as the state of the experiment, which we denote by $|\Psi^R(t)\rangle$, as well as define a sampling protocol that takes a distribution as input and give a scalar quantity, which we denote by $\langle \Psi^L(t)|$. \footnote{Probability distribution and quantum amplitudes are both distributions, formally defined as vectors in $L_1$ and $L_2$ normed Hilbert space, respectively \cite{baez2012quantum}.}

Following Eyink's construction in \cite{eyink1996action}, given a Hamiltonian operator $\hat{H}$ we define the non-equilibrium action (NEA) functional $\Gamma$ to be 
\begin{align*}
    \Gamma[\Psi^L(t),\Psi^R(t)] &= \int  \langle \Psi^L(t),\big(\partial_t - \hat{H}\big)\Psi^R(t) \rangle \,dt  
\end{align*}
that takes a time-evolving state $|\Psi^R\rangle$ and sampling protocol $\langle \Psi^L|$ as input and yields a scalar quantity. If we expand the NEA around Hilbert-space vectors $\phi^R$ and $\phi^L$, 
\begin{align*}
    \Psi^R &= \phi^R + \delta \Psi^R\\
    \Psi^L &= \phi^L + \delta \Psi^L
\end{align*}
we get the variation in the NEA to be
\begin{align*}
    & \phantom{=} \Gamma[\Psi^L(t),\Psi^R(t)] = \Gamma[\phi^L(t),\phi^R(t)] \\
    & +  \int  \langle \delta \Psi^L(t),\big(\partial_t - \hat{H}\big)\phi^R(t) \rangle \,dt  \\
    & + \int  \langle \big(-\partial_t - \hat{H}^\dagger\big)\phi^L(t),\delta \Psi^R(t) \rangle \,dt  + \mathcal{O}(\delta^2).
\end{align*}

The stationary condition then yields the following Hamilton's equations of motion on the Hilbert space vectors
\begin{align}
    \partial_t \phi^R(t) &= \phantom{-} \hat{H}\phi^R(t),\nonumber\\
    \partial_t \phi^L(t) &=- \hat{H}^\dagger \phi^L(t). \label{eq:NEA_Ham_eqn}
\end{align}

The norm of the state and the sampling protocol must be preserved during the dynamics, which yields extra conditions on the Hamiltonian operator $\hat{H}$. In particular, when $\hat{H}$ preserves the $L_1$ norm, then we get the Master equation from Eq.\ \ref{eq:MasterEquation}. 

It is also straightforward to verify that the Hamiltonian operator that preserves the $L_2$ norm must correspond to an anti-Hermitian operator, implying $\hat{H}=-\hat{H}^\dagger$. Since the dual of an $L_2$ function is an $L_2$ function, there is no difference between the evolution of the state or the sampling protocol, thus reducing the two equations in Eq.\ \ref{eq:NEA_Ham_eqn} to one, yielding the Schr\"{o}dinger equation. We summarize these observations in table \ref{tab:NEA_setup}.

The two equations \ref{eq:NEA_Ham_eqn} can be collected in one by defining a density operator $\hat{\phi} = |\phi^R\rangle \langle \phi^L|$ and finding its total time derivative
\begin{align}
    \dv{\hat{\phi}}{t} \equiv \dv{|\phi^R\rangle \langle \phi^L|}{t} 
    &=
    \dv{|\phi^R\rangle}{t} \langle \phi^L| +     |\phi^R\rangle \dv{\langle \phi^L|}{t} \nonumber \\
    &= \hat{H}|\phi^R\rangle \langle \phi^L| - |\phi^R\rangle \langle \phi^L|\hat{H}^\dagger \nonumber \\
    &= \left[\hat{H},\hat{\phi}\right]. \label{eq:Von_neumann_eqn}
\end{align}
This is also known as the Von-Neumann equation and the operator $[\hat{H},.]$ is also referred to as the Liouville operator \cite{manzano2020short}. The statistical observation of the state of an experiment under a sampling protocol is given by $\langle O \rangle_\phi \equiv \Tr{\hat{O}\hat{\phi}}$. The time evolution of the statistical observable $O$ along the variational solutions of the non-equilibrium action is then given by 
\begin{align}
    \dv{\langle O \rangle_\phi }{t} &= \Tr{\left[\hat{O},\hat{H}\right]\hat{\phi}} + \Tr{\pdv{\hat{O}}{t}\hat{\phi}}.\label{eq:time_evo} 
\end{align}
For an excellent introduction to operator techniques in quantum statistical mechanics and quantum computation, see \cite{kadanoff2000statistical,nielsen2002quantum}.

\subsection{Derivation of the path integral formula}
\label{app:PathIntegral}
In this section we will answer the question, given the state of an experiment (PDF) at time $t=0$ and a stochastic Hamiltonian with which the system evolves following Eqs. \ref{eq:NEA_Ham_eqn} and \ref{eq:MasterEquation}, what is its state at an arbitrary time $t=T$? 

Let us revisit the time-evolution, from Eq.\ \ref{eq:TimeEvoln}, for a PDF $\rho(n,t)$, where $n\in\mathbb{Z}^D_{\geq 0}$ denotes the position in state space and $t$ is the time,   
\begin{align}
    \frac{\partial}{\partial t} \rho(n,t) 
    &=
    \hat{\underbar{H}}\left(-\pdv{}{n},n\right)\rho(n,t). \label{eq:Time_Evo_app}
\end{align}
Since $n$ takes only nonnegative integer values, we can consider the moment-generating function (MGF) or $\mathcal{Z}$-transform of the PDF $\rho(n,t)$, given by $\tilde{\rho}(z,t)$, and its inverse 
\begin{align}
    \tilde{\rho}(z,t) &= \sum_{n} z^n \rho(n,t) \label{eq:z_trans}\\
    \rho(n,t) &= \frac{1}{2\pi i }\oint \,dz \frac{1}{z^{n+1}}\tilde{\rho}(z,t). \label{eq:inv_z_trans}
\end{align}
We refer to the first line as $\mathcal{Z}$-transform and the second as inverse $\mathcal{Z}$-transform (for an introduction, see \cite{oppenheim1997signals}). Here $z\in \mathbb{C}^D$ and the contour integral is done over any contour that encloses $z=0$.

Since Eq.\ \ref{eq:Time_Evo_app} in general defines the time evolution of any distribution, in particular we can consider the time evolution of the MGF or $\mathcal{Z}-$transform of a distribution 
\begin{align}
        \frac{\partial \tilde{\rho}(z,t)}{\partial t} &=
        \sum_{n} z^n \hat{\underbar{H}}\left(-\frac{\partial}{\partial n},n\right)\rho(n,t) \nonumber
        \\
&=        \hat{\tilde{\underbar{H}}}\left(z,\frac{\partial}{\partial z}\right) \tilde{\rho}(z,t), \label{eq:Time_Evo_Z_trans}
\end{align}
where $\hat{\tilde{\underbar{H}}}$ is the time evolution operator for the $\mathcal{Z}$-transformed distribution, which we will henceforth refer to as the $\mathcal{Z}$-Hamiltonian operator. 

Using the observations
\begin{align*}
    \sum_{n} z^n e^{-y \frac{\partial}{\partial n}}\rho(n,t) &= z^y \sum_{n} z^n \rho(n,t)\\
    \sum_{n} z^n \frac{n!}{(n-y)!}\rho(n,t) &= z^y\bigg(\frac{\partial}{\partial z}\bigg)^y \sum_{n} z^n \rho(n,t),
\end{align*}
it is easy to see that the $\mathcal{Z}$-Hamiltonian operator for CRN from Eq.\ \ref{eq:Ham_CRN_op}
\begin{align}
    & \phantom{=} \hat{H}_{\text{CRN}}\left(-\pdv{}{n},n\right) \nonumber\\
    &=
    \sum_{y_{\alpha},y_{\beta}}\left( e^{-(y_{\beta}-y_{\alpha})\cdot\frac{\partial}{\partial n}} - 1\right) 
    \frac{k_{y_{\alpha}\to y_{\beta}}}{V^{y_{\alpha}-1}}\frac{n!}{(n-y_{\alpha})!}.  \label{eq:H_CRN_app}
\end{align}

takes the following form (for a more insightful derivation, see \cite{baez2012quantum})
\begin{align}
    \hat{\tilde{\underbar{H}}}_{\text{CRN}}\left(z,\frac{\partial}{\partial z}\right) 
    &=
    \sum_{y_{\alpha},y_{\beta}}\left(z^{y_{\beta}}-z^{y_{\alpha}}\right) 
\frac{k_{y_{\alpha}\to y_{\beta}}}{V^{y_{\alpha}-1}}\left(\pdv{}{z}\right)^{y_{\alpha}}. 
\label{eq:Z_ham_CRN}
\end{align}

Returning to the question posed in the beginning of this subsection, given a Hamiltonian operator $\hat{\underbar{H}}$ and $\rho(n,0)$, we wish to find $\rho(n,T)$. The approach that we will take is the following (for a pedagogical introduction to these methods, see \cite{altland2010condensed,peskin2018introduction}). First, we discretize time from $0$ to $T$ into $N$ intervals of length $\Delta t$, such that $N\Delta t= T$. Next, we label the random variable denoting the count of the system at time $i \Delta t$ by $n_i$. Since, we know that the system is evolving with the given Hamiltonian, there is a relation between $\rho(n_{i+1},(i+1)\Delta t)$ and $\rho(n_i,i\Delta t)$. To find $\rho(n_{i+1},(i+1)\Delta t)$, we first compute the $\mathcal{Z}$-transform of $\rho(n_i,i\Delta t)$, evolve it for $\Delta t$ by \ref{eq:Time_Evo_Z_trans} and compute the inverse $\mathcal{Z}$-transform at $n_{i+1}$.

\begin{figure}[!h]
\centering
\begin{tikzcd}
\rho(n_i,i\Delta t)  \arrow[r, "\mathcal{Z}\text{-Tr.}"] & \tilde{\rho}(z_i,i\Delta t) \arrow[d, "\text{Time evolve}"]  \\
\rho(n_{i+1},(i+1)\Delta t)        & e^{\hat{\tilde{\underbar{H}}}\Delta t} \tilde{\rho}(z_i,i\Delta t) \arrow[l, "\text{Inv}\mathcal{Z}\text{-Tr.}"] 
\end{tikzcd}
\caption{Evolving distribution by $\Delta t$} \label{fig:Evolving_Delta_t}
\end{figure}

Using the above procedure, which we represent in diagrammatic form in Figure  \ref{fig:Evolving_Delta_t}, we can read the distribution at time $(i+1)\Delta t$ to be 
\begin{align}
    &\rho(n_{i+1},(i+1)\Delta t)	\\
    &=\frac{1}{2\pi i}\oint\,dz_{i}\frac{1}{z_{i}^{n_{i+1}+1}}e^{\Delta t\hat{\tilde{\underbar{H}}}\left(z,\frac{\partial}{\partial z}\right)}\sum_{n_{i}}z_{i}^{n_{i}}\rho(n_{i},i\Delta t) \nonumber\\
    &=
\frac{1}{2\pi i}\oint\frac{\,dz_{i}}{z_{i}}\sum_{n_{i}}z_{i}^{-n_{i+1}}e^{\Delta t\hat{\tilde{\underbar{H}}}\left(z,\frac{\partial}{\partial z}\right)}z_{i}^{n_{i}}\rho(n_{i},i\Delta t) \nonumber\\
&=
\frac{1}{2\pi i}\oint\frac{\,dz_{i}}{z_{i}}\sum_{n_i} z_{i}^{-n_{i+1}+n_{i}}e^{\Delta t \underline{\mathcal{H}}(z_i,n_i)}\rho(n_{i},i\Delta t) \nonumber
\end{align}
where in the last line, the function $\underline{\mathcal{H}}(z_i,n_i)$ is obtained by an application of the operator $\hat{\tilde{\underbar{H}}}\left(z,\pdv{}{z}\right)$ on $z_i^{n_i}$. In particular for CRN, using \ref{eq:Z_ham_CRN}, we obtain
\begin{align}
    \underline{\mathcal{H}}_\text{CRN}(z,n) 
    &=
    \sum_{y_{\alpha},y_{\beta}}\left(z^{y_{\beta}-y_\alpha}-1\right) 
\frac{k_{y_{\alpha}\to y_{\beta}}}{V^{y_{\alpha}-1}}\frac{n!}{(n-y_\alpha)!}. \label{eq:Ham_CRN_mixed} 
\end{align}

Finally, by repeated application of the above procedure at the initial PDF $N$ times and taking the limit $N\to \infty$ and $\Delta t \to 0$, while their product is held fixed $N\Delta t=T$, we obtain
\begin{align*}
    &\rho(n_{N},N\Delta t)\\
    &=\prod_{i=0}^{N-1}\bigg(\frac{1}{2\pi i}\oint\frac{\,dz_{i}}{z_{i}}\sum_{n_{i}}\hspace{0.5em}z_{i}^{-n_{i+1}+n_{i}}e^{\Delta t \underline{\mathcal{H}}(z_i,n_i)}\bigg)\rho(n_{0},0)\\
    &\text{using }z_{i}=e^{p_{i}}\\
&	=\prod_{i=0}^{N-1}\bigg(\int\frac{\,dp_{i}}{2\pi i}\sum_{n_{i}}e^{-p_{i}(n_{i+1}-n_{i})}e^{\Delta t \underbar{H}(p_i,n_i)}\bigg) \cdot \rho(n_{0},0)\\
& \xrightarrow[\substack{N\to \infty\\\Delta t\to 0}]{\lim}\int\left[\frac{\,d p}{2\pi i}\right]\sum \left[\,d n\right] e^{-\int_{0}^{T}\,dt(p\dot{n}-\underbar{H}(p,n)\,dt)}\rho(n_{0},0),
\end{align*}
where $[\,d p]$ and $[\,d n]$ are the path-integral measures and $\underbar{H}(p,n)$. The last line yields us the path-integral formula and we define the \textbf{action functional} as
\begin{align}
    \mathcal{A}[n(t),p(t)] &= \int_{0}^{T}\,dt(p\dot{n}-\underbar{H}(p,n)\,dt). \label{eq:app_action}
\end{align}
For an alternative derivation of the functional integral in the Doi-Peliti formalism using coherent states, see \cite{smith2011large,smith2015symmetry,smith2019information}.

Note that For CRN, using \ref{eq:Ham_CRN_mixed}, the Hamiltonian function takes the particular form
\begin{align}
    \underbar{H}_\text{CRN}(p,n) &= 
        \sum_{y_{\alpha},y_{\beta}}\left(e^{p \cdot(y_{\beta}-y_\alpha)}-1\right) 
\frac{k_{y_{\alpha}\to y_{\beta}}}{V^{y_{\alpha}-1}}\frac{n!}{(n-y_\alpha)!}. 
\end{align}
which is the same as the functional form of the Hamiltonian operator in Eq.\ \ref{eq:H_CRN_app}, with the operator $-\pdv{}{n}$ replaced by the momentum variable $p$, thus justifying our notation. 

\subsection{Optimality condition: Hamilton's equations of motion and their relation to the Hamilton-Jacobi equation}
\label{sec:Equiv_HJ_HamEoM}
Analogous to Eq.\ \ref{eq:app_action}, in Eq.\ \ref{eq:PathIntegral_q_coord} we defined an action functional in the concentration and momentum coordinates $(q,p)$ to be
\begin{align*}
   \mathcal{A}\left[q(t),p(t)\right] 
    &=
    \int_0^T \left[ p\cdot \dv{q}{t} - H(p,q)\right]\,dt, 
\end{align*}
where the relation of the Hamiltonian function $H(p,q)$ to $\underbar{H}(p,n)$ was discussed in Eq.\ \ref{eq:Convexity_of_H}. As discussed around Eq.\ \ref{eq:StationaryAction}, the path integral is dominated by the value of the action functional around the optimal or stationary path, i.e. the path for which the first variation of the action is zero.

To calculate the first variation of the action, we consider the difference of the action functional between a path $(q(t),p(t))$ and $(q(t)+\delta q(t), p(t)+\delta p(t))$. In the following equation, we will suppress the time dependence to simplify notation and keep only the first order terms.
\begin{align}
    & \phantom{-} \delta \mathcal{A}[q,p] \nonumber \\
    &=  
    \mathcal{A}\left[ q+\delta q, p+\delta p \right] - \mathcal{A}\left[ q,p \right] \nonumber
    \\
    &= 
    \int \,dt \left\{ \delta p \left(\dv{q}{t} - \pdv{H}{p} \right) + \left(p\dv{\delta q}{t} - \pdv{H}{q}\delta q  \right) \right\} \nonumber\\
    &= 
    \int \,dt \left\{ \delta p \left(\dv{q}{t} - \pdv{H}{p} \right) - \delta q\left(\dv{p}{t} - \pdv{H}{q}\delta q  \right) + p \delta q  \bigg|_0^T \right\} \label{eq:1st_variation_Action}
\end{align}
where the last line is obtained by integrating by-parts. If we fix the end-points of the path in configuration space, then we can set $\delta q(0)=\delta q(T)=0$, getting rid of the last term in the above equation. 

Using Eq.\ \ref{eq:1st_variation_Action} and
\begin{equation*}
    \delta \mathcal{A}[q(t),p(t)] =
    \frac{\delta \mathcal{A}}{\delta q(t)} \delta q(t) + \frac{\delta \mathcal{A}}{\delta p(t)} \delta p(t)
\end{equation*}
     we can read off the variation of the action functional in configuration and momentum around $(q(t),p(t))$ to be
\begin{align}
\frac{\delta \mathcal{A}}{\delta q(t)} 
    &= 
    -\left( \dv{p}{t} + \pdv{H}{q} \right), \nonumber\\
    \frac{\delta \mathcal{A}}{\delta p(t)} 
    &= 
    \phantom{-}\left( \dv{q}{t} - \pdv{H}{p}\right). \label{eq:FuncGradient}
\end{align}

The optimality condition $\mathcal{A}[q^*,p^*] = 0$ for a trajectory $(q^*(t),p^*(t))$ then yields 
\begin{align}
    \dv{p^*}{t} &= -\pdv{H}{q^*}, \nonumber\\
    \dv{q^*}{t} &= \phantom{-} \pdv{H}{p^*}.    \label{eq:Hamilton'sEoM}
\end{align}
The last two lines are also referred to as Hamilton's equations of motion (EoM) and are the equations that any optimal trajectory must satisfy. 

There is a rich and deep mathematical structure underlying these equations, which is the subject of symplectic geometry. For instance, the change in value of an observable function $f(p,q)$ along an optimal trajectory $(q^*,p^*,t)$ can be given in terms of its Poisson-bracket commutator with the Hamiltonian,
\begin{align}
    \dv{f}{t} &= \pdv{f}{t} + \pdv{f}{q}\dv{q}{t} + \pdv{f}{p}\dv{p}{t} \nonumber\\
            &= \pdv{f}{t} + \pdv{f}{q}\pdv{H}{p} - \pdv{f}{p}\pdv{H}{q} \nonumber\\
            &= \pdv{f}{t} + \left\{ f,H \right\}.
\end{align}
This immediately shows that the value of a time-independent Hamiltonian as well as any time independent operator that commutes with the Hamiltonian stays constant along equations of motion (notice the resemblance of the above equation with Eq.\ \ref{eq:time_evo}). For a classical introduction to the subject, see \cite{goldstein2002classical,arnol2013mathematical}.

To end this subsection,  following \cite{courant2008methods}, we provide a proof of the equivalence between Hamilton's equations of motion and Hamilton-Jacobi equation on paths where the velocity and the momentum are related by a Legendre transform.
\begin{align*}
    p & = \pdv{S} {q}\\
    \dot{p}&=\dv{}{t}\bigg(\pdv{S} {q}\bigg) \\
    &=\pdv{S} {q} {q}\dot {q} + \pdv{S} {q}{t}\\
    &=\pdv{S} {q} {q}\pdv{H}{p} - \pdv{H} {q}-\pdv{H}{p}\pdv{p} {q}\\
    &=\pdv{p} {q}\pdv{H}{p} -\pdv{H}{p}\pdv{p} {q} - \pdv{H} {q}\\
    &=-\pdv{H} {q}
\end{align*}
where the first and fourth line use the Hamilton-Jacobi equation. The fourth line also makes use of the relation between optimal momentum and velocity, or the Legendre-transform condition. 

\subsection{Second variational derivative of action functional: Onsager-Machlup action and convexity of instanton}

\label{sec:2nd_var_der}
In this subsection, we will first calculate the second variational derivative of the action functional. Next, we provide a derivation of the Onsager-Machlup action, followed by comments on the convexity of the action functional around any optimal trajectory. 

To simplify notation we use the following convention. For vectors $v,w$, we represent their dot product $v^T w$ as $vw$. For a quadratic form $M$, we represent $M(v,v) \equiv v^T M v$ simply as $Mv^2$. Finally, the time derivative and perturbation of a function $x$ is given by 
\begin{align}
    \dot{x} &= \dv{x}{t}, \nonumber \\
    x^+ &= x + \delta x, \nonumber
\end{align}
respectively.

Recall, in Eq.\ \ref{eq:PathIntegral_q_coord}, the action functional was defined to be 
\begin{align*}
    \mathcal{A}[q,p] &= \int_0^T \,dt \left\{ p \dot{q} - H \right\}.
\end{align*}
Thus, the variation in the action functional up to terms second order in the variation is
\begin{widetext}
\begin{align}
    &\delta \mathcal{A} 
    =  
    \mathcal{A}\left[ q^+,p^+\right] - \mathcal{A}\left[ q,p \right] \nonumber
    \\
    &= 
    \int \,dt \left\{ \delta p \left(\dot{q} - \pdv{H}{p} \right) - \delta q \left(\dot{p} + \pdv{H}{q} \right) \right\}
     + \delta p \delta \dot{q} - \frac{1}{2} \begin{bmatrix}
    \delta p &  \delta q
    \end{bmatrix}
    \begin{bmatrix}
    \pdv{H}{p}{p}  & \pdv{H}{p}{q} \\
    \pdv{H}{q}{p} & \pdv{H}{q}{q}
    \end{bmatrix}
    \begin{bmatrix}
    \delta p \\ \delta q
    \end{bmatrix} \nonumber + p\delta q\bigg|_0^T\\
    &= 
     \int \,dt \left\{ \delta p \left(\dot{q} - \pdv{H}{p} \right) - \delta q \left(\dot{p} + \pdv{H}{q} \right) 
      - \frac{1}{2} \begin{bmatrix}
    \delta p &  \delta q
    \end{bmatrix}
    \begin{bmatrix}
    \pdv{H}{p}{p}  & \left( - \,d_t + \pdv{H}{p}{q}\right) \\
    \left(\,d_t + \pdv{H}{q}{p}\right) & \pdv{H}{q}{q}
    \end{bmatrix}
    \begin{bmatrix}
    \delta p \\ \delta q
    \end{bmatrix}\right\}  + \left(p\delta q+\frac{1}{2}\delta p\delta q\right)\bigg|_0^T\nonumber \\
    &=      \int \,dt \Bigg\{ \delta p \left(\dot{q} - \pdv{H}{p} \right) - \delta q \left(\dot{p} + \pdv{H}{q} \right) 
    - \frac{1}{2}\pdv{H}{p}{p}\left[    \delta p - \left(\pdv{H}{p}{p}\right)^{-1} \left( \delta \dot{q} - \pdv{H}{p}{q}\delta q\right) \right]^2  \nonumber \\
    & \phantom{=\int \,dt\{} +\frac{1}{2}\left[\left(\pdv{H}{p}{p}\right)^{-1} \left[ \left( \,d_t-\pdv{H}{p}{q}\right)\delta q\right]^2 - \pdv{H}{q}{q}\delta q^2 \right] \Bigg\} + \left(p\delta q+\frac{1}{2}\delta p\delta q\right)\bigg|_0^T, \label{eq:deltaAction}
\end{align}
\end{widetext}

where we have implicitly assumed that the original and perturbed phase space trajectories lie in the same $H=h$ submanifold. 

Now, if we assume that the two trajectories are constrained to the same end points in configuration space, i.e. $\delta q(0)=\delta q(T) = 0$, then the boundary terms vanish. Moreover, if we assume that the momentum assignment before and after the perturbation is optimal, i.e. $p$ and $p^+$ are assigned to $q$ and $q^+$ such that the phase space point is at its Legendre transform before and after the perturbation, then we have 
\begin{align}
&\text{For }(q,p) \nonumber\\
    0 &= \left(\dot{q} - \pdv{H}{p} \right)  \label{eq:LegTransformb4}\\
&\text{For }(q^+,p^+) \nonumber\\
    \dot{q}^+ 
    &= 
    \pdv{H(p^+,q^+)}{p} \nonumber\\
    \delta \dot{q} 
    &=
    \pdv{H}{p}{q}\delta q +  \pdv{H}{p}{p}\delta p \nonumber\\
    0  &=
    \delta p - \left(\pdv{H}{p}{p}\right)^{-1} \left( \delta \dot{q} - \pdv{H}{p}{q}\delta q\right)  \label{eq:LegTransform_after}
\end{align}

Substituting eqs \ref{eq:LegTransformb4} and \ref{eq:LegTransform_after} in Eq.\ \ref{eq:deltaAction} and making use of the assumptions, we get the variation in the action functional to be
\begin{widetext}
\begin{align}
    \delta \mathcal{A} 
    &=      \int \,dt \left\{ - \delta q \left(\dot{p} + \pdv{H}{q} \right)  
     +\frac{1}{2}\left[\left(\pdv{H}{p}{p}\right)^{-1} \left[ \left( \,d_t-\pdv{H}{p}{q}\right)\delta q\right]^2 - \pdv{H}{q}{q}\delta q^2 \right] \right\} \nonumber \\
     &=
     \int \,dt \left\{ - \delta q \left(\dot{p} + \pdv{H}{q} \right)  
     +\frac{1}{2}\left[\left(\pdv{H}{p}{p}\right)^{-1} \left[ \,d_t\delta q+\pdv{H}{p}-\pdv{H}{p}-\pdv{H}{p}{q}\delta q\right]^2 - \pdv{H}{q}{q}\delta q^2 \right] \right\} \nonumber\\
     &=
     \int \,dt \left\{ - \delta q \left(\dot{p} + \pdv{H}{q} \right)  
     +\frac{1}{2}\left[\left(\pdv{H}{p}{p}\right)^{-1} \left[\dot{q}^+ - \dot{q}-\pdv{H}{p}{q}\delta q \right]^2 - \pdv{H}{q}{q}\delta q^2 \right] \right\},
     \label{eq:deltaAction_2}
\end{align}
\end{widetext}

where we add and subtract the same quantity in the second line and make use of the Legendre transform condition in the third line.

\subparagraph{Onsager-Machlup Action}
$\delta \mathcal{A}$ measures the log-conditional improbability of trajectory $q^+$ given base trajectory $q$. If $q$ is Hamilton's equation of motion along $p=0$, then we have
\begin{align}
  \left(\dot{p} + \pdv{H}{q} \right) &= 0, \nonumber\\
  \pdv{H}{q}{q}\bigg|_{p=0} &= 0, \nonumber\\
  \pdv{H}{q}{p}\bigg|_{p=0} &= 0, \label{eq:Conditions_OM}
\end{align}
and we thus recover the Onsager-Machlup action
\begin{align}
  &\phantom{=} \mathcal{A}_\text{OM}[q^+] \equiv \mathcal{A}[(q^+,0)]-\mathcal{A}[(q,0)] \nonumber \\
   &= 
     \int \,dt \left\{  
     \frac{1}{2}\left[\left(\pdv{H}{p}{p}\right)^{-1} \left[\dot{q}^+ - \dot{q}\right]^2 \right] \right\}.
     \label{eq:Onsager_Machlup}
\end{align}
Recall from the discussion around Eq.\ \ref{eq:MassActionKinetics} that the equations of motion along which $p=0$ are the deterministic trajectories of the system. Thus the Onsager-Machlup action measures the log improbability of a trajectory $q^+$ conditioned on the deterministic or relaxation trajectory $q$ for the stochastic system. For a similar derivation and applications to population biology, see \cite{smith2021beyond}.

\subparagraph{Convexity of action functional around an optimal path}
\label{app:convexity}
If $q$ is an escape curve for a CRN Hamiltoninan, only the first equation of Eqs. \ref{eq:Conditions_OM} holds since $p(q_\text{esc})\neq 0$ (see Eqs. \ref{eq:p_neq_0}). The variation in action around the optimal path then becomes
\begin{align}
    &\phantom{-} \delta \mathcal{A} =\nonumber\\
    &
    \int \,dt \left\{ 
    \frac{1}{2}\left[\left(\pdv{H}{p}{p}\right)^{-1} \left[\dot{q}^+ - \dot{q}-\pdv{H}{p}{q} \delta q \right]^2 - \pdv{H}{q}{q}\delta q^2 \right] \right\}. \label{eq:convexity}
\end{align}
Notice that due to convexity of the Hamiltonian in $p$, the first term is positive definite. However, nothing can be said a priori about the second term since the Hessian of the Hamiltonian in $q$ for multi-stationary networks is in general non-convex. For the applications demonstrated in the paper, the variation is indeed positive definite and a global minimum exists in all cases. Generically however, as can be seen from the equation above, not all optimal curves are global minimizers of the action. For an example of an infinite dimensional stochastic system exhibiting a saddle optimal solution, we refer the readers to Chapter 8, \cite{kamenev2011field}. The authors are not aware of a similar example for a finite dimensional chemical reaction network at this point, and leave it as a question for future investigation.

\section{Details of AFGD}
\label{app:Details_AFGD}

\subsection{Notation}

As noted in Table \ref{tab:notation}, starting Section \ref{sec:MinMax}, we use the following notation. 
\begin{align*}
    & \text{Configuration space} : Q\subset \mathbb{R}^{D} \\
    & \mathcal{P}_Q = \left\{ q: [0,1] \to Q : q(0)=q_I, q(1) = q_F \right\}\\
    & \text{Configuration space curve} : q \in \mathcal{P}_{Q}\\
    & \text{Phase space} : T^*Q \subset \mathbb{R}^{2D}\\
    & \mathcal{P}_{T^*Q} = \left\{ (q,p): [0,T] \to T^*Q : q^*(0)=q_I, q^*(T) = q_F \right\}\\
    & \text{Phase space trajectory} : \gamma \in \mathcal{P}_{T^*Q}
\end{align*}
We refer to paths in configuration space as `curves', and paths in phase space as `trajectories'. This is an unusual choice intended to avoid confusion and simplify terminology, and will be used consistently through this section. Notice that curves are parametrized on the unit interval $[0,1]$, which is the normalized arc length. On the other hand, trajectories are parametrized with time, on the interval $[0,T]$ (for more details, see Section \ref{sec:MinMax}).  

Since computationally we have to work with discrete paths, we now introduce notation for discretization. Say we sample a continuous path $x \in X$ on $N$ points, we denote its discrete counterpart with $\tilde{x} \in \tilde{X}$ where
\begin{align*}
    \tilde{x}(n) &= x\left( \frac{n}{N} \right) \hspace{1em} \text{where } n \in [1,N] \subset \mathbb{Z}.
\end{align*}

We denote the $I^\text{th}$ iteration of the algorithm with a superscript $I$. For instance, the configuration space curve in the first iteration is denoted by $\tilde{q}^1$, where
\begin{align*}
        \tilde{q}^1(n) &= q^{\text{IC}}\left( \frac{n}{N} \right). 
\end{align*}

\begin{figure*}[htb]
    \centering 
\begin{subfigure}{0.5\textwidth}
  \includegraphics[width=\linewidth]{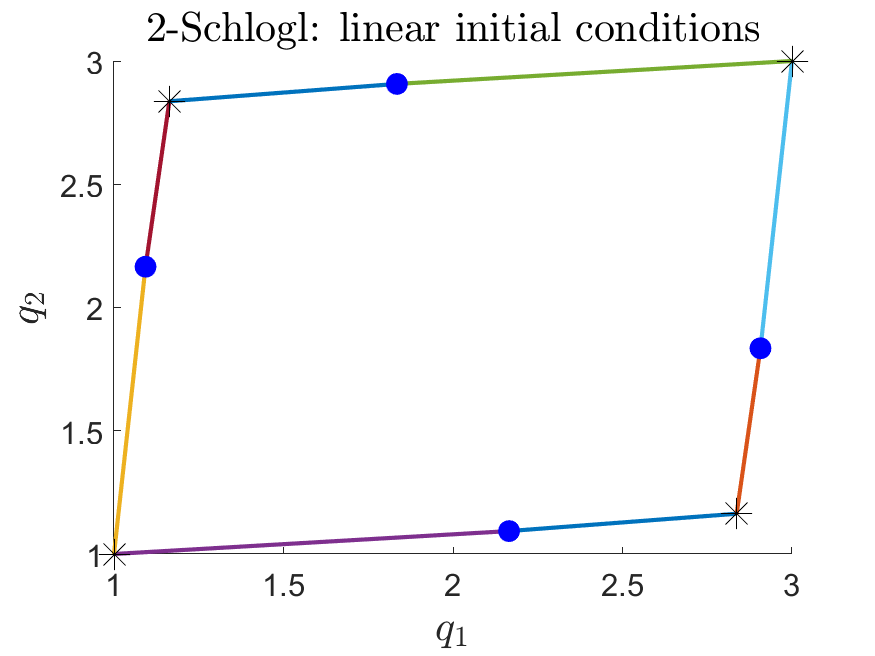}
  \caption{2-Schl{\"{o}}gl model}
  \label{fig:1}
\end{subfigure}\hfil 
\begin{subfigure}{0.5\textwidth}
  \includegraphics[width=\linewidth]{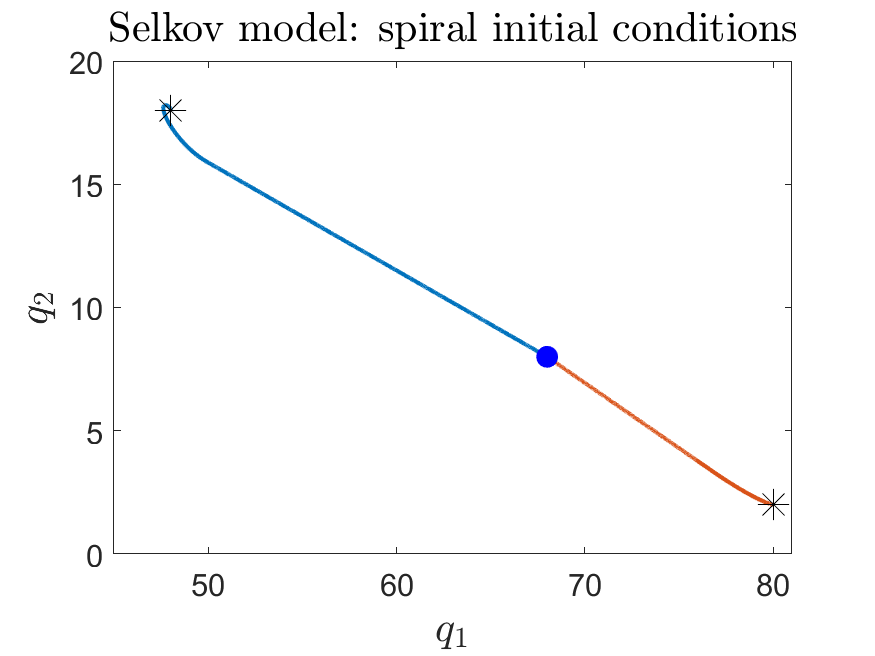}
  \caption{Selkov model}
  \label{fig:2}
\end{subfigure}
\caption{Initial conditions for the AFGD algorithm.} \label{fig:initial_conditions}
\end{figure*}

\subsection{Initial condition}
\label{app:initial_cond}
To begin the descent towards an optimal trajectory, in principle, one can start with an arbitrary curve connecting the end-points $q_I$ and $q_F$ that does not pass through or go around another fixed point and remains in $q_I$'s basin of escape. In practice however, the descent might require a large number of sample points and high numerical accuracy to obtain the optimal trajectory. In this subsection we provide a method to classify the optimal trajectory and pick an appropriate initial condition with the desired number of sample points effectively.

We classify the optimal trajectory near a fixed point by linearizing Hamilton's equations of motion (EoM) and analyzing the eigenvalues of the Hessian. If the eigenvalues have a non-zero imaginary part then the optimal trajectory spirals into or out of the fixed point. An exposition of this technique can be found in Chapter 4 of \cite{kamenev2011field} (their `activation trajectory' is our `escape trajectory'), but we repeat the relevant construction here.

Let the fixed point be denoted by $\underline{q}$. Recall that the momentum value at the fixed point in the $H(p,q)=0$ submanifold is identially zero i.e. $p(\underline{q}) = 0$. Linearizing Hamilton's equations of motion around the fixed point, we get
\begin{align}
\begin{bmatrix}
\dfrac{\,d q}{\,d t} \\ \\
\dfrac{\,d p}{\,d t}
\end{bmatrix} 
&=
\begin{bmatrix}
\vphantom{-} \dfrac{\partial H}{\partial p} \\ \\
- \dfrac{\partial H}{\partial q}
\end{bmatrix} _{(\underline{q},0)} 
+
\begin{bmatrix}
\dfrac{\partial H}{\partial p \partial q} & \dfrac{\partial H}{\partial p^2} \\ \\
- \dfrac{\partial H}{\partial q^2} & -\dfrac{\partial H}{\partial q \partial p} 
\end{bmatrix} _{(\underline{q},0)} 
\begin{bmatrix}
\vphantom{\dfrac{\,d q}{\,d t}} (q-\underline{q})\\ \\
\vphantom{\dfrac{\,d q}{\,d t}}(p-0)
\end{bmatrix}  \nonumber \\
&= \begin{bmatrix}
\dfrac{\partial H}{\partial p \partial q} & \dfrac{\partial H}{\partial p^2} \\ \\
0 & -\dfrac{\partial H}{\partial q \partial p} 
\end{bmatrix} _{(\underline{q},0)} 
\begin{bmatrix}
\vphantom{\dfrac{\,d q}{\,d t}} (q-\underline{q})\\ \\
\vphantom{\dfrac{\,d q}{\,d t}}(p-0)
\end{bmatrix}  
\end{align}
where the second line is obtained by using the fact that $\underline{q}$ is a fixed point and $p(\underline{q})=0$. To simplify notation, we denote the column vector $(q-\underline{q},p-0)^T$ by $\Delta \gamma$. Then we can rewrite the above equation as
\begin{align}
    \dv{\Delta \gamma}{t} &= M \Delta \gamma \nonumber \\
    M 
    &= 
    \sum_{i=1}^{2n} \lambda_i v_iv_i^T
\end{align}
where $M$ is the Hessian with eigenvectors $v_i$ and their corresponding eigenvalues $\lambda_i$.

Let the initial condition be denoted by $q^{\text{IC}}$. 
\paragraph{Systems without voriticity}
If all $\lambda_i$s are real then we know that the optimal trajectory will not spiral into the saddle or out of the stable fixed point. In this case, we pick the initial condition to be a straight line starting from the stable fixed point towards the saddle fixed point.
\begin{align}
    q^{\text{IC}}(s) 
    &=
    q_F + (1-s)(q_I-q_F) \label{eq:st_line}
\end{align}
The discrete initial curve for the algorithm will then be given by,
\begin{align}
        \tilde{q}^1(n) &= q^{\text{IC}}\left( \frac{n}{N} \right). 
\end{align}

\paragraph{Systems with vorticity}

If some $\lambda_i$s are complex, then the trajectory spirals outwards or inwards around the stable or saddle fixed point respectively. For the purpose of this subsection, let us assume that the eigenvalues are complex around the stable fixed point $q_I$ and real around the saddle fixed point $q_F$.

All the trajectories that depart from the fixed point, including the escape trajectory in the $H(p,q)=0$ submanifold, lie in the $D$-dimensional submanifold \footnote{Also called the Lagrangian submanifold \cite{kamenev2011field}} (in the $2D$-dimensional phase space) given by the superposition of the eigenvectors with non-zero eigenvalues. Thus, we can write the escape trajectory near the fixed point as 
\begin{align}
    \Delta \gamma(0) 
    &= 
    \sum \mu_i v_i \nonumber \\
    \Delta \gamma(t) 
    &=
    \sum_i \mu_i e^{\lambda_i t} v_i
\end{align}
where we must pick $\mu_i$ such that 
\begin{align*}
    \mu_i \in [-1,1] \subset \mathbb{R} \hspace{1em} &\text{for all} \hspace{0.25em} i, \\ 
    \mu_i = 0 \hspace{1em} &\text{if Re($\lambda_i$)}<0, \\
    \text{and } |\mu_i| = |\mu_j| \hspace{1em} &\text{if } \overline{\lambda_i} = \lambda_j
\end{align*}
which ensures that we only consider the repelling subspace and the trajectory has purely real coordinates. There is still some arbitrariness in the selection of the magnitude and sign of the coefficients which must be resolved through experimentation on a case-by-case basis, the goal being to select coefficients such that the curve spirals outwards (Re($\lambda_i$)$>0$) from $q_I$ towards $q_F$.

We know that the trajectory obtained by integrating the linearization is only valid in some small distance around the fixed point. To end the curve at the saddle fixed point, we find time $t$ such that the first derivative of the spiral matches the slope of the joining line, i.e. solve for $t=t^*>0$ such that
\begin{align}
    \dfrac{\dv{q_i}{t}}{\dv{q_1}{t}} 
    &=
    \dfrac{(q(t)-q_F)_i}{(q(t)-q_F)_1} \hspace{1em}\text{for all} \hspace{0.25em} i \in [2,n] \nonumber .\\
\end{align}

We obtain the complete initial condition by taking the union of the spiral part and straight part,
\begin{align}
    q_{\text{spiral}}(u) 
    &= 
    q_I + \Delta \gamma(u) \hspace{1em} \text{for } u\in[-T,t^*] \nonumber \\
    q_{\text{straight}}(u)
    &= 
    q_F + (1-u)(q_I+\Delta \gamma(t^*) -q_F) \hspace{1em} \text{for } u\in[0,1] \nonumber \\
    q^{IC} 
    &=
    q_{\text{spiral}} \bigcup q_{\text{straight}}.
\end{align}
We then use the length along the curve to parametrize $q^{\text{IC}}$, and obtain the initial condition 
\begin{align}
        \tilde{q}^1(n) &= q^{\text{IC}}\left( \frac{n}{N} \right). 
\end{align}

\subsection{Lift curve to trajectory}
\label{sec:lift_curve_to_traj}

\begin{figure*}[htb]
    \centering 
\begin{subfigure}{\textwidth}
  \includegraphics[width=\linewidth]{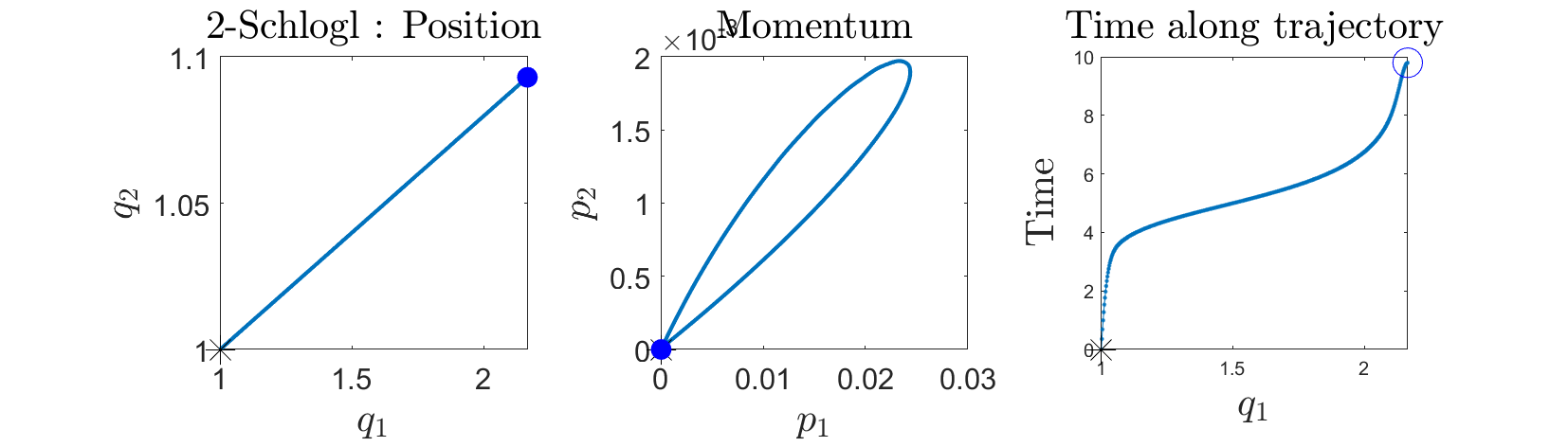}
  \caption{2-Schl{\"{o}}gl model initial condition phase space trajectory}
\end{subfigure}\hfil 
\medskip
\begin{subfigure}{\textwidth}
  \includegraphics[width=\linewidth]{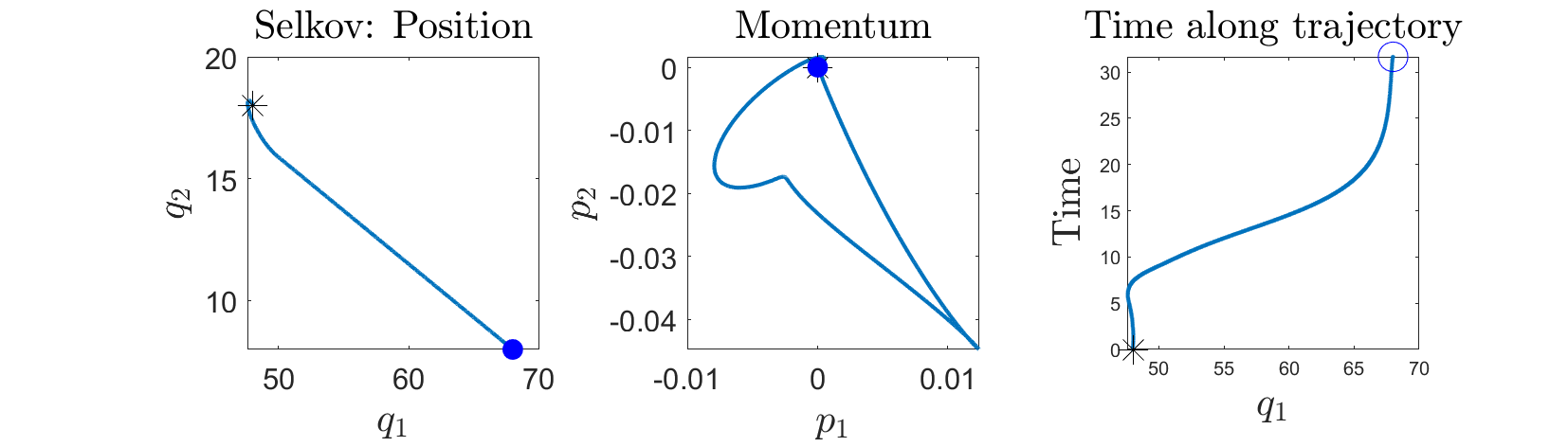}
  \caption{Selkov model initial condition phase space trajectory}
\end{subfigure}
\caption{Phase space trajectory obtained by lifting the initial conditions shown in Figure \ref{fig:initial_conditions}.} \label{fig:phase_space_initial_cond}
\end{figure*}

In every iteration, we begin with a configuration space curve $q^I$ and lift it to a phase space trajectory $(q,p,t)^I$, which we then use to obtain a variation curve $\delta q^I$, which gives us a curve for the next iteration $q^{I+1}= q^I+\delta q^I$. In this subsection, we define the \textit{lift} map and comment on methods for its implementation.  

For a discretized curve $\tilde{q}$, for each segment $\tilde{q}(i)$ to $\tilde{q}(i+1)$ we assign an optimal momentum value $\mathfrak{p}(i)$ at the center of the segment and a time interval $\Delta t(i)$ denoting the time taken for transition from $q(i)$ to $q(i+1)$ (see Figure \ref{fig:Leg_trans}), using the following equations. The process of assigning an optimal momentum value using a convex Hamiltonian is an instance of a Legendre-transformation, for a detailed exposition see \cite{touchette2005legendre}. 
\begin{align}
   0 &=  H\left(\mathfrak{p}(n),\tilde{q}(n)+\frac{\Delta \tilde{q}(n)}{2} \right)  \nonumber \\
    \dfrac{\Delta \tilde{q}(n)}{\Delta t} 
    &= 
    \pdv{H}{p} \bigg|_{\left(\mathfrak{p}(n),\tilde{q}(n)+\frac{\Delta \tilde{q}(n)}{2}\right)}
    \label{eq:LiftSolve}
\end{align}
or equivalently 
\begin{align}
&\mathfrak{p}(n), \Delta t(n) =      \arg \max_p \min_{\Delta t} \nonumber\\
&\left( p\cdot \Delta \tilde{q}(n) - \left(H\left(p,\tilde{q}(n)+\frac{\Delta \tilde{q}(n)}{2}\right)-0\right)\Delta t\right) \label{eq:LiftOptimization} 
\end{align}
The $D+1$ values are assigned so as to simultaneously satisfy the $H(p,q)=0$ constraint ($1$ equation) and the Legendre transform condition ($D$ equations), as shown in \ref{eq:LiftSolve}. In fact, the two problems can equivalently be assimilated in a single optimization problem, as shown in Eq.\ \ref{eq:LiftOptimization}, where $\Delta t$ plays the role of a Lagrange-multiplier. Since the Hamiltonian is convex in momentum, the objective function is concave in $p$, so a unique $\max$ exists. At the optimal momentum assignment, the gradient of $H$ in $p$ and $\,dq$ are in the same direction, and the magnitude of the two vectors is made equal via $\Delta t$.

Finally, we appropriately assign the momentum and time assignment along the configuration curve as shown below and obtain a discrete phase space trajectory.
\begin{align}
    t(1) &= 0 \nonumber \\
    t(n) &= \sum_{m=1}^{n-1} \Delta t(m) \nonumber\\
    \tilde{p}(1) &= 0 = \tilde{p}(N) \nonumber \\
    \tilde{p}(n) &= \frac{\mathfrak{p}(n-1) + \mathfrak{p}(n)}{2} \nonumber\\
    \tilde{\gamma}(n) &= (\tilde{q},\tilde{p},t)(n) \label{eq:PhaseSpaceLift} 
\end{align}

We call this procedure as the \textit{lift} of a configuration curve to phase space trajectory. Denoting the lift map by $\Lambda$, we define it as
\begin{align}
    \Lambda  &:\tilde{\mathcal{P}}_Q \to \tilde{\mathcal{P}}_{T^*Q} \nonumber\\  
    & \phantom{\mathcal{P}12} \tilde{q} \mapsto \tilde{\gamma} \equiv (\tilde{q},\tilde{p},t) \label{eq:LiftMap} 
\end{align}
We denote the inverse or the projection map from phase space trajectory to configuration space curve by $\Pi$, and define it to be
\begin{align}
    \Pi  &:\tilde{\mathcal{P}}_{T^*Q} \to \tilde{\mathcal{P}}_{Q} \nonumber\\  
    & \phantom{\mathcal{P}1234} \tilde{\gamma} \mapsto \tilde{q}.  \label{eq:ProjectionMap}         
\end{align}

\subparagraph{Comments on implementation}

Eqs. \ref{eq:LiftSolve} can be solved numerically using a nonlinear least-squares problem solver, such as MATLAB's \textit{lsqnonlin}. Eq.\ \ref{eq:LiftOptimization} can be solved using a constrained optimization solver, such as MATLAB's \textit{fmincon}. 

Eq.\ \ref{eq:LiftSolve} can also be solved analytically by first doing a change of coordinates, like the one in Eq.\ \ref{eq:Ham_CRN_mixed}, in which the Hamiltonian function and its derivatives are polynomials rather than consisting of exponential terms.
\begin{align*}
    p &\to  z = e^p \\
    \mathcal{H}(q,z) &= 0\\
    \frac{\Delta q_i}{\Delta q_1} &= \dfrac{\pdv{\mathcal{H}}{z_i}}{\pdv{\mathcal{H}}{z_1}}\frac{z_1}{z_i} \hspace{1em} i\in[2,n]\subset\mathbb{Z}
\end{align*}
This gives $D$ equations for $D$ components of $z$. We can then use this to solve for $\Delta t$
\begin{align*}
    \Delta t &= \Delta q_1 \left(z_1\pdv{\mathcal{H}}{z_1}\right)^{-1}
\end{align*}
It must be noted that numerically solving for roots of polynomials can get costly in high dimensions. Also, the numerical precision decreases as we go from $z$ to $p$ coordinates, and it is the latter that we need for integrating Hamilton's equations of motion. 

Irrespective of the implementation, for calculating momentum and time assignments along the complete trajectory, the problem  \ref{eq:LiftOptimization} or  \ref{eq:LiftSolve} must be solved individually for each segment. Thus, a parallel implementation over multiple cores is natural and can save a lot of computational time. Also, we recommend smoothing the momentum values along the trajectory using a moving average at this step in order to partially reduce noise introduced by discretization and even out the effect of numerical errors. For a concrete MATLAB implementation, see \cite{Gagrani_AFGD-for-CRN-escapes_2022}.

\subsection{Functional gradient - obtaining and filtering}
\label{sec:func_gradient}
\begin{figure*}[htb]
    \centering 
\begin{subfigure}{\textwidth}
  \includegraphics[width=\linewidth]{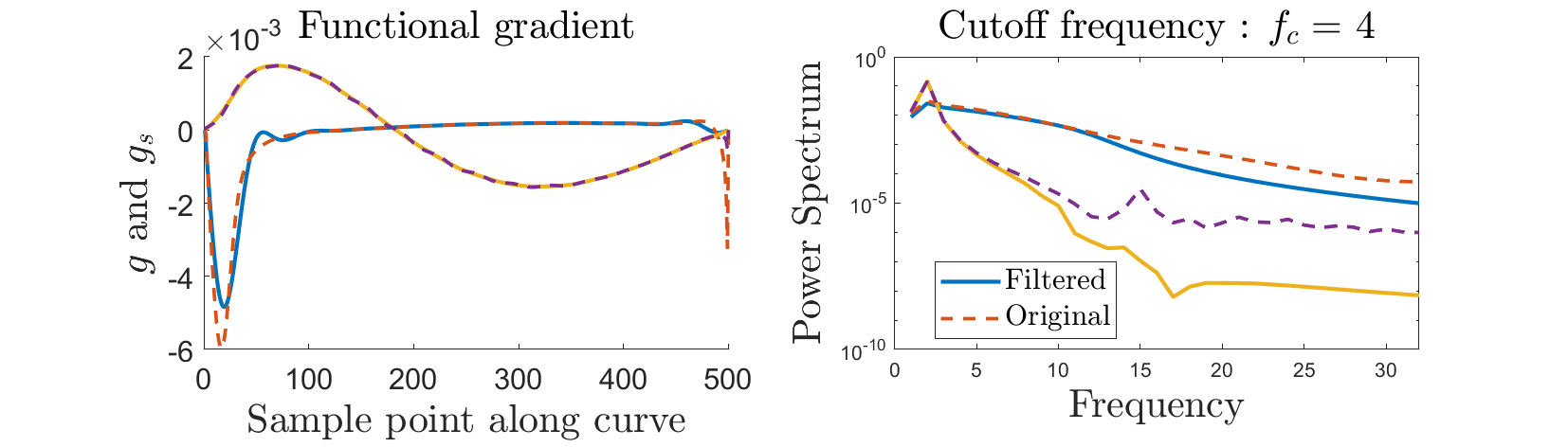}
  \caption{2-Schl{\"{o}}gl model original and filtered functional gradient at the first iteration.}
\end{subfigure}\hfil 
\medskip
\begin{subfigure}{\textwidth}
  \includegraphics[width=\linewidth]{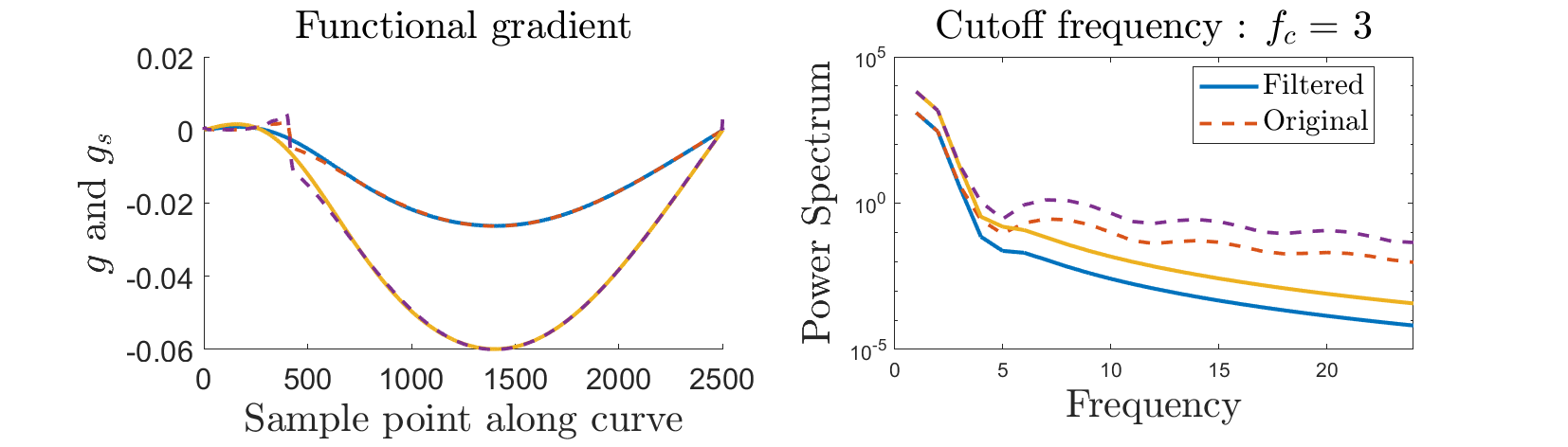}
  \caption{Selkov model original and filtered functional gradient at the first iteration.}
\end{subfigure}
\caption{Extracting functional gradient from phase space trajectory and smoothing it. In all panels, the unfiltered and filtered curves are represented by dashed and solid lines, respectively. } \label{fig:functional_gradient}
\end{figure*}

In the action functional gradient descent (AFGD) algorithm, we perform a functional gradient descent using negative of the functional gradient of the action functional, derived in Eq.\ \ref{eq:FuncGradient}. Denoting the functional gradient by $g$, we get  
\begin{align}
    g
    &=
     -\frac{\delta \mathcal{A}}{\delta q} \nonumber\\
     &= -\left( \dv{p}{t} + \pdv{H}{q}\right) \nonumber
 \end{align}
Thus we define the discrete functional gradient $\tilde{g}$ as
 \begin{align}
    \tilde{g} (m) 
    &=
    \left( 2\frac{\mathfrak{p}(m)-\mathfrak{p}(m-1)}{\Delta t(m)+\Delta t(m-1)} + \pdv{H}{q}\bigg|_{(\tilde{q}(m),\tilde{p}(m))}\right) \label{eq:Disc_func_gradient}
\end{align}

The result of naively performing a functional gradient descent using the above gradient is displayed in Figure \ref{fig:unfiltered_2Schlogl}. It can be seen that any numerical inaccuracies in solving for the momentum will be amplified by taking the time derivative, resulting in self-amplifying noise and instability. In order to smooth the noisy signal thus produced, we employ a filtering routine, that we explain in the next subsection, and obtain a smooth discrete function $\tilde{g}_s$ that we use for updating the algorithm.

\subsection{Filtering and resampling routines}
\label{sec:Filter}

\begin{figure*}[htb]
    \centering 
\begin{subfigure}{\textwidth}
  \includegraphics[width=\linewidth]{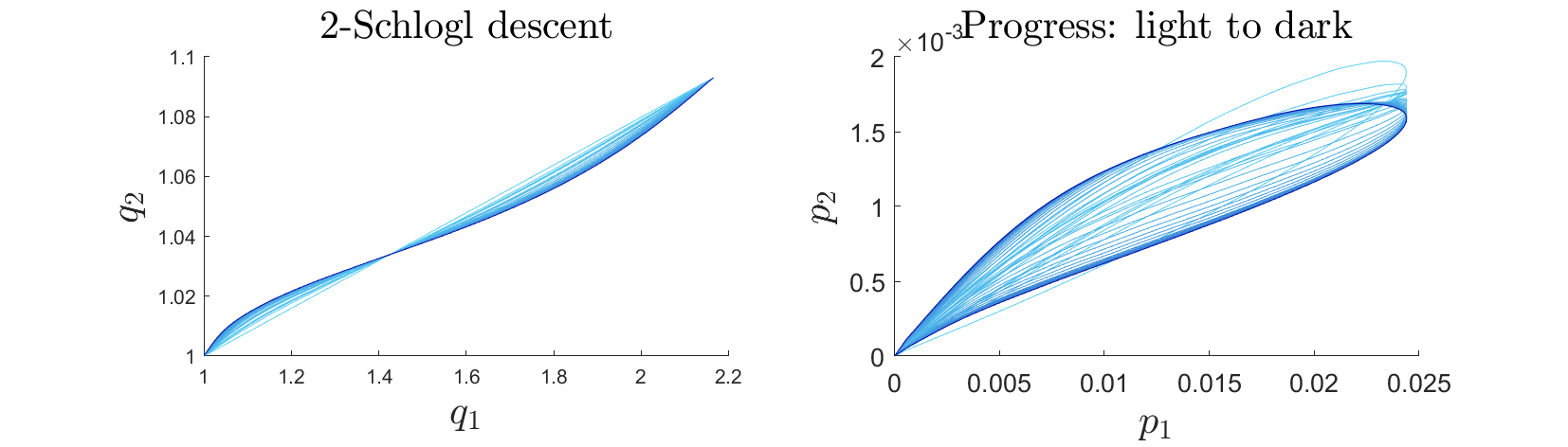}
\end{subfigure}
\medskip
\vspace{-1.2em}
\begin{subfigure}{\textwidth}
  \includegraphics[width=\linewidth]{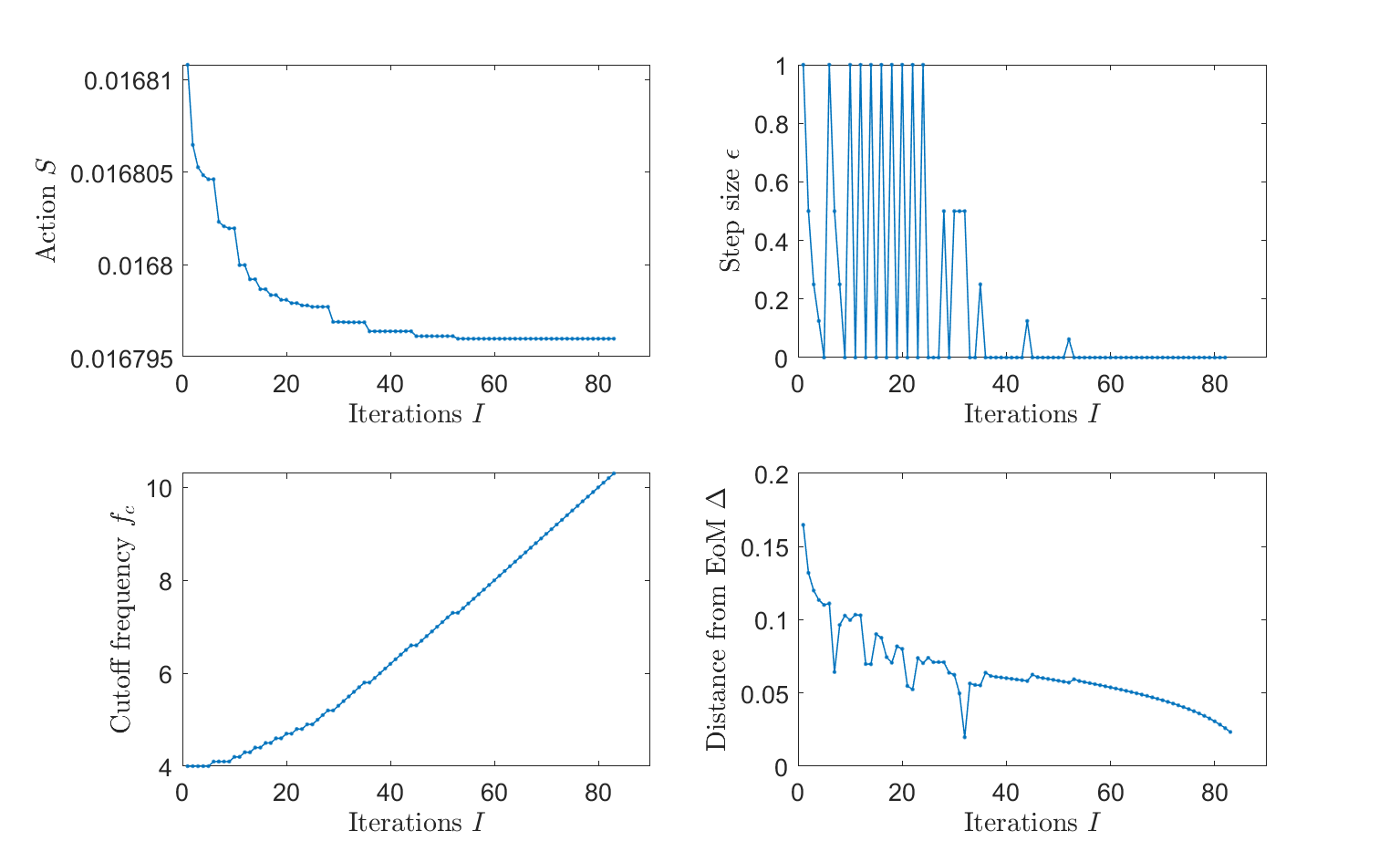}
\end{subfigure}\hfil 
\caption{Descent progress and summary for the 2-Schl{\"{o}}gl model for a trajectory with 500 points.}
 \label{fig:2-Schlogl_descent_progress}
\end{figure*}

\subsubsection{Filtering}
In order to filter a discrete function $\tilde{y}$ which takes value at $N$ points, we employ the following routine. 

First, we subtract a straight line joining the end points from the function, to get a new function $\delta \tilde{y}$ which is identically zero at the end point. 
\begin{align}
    \tilde{y}_\text{st}(n) 
    &= 
    \tilde{y}(N) + \left(1-\frac{n-1}{N-1}\right)\left(\tilde{y}(1) - \tilde{y}(N)\right) \nonumber\\ 
    \delta \tilde{y} 
    &=
    \tilde{y} - \tilde{y}_\text{st} \nonumber 
\end{align}

Next, we define a concatenated function $\delta \tilde{y}_c$ of size $2N-1$ which is obtained by juxtaposing a flipped copy with a negative sign next to the original signal and by removing the duplicate point at $N$. 
\begin{align*}
        \delta \tilde{y}_c(u)
    &= \begin{cases}
        -\delta \tilde{y}(N-u+1) & \text{if } u \in [1,N-1]\\
        \phantom{-}\delta \tilde{y}(u-N+1) & \text{if } u \in [N,2N-1]
    \end{cases}
\end{align*}
Next, we apply a Butterworth lowpass filter $\mathcal{B}(\cdot,f_c)$\footnote{For our implementation we choose Butterworth filter of order 4, but this choice is arbitrary and can be experimented with.}, with some cutoff frequency $f_c$, to the concatenated function and obtain a filtered function $\delta \tilde{y}_\text{cs}$. Notice $\delta \tilde{y}_\text{c}$ is an odd function across the mid-point $N$, and thus $\delta \tilde{y}_\text{cs}$ will also be of the same form. Concatenation before filtering is a common technique employed in signal processing, without which the end-points are not guaranteed to remain at zero after applying the lowpass filter. 
\begin{align*}
    f_c &:= \text{Cutoff frequency} \nonumber \\
   \delta \tilde{y}_\text{cs} &= \mathcal{B}\left(\delta \tilde{y}_c,f_c\right) 
\end{align*}
Finally, we  obtain the desired signal $\tilde{y}_s$ by picking out only the second half of the smoothed concatenated function $ \delta \tilde{y}_\text{cs}$ and adding that back to the straight line from the first step
\begin{align}
       \delta \tilde{y}_s(n) &= \delta \tilde{y}_\text{cs}(n+N-1) \hspace{1em}\text{for }n\in[1,N] \nonumber\\  
   \tilde{y}_s &=  \tilde{y}_\text{st} + \delta \tilde{y}_s.
\end{align}

\subsubsection{Resampling}
To resample the discrete curve $\tilde{q}$, we first need a parametrization. Consider the continuous curve $q$ obtained by interpolation. Then there are two choices of parametrization canonically available to us, namely
\begin{enumerate}
    \item Arc-length parametrization 
    \begin{align}
    s(x) &= \frac{\int_{q_I}^{x} |\,dq| }{\int_{q_I}^{q_F} |\,dq| } \label{eq:space_uni}
\end{align}
    \item Time parametrization
    \begin{align}
    s(x) &= \frac{t(x) }{t(q_F)} \label{eq:time_uni}
\end{align}
\end{enumerate}

Now we obtain a new discrete function $\tilde{q}_u$ uniform in a chosen parametrization $s(x)$ by finding a point $x$ where $s(x)=m/N$, i.e.

\begin{align}
    \tilde{q}_u(m)   
    &=
     \bar{x} \hspace{1em}\text{  such that } s(\bar{x}) =\frac{m}{N} \text{ for } m\in[1,N].
\end{align}

We will call a discrete curve $\tilde{q}_u$ space-uniform or time-uniform sampled if we use the arc-length parametrization in Eq.\ \ref{eq:space_uni} or time parametrization in Eq.\ \ref{eq:time_uni} respectively.

\subsection{Pick step size}
\label{sec:step_size}
Once we have a descent direction $\tilde{g}_s$ obtained by filtering the functional gradient, we need to pick a step size that ensures that the value of the action functional is strictly decreasing. More precisely, we need to pick $\epsilon>0$ such that
\begin{align}
\mathcal{A}[\tilde{\gamma}^\epsilon] 
&<
\mathcal{A}[\tilde{\gamma}],\\
\text{where }\hspace{1em} \tilde{\gamma} 
&=
\Lambda\left(\tilde{q}\right) \nonumber\\
\text{and }\hspace{1em}\tilde{\gamma}^\epsilon 
&=
\Lambda\left(\tilde{q} + \epsilon \tilde{g}_s\right). \nonumber
\end{align}

Ideally, in order to maximize descent, we want to pick the largest $\epsilon>0$ such that the above conditions are satisfied. In practice however, this will require us to solve another optimization problem which can be rather time consuming. Thus, for ease of implementation we employ the \textit{backtracking line search} method in which one starts from a large value for $\epsilon$ and keeps making it smaller until the conditions are satisfied. For an exposition of the method and more sophisticated `line search' algorithms for picking a step size, see \cite{wright1999numerical}.

If the step size is below a threshold $\epsilon<\epsilon_\text{thresh}$ or change in the value of the action is too small $|\mathcal{A}[\tilde{\gamma}^\epsilon] - \mathcal{A}[\tilde{\gamma}]| < \Delta S_\text{thresh}$, then we end the search and assign a step size $\epsilon = 0$. This indicates that the algorithm can not descend further with the given conditions and takes it to the next phase of either updating cutoff frequency or increasing the number of sample points.

\begin{figure*}[htb]
    \centering 
\begin{subfigure}{\textwidth}
  \includegraphics[width=0.9\linewidth]{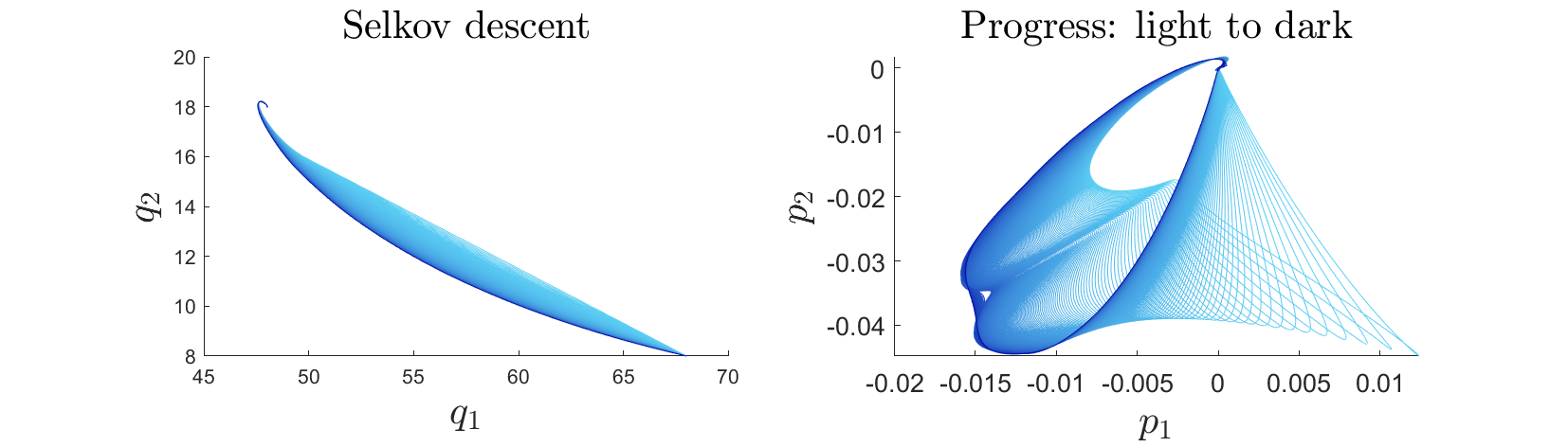}
\end{subfigure}
\medskip
\vspace{-1.2em}
\begin{subfigure}{\textwidth}
  \includegraphics[width=0.9\linewidth]{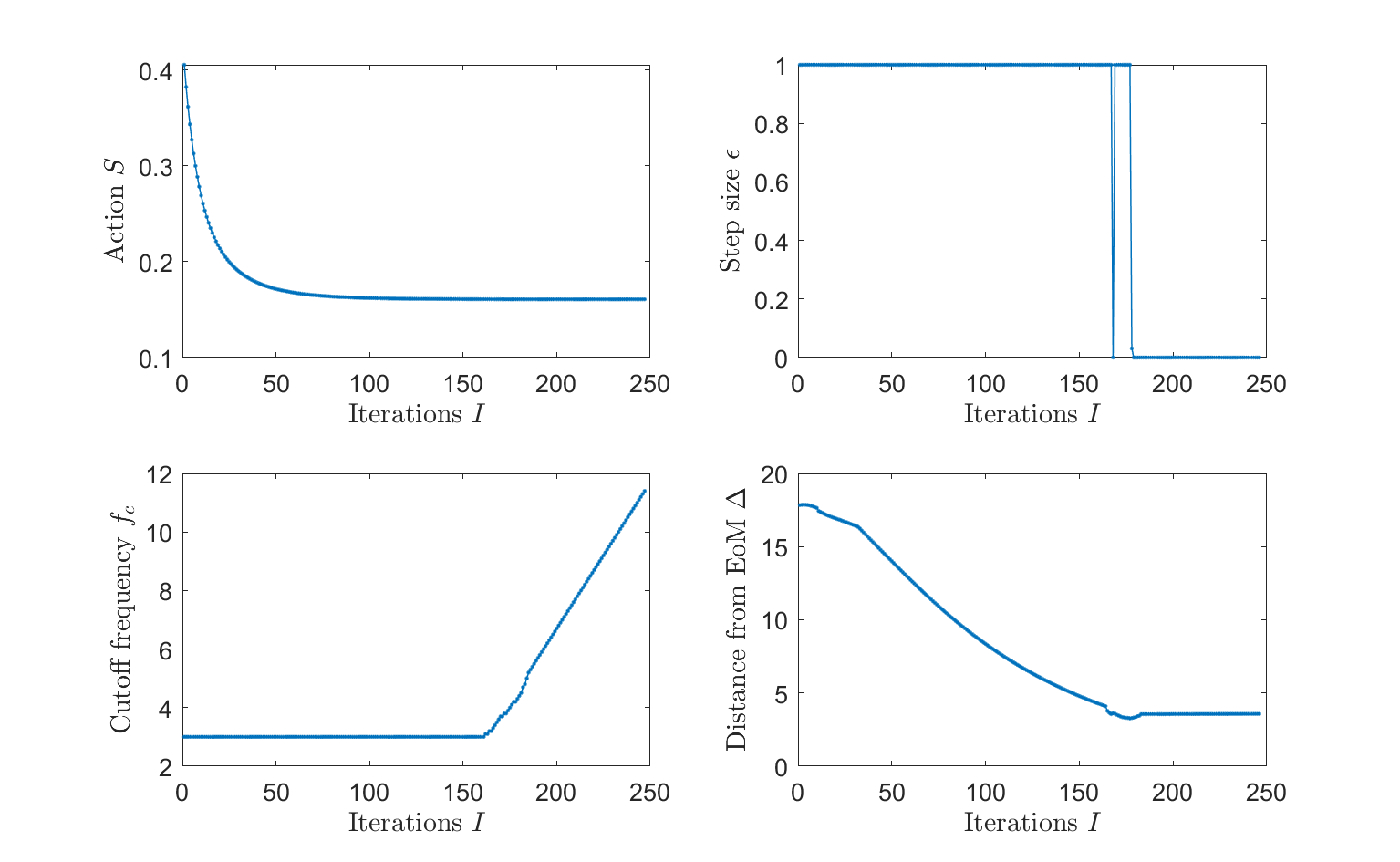}
\end{subfigure}\hfil 
\caption{Descent progress and summary for the Selkov model for a trajectory with 2500 points.}
 \label{fig:Selkov_run_2500}
\end{figure*}

As an illustration of convergence, see the fourth panel in Figures \ref{fig:2-Schlogl_descent_progress}, \ref{fig:Selkov_run_2500} and \ref{fig:Selkov_run_4000}. Notice that in all figures, the step size is identically zero for a few iterations, thus indicating that the algorithm must either be terminated or taken to the next phase.

\subsection{Updating cutoff frequency and increasing sample points during descent}
\label{sec:cutoff_update}

\subsubsection{Updating cutoff frequency}
The low-pass filter on the functional gradient serves the purpose of controlling noise due to discretization and numerical solving. However, it also cuts off meaningful signal in the gradient, especially in the beginning of the descent when the cutoff frequency $f_c$ takes a very small value. To remedy this, at a given cutoff frequency, we let the algorithm converge till it cannot take a further step in the descent direction, and then increase the cutoff frequency to $f_c+\Delta f$. The value of $\Delta f$ can be chosen by experimentation and we choose it to be $0.1$ for our implementation.

We also define a maximum cutoff frequency $f_\text{Max}$ as a considerable fraction of the Nyquist frequency. However, it must be noted that, in practice, the algorithm will stop descending at a much lower cutoff frequency than $f_\text{Max}$. In other words, for no step size will the filtered gradient yield a smaller value of the action functional than its current value. A reason for this might be that the gradient is dominated by noise which is being allowed to go through the pass-band. This is when, for descending further, we employ the \textit{annealing} subroutine.  

For an illustration of how the cutoff frequency updates with with iterations, see the fifth panel in Figures \ref{fig:2-Schlogl_descent_progress}, \ref{fig:Selkov_run_2500} and \ref{fig:Selkov_run_4000}. Notice that in all of these, there are a few iterations where the cutoff frequency remains the same before increasing to a slightly higher value. As explained earlier, during these iterations the algorithm takes a non-zero step size and the value of the action functional steadily decreases.

\subsubsection{Annealing or increasing sample points} 
\label{subsec:anneal}
At a given iteration the curve only has a finite number of points, say $N$. However, in principle, the optimal trajectory that minimizes the action functional is a continuous function i.e. we need the limit $N\to\infty$ to accurately represent it. We get around this problem by first descending with a small value for $N$ until the algorithm converges, and then updating the number of points in the trajectory to $N+\Delta N$.

To illustrate the importance of this subroutine, we consider an application of the AFGD algorithm to the Selkov model. In Figure \ref{fig:Selkov_run_2500} it can be seen that the algorithm converges to some trajectory that exhibits non-differentiability in the momentum coordinates. We know that the true optimal path must be a smooth function, and thus the algorithm has not converged to the true solution. To remedy this, we take the converged trajectory and resample in the time-uniform sampling, defined in Section \ref{sec:Filter}, with $4000$ points. As is evident from Figure \ref{fig:Selkov_run_4000} and the discussion in Section \ref{sec:Results}, this indeed takes the algorithm towards the optimal trajectory.  

In principle, one must employ the annealing subroutine infinitely many times, since the true optimal trajectory is a continuous function. In practice however, since we can monitor the progress of the algorithm by integrating Hamilton's EoM, as we explain below, after a desired accuracy is reached one can terminate the AFGD algorithm and, if needed, use the shooting-method \cite{press2007numerical,dykman1994large}.

\subsection{Integrating Hamilton's equations of motion and convergence criteria}
\label{sec:int_Ham_EoM}

\begin{figure*}[t]
    \centering 
\begin{subfigure}{\textwidth}
  \includegraphics[width=\linewidth]{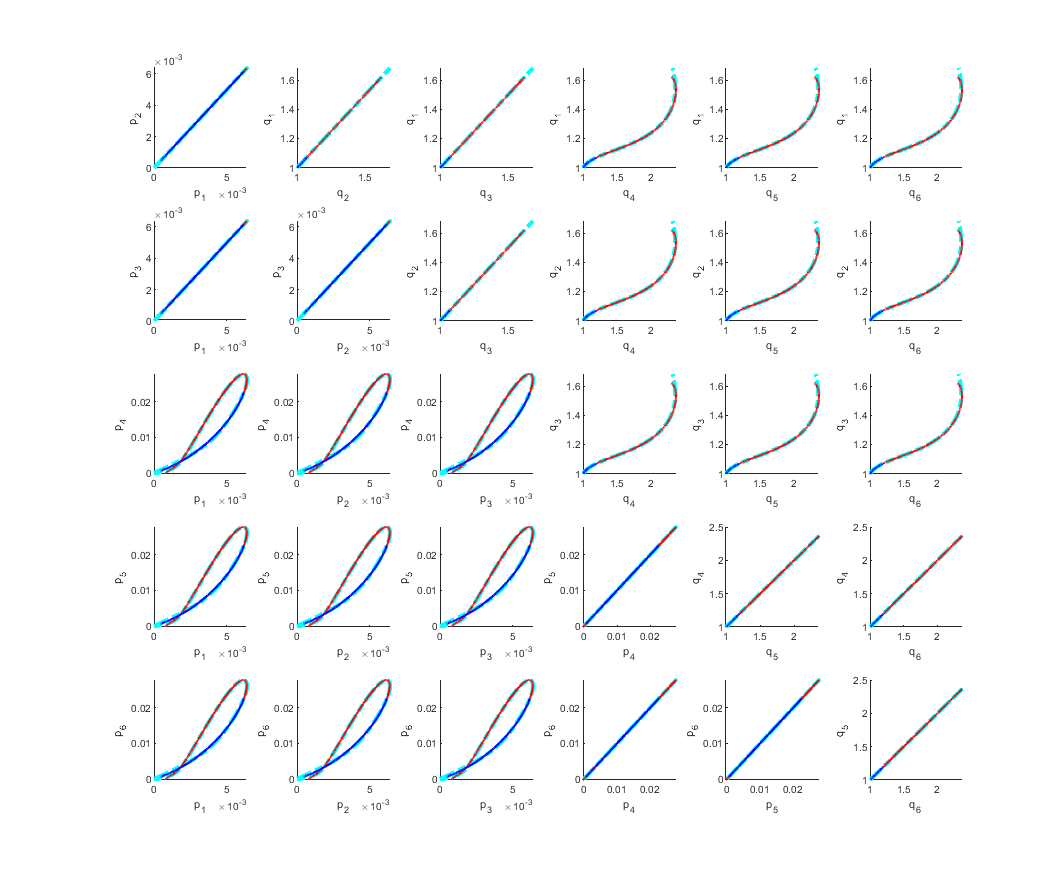}
\end{subfigure}\hfil 
\caption{Integrating Hamilton's equations of motion against AFGD algorithm output for the 6-D Schl{\"{o}}gl model after 300 iterations. The convention is the same as used in Figure \ref{fig:2-Schlogl_HamEoM}, and is omitted for neatness.}
\label{fig:6-Schlogl_HamEoM}
\end{figure*}

The objective of the AFGD algorithm is to converge at the optimal curve constrained at the end points, such that the lifted trajectory is a solution to the Hamilton-Jacobi PDE in Eq.\ \ref{eq:HJ_NEP}. In App.\ \ref{sec:Equiv_HJ_HamEoM}, we prove that any solution of Hamilton-Jacobi must also satisfy Hamilton's equations of motion (EoM), thus the optimal lifted trajectory must also satisfy them. Since Hamilton's EoM is a system of coupled ODEs, it is easier to find their solution starting from an initial condition as opposed to solving HJ equations. We will now use this property of the optimal trajectory to define a `distance' from the true solution at a given iteration and a convergence criterion, as explained below.        

At a given iteration $I$, let us denote the discrete phase space trajectory as $\tilde{\gamma}$ and parametrize it with discrete index $m \in [1,N]$. Now, integrate Hamilton's EoM forwards and backwards starting from each point $m$ until the configuration space distance from the saddle and stable fixed points, respectively starts diverging, having passed through its point of closest approach. We take the minimum Euclidean distance in configuration space near the stable and saddle points, and add them to obtain a $\Delta(m)$ for each point $m$. Finally, we find the minimum over all $m$ to assign a distance of the phase space trajectory, denoted by $\Delta^I =\min_m \Delta(m)$, and use it as a `measure' for distance from optimality. 

Note that since the optimal curve passes through both the saddle and stable fixed point, the distance $\Delta$ of the optimal trajectory must be zero by definition. In practice, however, one can define a $\Delta_\text{thresh}$ such that when $\Delta \leq \Delta_\text{thresh}$, we will declare the algorithm to have converged.

For an illustration of how this measure changes with iterations for the Selkov model, see the bottom right panel in Figures \ref{fig:Selkov_run_2500}, \ref{fig:2-Schlogl_descent_progress} or last three panels in Figure \ref{fig:Selkov_run_4000}. To see how $\Delta$ changes during the descent of the 2-Schl{\"{o}}gl model, see Figure \ref{fig:2-Schlogl_HamEoM}. In some of these figures, it can be seen that $\Delta$ continues to decrease even though the step size is $0$. The reason for this is the additional time-uniform filtering on the curve each time the cutoff frequency is updated. Since distance from Hamilton's EoM is a theoretically rigorous measure of convergence, we can use this to convince ourselves of convergence for higher dimensional models, where alternative verification methods such as the shooting method or Gillespie simulations can be rather expensive. For example of the application of this measure to a 6-D model, see Figure \ref{fig:6-Schlogl_HamEoM}.

\bibliographystyle{unsrt} 
\bibliography{bibliography}
\end{document}